\documentclass[preprint,10pt]{elsarticle}
\pdfoutput=1
\pdfcatalog{/PageLayout /SinglePage}



\makeatletter
\def\ps@pprintTitle{%
   \let\@oddhead\@empty
   \let\@evenhead\@empty
   \def\@oddfoot{\reset@font\hfil\thepage\hfil}
   \let\@evenfoot\@oddfoot
}
\makeatother

\usepackage{titlesec}
\usepackage{xcolor}
\usepackage{enumitem}
\usepackage{hyperref}
\hypersetup{colorlinks=true,urlcolor=black,pdfpagemode=UseNone,pdfstartview=Fit}
\usepackage[T1]{fontenc}\usepackage{lmodern}

\usepackage{mathtools,amsfonts,amssymb}
\usepackage{old-arrows}
\usepackage{caption}
\usepackage{subcaption}
\usepackage{booktabs}
\usepackage{pifont}
\usepackage{placeins}

\definecolor{green}{rgb}{0,0.5,0}

\usepackage{tikz}
\usetikzlibrary{calc}
\usepackage{tkz-euclide}
\usepackage{calc}
\usepackage{tikz,pgfplots}
\pgfplotsset{compat=1.12}
\usetikzlibrary{arrows.meta,3d,matrix,positioning,decorations.markings,math}

\usepackage{xfrac}

\newcommand{\bnlra}{$\scalebox{.925}{\normalsize${}\nleftrightarrow{}$}$}
\newcommand{\bra}{  $\scalebox{.925}{\normalsize${}\rightarrow     {}$}$}
\newcommand{\bla}{  $\scalebox{.925}{\normalsize${}\leftarrow      {}$}$}
\newcommand{\blra}{ $\scalebox{.925}{\normalsize${}\leftrightarrow {}$}$}

\newcommand{\pz}{\phantom{0}}
\newcommand{\pn}{\phantom{-}}
\newcommand{\vpad}{\vphantom{\bigg(}}
\newcommand{\ntriangles}{n_t}

\newcommand{\nbasis}{n_b}
\newcommand{\srcidx}{j}
\newcommand{\testidx}{i}

\newcommand{\aee}{a_{\mathcal{E},\mathcal{E}}} % a_1^\mathcal{E}
\newcommand{\amm}{a_{\mathcal{M},\mathcal{M}}} % a_3^\mathcal{M}
\newcommand{\ame}{a_{\mathcal{M},\mathcal{E}}} % a_2^\mathcal{E}
\newcommand{\aem}{a_{\mathcal{E},\mathcal{M}}} % a_*^\mathcal{M}
\newcommand{\be}{b_\mathcal{E}} % b^\mathcal{E}

\newcommand{\lint}{L^\text{int}}
\newcommand{\lext}{L^\text{ext}}
\newcommand{\aint}{a^\text{int}}
\newcommand{\bint}{b^\text{int}}
\newcommand{\cint}{c^\text{int}}

\newcommand{\aslot}{a^\text{slot}}
\newcommand{\bslot}{b^\text{slot}}
\newcommand{\lslot}{L}
\newcommand{\xslot}{x_w^\text{int}}
\newcommand{\solvec}{\boldsymbol{\mathcal{J}}^h}

\newcommand{\WidestEntry}{$\beta$}%
\newcommand{\SetToWidest}[1]{\makebox[\widthof{\WidestEntry}]{$#1$}}%
\newcommand{\WidestEntryB}{$\xslot-\alpha(\lint-\xi_1)$}%
\newcommand{\SetToWidestB}[1]{\makebox[\widthof{\WidestEntryB}]{$#1$}}%

\DeclareMathOperator{\csch}{csch}

\makeatletter
\tikzoption{canvas is xy plane at z}[]{%
  \def\tikz@plane@origin{\pgfpointxyz{0}{0}{#1}}%
  \def\tikz@plane@x{\pgfpointxyz{1}{0}{#1}}%
  \def\tikz@plane@y{\pgfpointxyz{0}{1}{#1}}%
  \tikz@canvas@is@plane
}
\makeatother

\makeatletter
\tikzoption{canvas is plane}[]{\@setOxy#1}
\def\@setOxy O(#1,#2,#3)x(#4,#5,#6)y(#7,#8,#9)%
  {\def\tikz@plane@origin{\pgfpointxyz{#1}{#2}{#3}}%
   \def\tikz@plane@x{\pgfpointxyz{#4}{#5}{#6}}%
   \def\tikz@plane@y{\pgfpointxyz{#7}{#8}{#9}}%
   \tikz@canvas@is@plane
  }
\makeatother 

\definecolor{orange}{rgb}{1,0.5,0}
\definecolor{green}{rgb}{0,0.5,0}
\definecolor{purple}{rgb}{0.5,0,0.5}
\newcommand{\reviewerTwo}[1]{#1}

\newcommand{\reviewerThree}[1]{#1}
\usepackage[margin=1in]{geometry}
\biboptions{sort&compress}
\begin{document}

\begin{frontmatter}
%\vspace{-3em}
\title{Manufactured Solutions for an Electromagnetic Slot Model}

\author[freno]{Brian A.~Freno}
\ead{bafreno@sandia.gov}
\author[freno]{Neil R.~Matula}
\author[freno]{Robert A.~Pfeiffer}
\author[freno]{Evelyn A.~Dohme}
\author[freno]{Joseph D.~Kotulski}
\address[freno]{Sandia National Laboratories, Albuquerque, NM 87185}
\begin{abstract}
The accurate modeling of electromagnetic penetration is an important topic in computational electromagnetics.  Electromagnetic penetration occurs through intentional or inadvertent openings in an otherwise closed electromagnetic scatterer, which prevent the contents from being fully shielded from external fields.  To efficiently model electromagnetic penetration, aperture or slot models can be used with surface integral equations to solve Maxwell's equations.  A necessary step towards establishing the credibility of these models is to assess the correctness of the implementation of the underlying numerical methods through code verification.  Surface integral equations and slot models yield multiple interacting sources of numerical error and other challenges, which render traditional code-verification approaches ineffective.  In this paper, we provide approaches to separately measure the numerical errors arising from these different error sources for the method-of-moments implementation of the electric-field integral equation with a slot model. We demonstrate the effectiveness of these approaches for a variety of cases.
\end{abstract}


\end{frontmatter}

\section{Introduction}

To model electromagnetic scattering and radiation, Maxwell's equations, together with appropriate boundary conditions, may be formulated as surface integral equations (SIEs).  
The most common SIEs for modeling time-harmonic electromagnetic phenomena are the electric-field integral equation (EFIE), which relates the surface current to the scattered electric field, and the magnetic-field integral equation (MFIE), which relates the surface current to the scattered magnetic field.  
At certain frequencies, the accuracy of the solutions to the EFIE and MFIE deteriorates due to the internal resonances of the scatterer.  Therefore, the combined-field integral equation (CFIE), which is a linear combination of the EFIE and MFIE, is employed to overcome this problem.

These SIEs are typically solved through the method of moments, wherein the surface of the electromagnetic scatterer is discretized using planar or curvilinear mesh elements, and four-dimensional integrals are evaluated over two-dimensional source and test elements.  
These integrals contain a Green's function, which yields singularities when the test and source elements share one or more edges or vertices, and near-singularities when they are otherwise close.  The accurate evaluation of these integrals is an active research topic, with many approaches being developed to address the (near-)singularity for the inner, source-element integral~\cite{graglia_1993,wilton_1984,rao_1982,khayat_2005,fink_2008,khayat_2008,vipiana_2011,vipiana_2012,botha_2013,rivero_2019}, as well as for the outer, test-element integral~\cite{vipiana_2013,polimeridis_2013,wilton_2017,rivero_2019b,freno_em}.

Aperture and slot models are commonly used to model electromagnetic penetration through otherwise closed conducting surfaces.  Practically every material interface yields an opportunity for an intentional or unintentional opening~\cite{butler_1978}.  Through electromagnetic penetration, the exterior and interior electromagnetic fields interact.  Rectangular apertures and slots are some of the most common antennas in practice~\cite[Chap.~8]{balanis_2012}.  \reviewerTwo{The highest fidelity approach to capturing the effects of a slot is to explicitly mesh the slot geometry and include its degrees of freedom in the SIE system.  However, because the slot dimensions are typically very small compared to the overall problem dimensions, a very fine mesh resolution is required in the neighborhood of the slot, which may require a prohibitive computational expense in assembling and solving the associated linear system.  As an alternative, the slot may be modeled by replacing it with a conceptually simpler geometry, such as a system of conducting wires embedded in the surrounding surface.  While the model geometry requires its own mesh elements, the simplified geometry yields a significant reduction in the overall resolution requirements.  Through appropriate boundary conditions, the degrees of freedom associated with the slot model may be coupled with those associated with the surface.  The coupled system may then be solved in either a monolithic or a segregated, iterative fashion.}

The development and validation of aperture and slot models are active research topics~\cite{schelkunoff_1952,cerri_1992,robinson_1998,araneo_2008,hill_2009,pozar_2011,campione_2020,illescas_2023}.  In this work, we focus on the \textit{thick} slot model described in~\cite{warne_1990,warne_1992,warne_1995,johnson_2002}, which captures penetration through an aperture of small electrical depth in a wall modeled with finite thickness.  When a slot connects an otherwise enclosed interior cavity to the exterior of the scatterer, it can be modeled by two thin wires at the apertures that carry magnetic current.  The surface currents on the exterior and interior interact with the respective wire instead of directly with each other, and the two wires interact with each other.  \reviewerTwo{This modeling approach is illustrated in Figure~\ref{fig:slot} and described in detail in Section~\ref{sec:equations}.}

Code verification plays an important role in establishing the credibility of results from computational physics simulations by assessing the correctness of the implementation of the underlying numerical methods~\cite{roache_1998,knupp_2022,oberkampf_2010}.  \reviewerTwo{Differential, integral, and integro-differential equations may be solved exactly only in special cases.  In the general case, the integral and differential operators must be approximated by discrete operators to yield a tractable system of equations.  The difference between the discrete and continuous operators is the truncation error.  As a result of the truncation error, even if the discretized equations are solved exactly, the resulting solution will only approximately satisfy the original continuous equations, introducing a discretization error -- the difference between the solution to the discrete equations and the solution to the continuous equations.} %The discretization of differential, integral, or integro-differential equations incurs some truncation error, and thus the approximate solutions produced from the discretized equations will incur an associated discretization error. 
If the discretization error tends to zero as the discretization is refined, the consistency of the code is verified~\cite{roache_1998}.  This may be taken a step further by examining not only consistency, but the rate at which the error decreases as the discretization is refined, thereby verifying the order of accuracy of the discretization scheme.  The correctness of the numerical-method implementation may then be verified by comparing the expected and observed orders of accuracy obtained from numerous test cases with known solutions.

To measure the discretization error, a known solution is required to compare with the discrete solution.  Exact solutions are generally limited and may not sufficiently exercise the capabilities of the code.  Therefore, manufactured solutions~\cite{roache_2001} are a popular alternative, permitting the construction of arbitrarily complex problems with known solutions.  %\reviewerTwo{To apply the method of manufactured solutions (MMS), a function with certain desirable properties is selected and substituted directly into the governing equations to yield a residual term.  If this term is added to the original governing equations as a source, the result is a new system that captures the character of the original, but for which an exact solution is known.  This source term may be trivially implemented as a run-time option within the source code, allowing the order of accuracy to be easily assessed for the manufactured problem, while still permitting the code to be used for general problems.}  
Through the method of manufactured solutions (MMS), a solution is manufactured and substituted directly into the governing equations to yield a residual term, which is added as a source term to the governing equations. \reviewerTwo{These modified equations are consequently satisfied by the manufactured solution, allowing the discretization error to be evaluated.}

For code verification, integral equations yield an additional challenge.  While analytical differentiation is straightforward, analytical integration is not always possible.  Therefore, the residual source term arising from the manufactured solution may not be representable in closed form, and its implementation may incur its own numerical errors.  Furthermore, for the EFIE, MFIE, and CFIE, the aforementioned (nearly) singular integrals can further complicate the numerical evaluation of the source term.  Therefore, many of the benefits associated with MMS are lost when applied straightforwardly to these integral equations.

There are many examples of code verification in the literature for different computational physics disciplines.  These disciplines include aerodynamics~\cite{nishikawa_2022}, fluid dynamics~\cite{roy_2004,bond_2007,veluri_2010,oliver_2012,eca_2016,hennink_2021,freno_2021}, solid mechanics~\cite{chamberland_2010}, fluid--structure interaction~\cite{etienne_2012,bukac_2023}, heat transfer in fluid--solid interaction~\cite{veeraragavan_2016}, multiphase flows~\cite{brady_2012,lovato_2021}, radiation hydrodynamics~\cite{mcclarren_2008}, plasma physics~\cite{riva_2017,tranquilli_2022,rueda_2023,rudi_2024}, electrodynamics~\cite{amormartin_2021}, and ablation~\cite{amar_2008,amar_2009,amar_2011,freno_ablation,freno_ablation_2022}.  
For electromagnetic SIEs, code-verification activities that employ manufactured solutions have been described for the EFIE~\cite{marchand_2013,marchand_2014,freno_em_mms_2020,freno_em_mms_quad_2021}, MFIE~\cite{freno_mfie_2022}, and CFIE~\cite{freno_cfie_2023}.

As described in~\cite{freno_em_mms_2020,freno_mfie_2022,freno_cfie_2023}, SIEs incur numerical error due to curved surfaces being approximated by planar elements (domain-discretization error), the solution being approximated as a linear combination of a finite number of basis functions (solution-discretization error), and the approximate evaluation of integrals using quadrature rules (numerical-integration error).

For the EFIE, Marchand et al.~\cite{marchand_2013,marchand_2014} compute the MMS source term using additional quadrature points.  Freno et al.~\cite{freno_em_mms_2020} manufacture the Green's function, permitting the numerical-integration error to be eliminated and the solution-discretization error to be isolated.  Freno et al.~\cite{freno_em_mms_quad_2021} also provide approaches to isolate the numerical-integration error.
For the MFIE and CFIE, Freno and Matula~\cite{freno_mfie_2022,freno_cfie_2023} isolate and measure the solution-discretization error and numerical-integration error.

\reviewerThree{It should be noted that other methods exist for generating exact solutions to scattering problems, such as constructing a field solution using a superposition of elementary current sources.  However, our MMS approaches have the distinct advantage of being able to yield an analytical solution for the surface current without singularities.  This property enables us to assess the linear system assembly and solution prior to postprocessing.  In the course of doing so, we can isolate and measure the solution-discretization error and numerical-integration error and detect problems that lead to suboptimal convergence rates.}

In this paper, we present code-verification techniques for the method-of-moments implementation of the EFIE with a thick slot model that isolate and measure the solution-discretization error and numerical-integration error.  We manufacture the electric surface current density, which yields a source term that we can treat as a manufactured incident field.  \reviewerTwo{Given the manufactured electric surface current, we can analytically solve the continuous slot equation to obtain an exact, known solution for the magnetic current.  As a result, unlike the EFIE, the slot equation does not require the MMS source term.}
For curved surfaces, the domain-discretization error cannot be completely isolated or eliminated, but methods are presented in~\cite{freno_mfie_2022} to account for it in the MFIE.  These methods can be applied to the other SIEs straightforwardly.  In this work, we avoid the domain-discretization error by considering only planar surfaces.  
As in~\cite{freno_em_mms_2020,freno_mfie_2022}, we isolate the solution-discretization error by manufacturing the Green's function in terms of even powers of the distance between the test and source points.  With this form, we can evaluate the integrals exactly, thereby avoiding numerical-integration error.  However, on each surface, the interaction between the wire and the surface introduces a line discontinuity, which contaminates convergence studies.  We present an approach to mitigate this problem and decouple the discretization errors.
We isolate the numerical-integration error on both sides of the equations by canceling the influence of the basis functions.  This approach has been demonstrated for the MFIE~\cite{freno_mfie_2022} and CFIE~\cite{freno_cfie_2023}.  
With \reviewerTwo{the solution-discretization error and numerical-integration error} isolated, we perform convergence studies for different manufactured Green's functions and slot depths, with and without discontinuities and coding errors.

This paper is organized as follows.  In Section~\ref{sec:equations}, we describe the EFIE and thick slot model.  In Section~\ref{sec:discretization}, we provide the details for their discretization. In Section~\ref{sec:mms}, we describe the challenges of using MMS with these equations, as well as our approaches to mitigating them.  In Section~\ref{sec:results}, we demonstrate the effectiveness of our approaches for several different configurations.  In Section~\ref{sec:conclusions}, we summarize this work.

\section{Governing Equations}
\label{sec:equations}

\begin{figure}%[!t]
\centering
\input{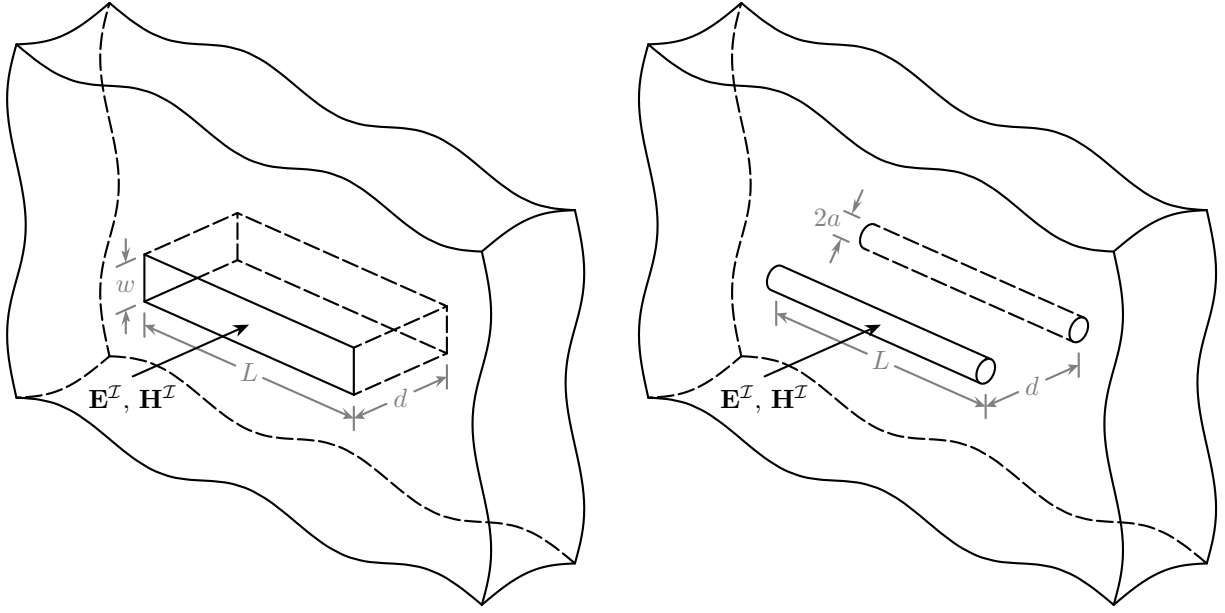}
\caption{Left: an excerpt of an exterior surface of an otherwise closed scatterer, which contains a slot.  The slot connects the exterior domain to an interior cavity.  Right: the slot is replaced with two wires located at the apertures of the slot.}
\vskip-\dp\strutbox
\label{fig:slot}
\end{figure}

We consider an electromagnetic scatterer that encloses a cavity.  The exterior of the scatterer is connected to the cavity through a narrow, rectangularly prismatic slot, as shown in Figure~\ref{fig:slot}.  The length of the slot $\lslot$ is much greater than the width $w$ and depth $d$ of the slot.  The scatterer is modeled as a closed surface with a finite thickness using the electric-field integral equation for a good, but imperfect, electric conductor.  The slot is modeled by circularly cylindrical wires at its apertures on the exterior and interior surfaces using transmission line theory.  These wires have a small but finite radius $a$ and carry magnetic current.  Through this approach, the interior and exterior surfaces do not directly interact.  Instead, the magnetic current on the exterior wire interacts with the electric current on the exterior surface and the magnetic current on the interior wire interacts with the electric current on the interior surface.  The wires additionally interact with each other.  In this paper, we consider a thick slot model, for which the depth is considered electrically small, such that the magnetic current on the wires is equal and flows in opposite directions.

\reviewerTwo{It is worth noting that these approximations are due to the slot model, rather than the code-verification process.  The code-verification approaches presented in Section~\ref{sec:mms} are tailored to this slot model.}

%The slot is modeled by two apertures, which are, in turn, modeled as thin filaments of magnetic current on the outer surface .

%The slot is modeled by two apertures of length $\lslot$, which are, in turn, modeled as thin filaments of magnetic current.  Along each filament at position $s$, there exists a contour $C(s)$ around that side of the surface that bounds a local region, for which the voltage is approximately constant.  

%===============================================================================
\subsection{The Electric-Field Integral Equation} %=============================
%===============================================================================

The EFIE is evaluated separately on the exterior and interior surfaces of the scatterer.  In time-harmonic form, the scattered electric field $\mathbf{E}^\mathcal{S}$ due to induced electric and magnetic surface currents on a scatterer can be computed by~\cite[Chap.~6]{balanis_2012}
\begin{align}
\mathbf{E}^\mathcal{S}(\mathbf{x}) &{}= -\biggl(j\omega\mathbf{A}(\mathbf{x})+\nabla\Phi(\mathbf{x}) + \frac{1}{\epsilon}\nabla\times\mathbf{F}(\mathbf{x})\biggr), 
\label{eq:Es}
\end{align}
where the magnetic vector potential $\mathbf{A}$ is defined by
\begin{align}
\mathbf{A}(\mathbf{x})= \mu \int_{S'} \mathbf{J} (\mathbf{x}')G(\mathbf{x},\mathbf{x}')dS',
\label{eq:A}
\end{align}
the electric vector potential $\mathbf{F}$ is defined by
\begin{align}
\mathbf{F}(\mathbf{x})= \epsilon \int_{S'} \mathbf{M} (\mathbf{x}')G(\mathbf{x},\mathbf{x}')dS',
\label{eq:F}
\end{align}
and, by employing the Lorenz gauge condition and the continuity equation, the electric scalar potential $\Phi$ is defined by 
\begin{align}
\Phi(\mathbf{x})=  \frac{j}{\epsilon\omega} \int_{S'} \nabla'\cdot\mathbf{J}(\mathbf{x}')G(\mathbf{x},\mathbf{x}')dS'.
\label{eq:Phi}
\end{align}
In~\eqref{eq:A}--\eqref{eq:Phi}, the integration domain $S'=S$ is the %closed 
exterior or interior surface of a scatterer, and the prime notation is introduced here to distinguish the source and test integration domains later in this section.  Additionally, $\mathbf{J}$ is the electric surface current density, $\mathbf{M}$ is the magnetic surface current density, $\mu$ and $\epsilon$ are the permeability and permittivity of the surrounding medium, and $G$ is the Green's function 
\begin{align}
G(\mathbf{x},\mathbf{x}') = \frac{e^{-jkR}}{4\pi R},
\label{eq:G}
\end{align}
where $R=\|\mathbf{R}\|_2$, $\mathbf{R}=\mathbf{x}-\mathbf{x}'$, and $k=\omega\sqrt{\mu\epsilon}$ is the wavenumber.  

The total electric field $\mathbf{E}$ is the sum of $\mathbf{E}^\mathcal{S}$ and the incident electric field $\mathbf{E}^\mathcal{I}$, which induces $\mathbf{J}$ and $\mathbf{M}$.  For an electric conductor with large but finite conductivity, the tangential component of the total electric field on $S$ is equal to the product of $\mathbf{J}$ and the resistive surface impedance $Z_s$~\cite[Chap.~1]{balanis_2012}, such that %~\cite[Chap.~3]{ramo_1993}, 
\begin{align}
\mathbf{n}\times\mathbf{E}=\mathbf{n}\times(\mathbf{E}^\mathcal{S}+\mathbf{E}^\mathcal{I})=Z_s\mathbf{n}\times\mathbf{J},
\label{eq:efie_bc}
\end{align}
where $\mathbf{n}$ is the unit vector normal to $S$.  Inserting~\eqref{eq:Es} into~\eqref{eq:efie_bc},
\begin{align}
\mathbf{n}\times\biggl(j\omega\mathbf{A}+\nabla\Phi + \frac{1}{\epsilon}\nabla\times\mathbf{F}+Z_s\mathbf{J}\biggr)=\mathbf{n}\times\mathbf{E}^\mathcal{I}.
\label{eq:efie_bc_2}
\end{align}
From~\eqref{eq:F} and noting that $\nabla\times\bigl[\mathbf{M}(\mathbf{x}')G(\mathbf{x},\mathbf{x}')\bigr]=\nabla G(\mathbf{x},\mathbf{x}')\times\mathbf{M}(\mathbf{x}')$ and $\nabla G(\mathbf{x},\mathbf{x}') = -\nabla' G(\mathbf{x},\mathbf{x}')$, in~\eqref{eq:efie_bc_2},
\begin{align*}
\frac{1}{\epsilon}\nabla\times\mathbf{F}(\mathbf{x}) = \int_{S'} \mathbf{M}(\mathbf{x}')\times\nabla' G(\mathbf{x},\mathbf{x}')dS'
%\label{eq:Hs2}
\end{align*}
when $\mathbf{x}$ is just outside of $S$.  Therefore, in~\eqref{eq:efie_bc_2} at $S$,
\begin{align}
\mathbf{n}\times\biggl(\frac{1}{\epsilon}\nabla\times\mathbf{F}(\mathbf{x})\biggr) = \lim_{\mathbf{x}\to S}\mathbf{n}\times \int_{S'} \mathbf{M}(\mathbf{x}')\times\nabla' G(\mathbf{x},\mathbf{x}')dS' = \frac{1}{2}\mathbf{M} + \mathbf{n}\times \int_{S'} \mathbf{M}(\mathbf{x}')\times\nabla' G(\mathbf{x},\mathbf{x}')dS',
\label{eq:limit}
\end{align}
where the final term is evaluated through principal value integration.  Inserting~\eqref{eq:A}, \eqref{eq:Phi}, and~\eqref{eq:limit} into~\eqref{eq:efie_bc_2} yields
\begin{align}
\mathbf{n}\times\biggl(j\omega\mu \int_{S'} \mathbf{J} (\mathbf{x}')G(\mathbf{x},\mathbf{x}')dS'
{}+{}
\frac{j}{\epsilon\omega} \int_{S'} \nabla'\cdot\mathbf{J}(\mathbf{x}')\nabla G(\mathbf{x},\mathbf{x}')dS'
& \nonumber \\
{}+{}
\int_{S'} \mathbf{M}(\mathbf{x}')\times\nabla' G(\mathbf{x},\mathbf{x}')dS'+Z_s\mathbf{J}\biggr)+\frac{1}{2}\mathbf{M}
&{}=\mathbf{n}\times\mathbf{E}^\mathcal{I}.
\label{eq:efie_bc_3}
\end{align}

We project~\eqref{eq:efie_bc_3} onto an appropriate space $\mathbb{V}$ containing vector fields that are tangent to $S$.  Noting that 
\begin{align}
-\bar{\mathbf{v}}\cdot\mathbf{n}\times(\mathbf{n}\times\mathbf{u})=\bar{\mathbf{v}}\cdot\mathbf{u}
\label{eq:identity}
\end{align}
and integrating by parts yields the variational form of the EFIE: find $\mathbf{J},\mathbf{M}\in\mathbb{V}$, such that
\begin{align}
j\omega\mu\int_S \bar{\mathbf{v}}(\mathbf{x})\cdot \int_{S'} \mathbf{J}(\mathbf{x}')G(\mathbf{x},\mathbf{x}')dS' dS 
{}-\frac{j}{\epsilon\omega}\int_S \nabla\cdot\bar{\mathbf{v}}(\mathbf{x}) \int_{S'} \nabla'\cdot \mathbf{J}(\mathbf{x}')G(\mathbf{x},\mathbf{x}')dS' dS 
& \nonumber\\ 
{}-\frac{1}{2}\int_S \bar{\mathbf{v}}\cdot (\mathbf{n}\times \mathbf{M}) dS
{}+\int_S \bar{\mathbf{v}}(\mathbf{x})\cdot  \int_{S'} \mathbf{M}(\mathbf{x}')\times\nabla' G(\mathbf{x},\mathbf{x}')dS' dS
{}+Z_s\int_S \bar{\mathbf{v}}\cdot \mathbf{J} dS
&{}=
\int_S \bar{\mathbf{v}} \cdot \mathbf{E}^\mathcal{I} dS
\label{eq:efie_variational}
\end{align} 
for all $\mathbf{v}\in\mathbb{V}$, where the overbar denotes complex conjugation.  

%===============================================================================
%\subsection{Equation for the Surfaces} %========================================
%===============================================================================

The magnetic current is limited to the vicinity of the slot aperture.  Along the length of the slot, at position $s\in[0,\,\lslot]$, there exists a contour $C(s)$ around that side of the surface that bounds a local region, for which the voltage is approximately constant.  
Each wire used to model the slot carries a filament line-source magnetic current $\mathbf{I}_m(s)=I_m(s)\mathbf{s}$, where $\mathbf{s}$ denotes the direction of the wire. $\mathbf{I}_m$ is related to $\mathbf{M}$ by~\cite{warne_1990}
\begin{align}
\mathbf{I}_m(s) = 2 \int_{C(s)} \hspace{-.7em}\mathbf{M}(\mathbf{x}) d\ell.
\label{eq:i_m}
\end{align}
Denoting the surface of the local region as $S_\text{local}$, and using~\eqref{eq:i_m},
\begin{align*}
\int_S \mathbf{M}(\mathbf{x}) dS = \int_{S_\text{local}} \hspace{-1em}\mathbf{M}(\mathbf{x}) dS = \int_0^\lslot \int_{C(s)}\hspace{-.7em} \mathbf{M}(\mathbf{x}) d\ell ds = \frac{1}{2}\int_0^\lslot \mathbf{I}_m(s) ds.
\end{align*}

Assuming the local region is small, such that there is no variation with respect to the contour coordinate $\ell$~\cite{warne_1990}, in~\eqref{eq:efie_variational}, we can write
\begin{align}
\frac{1}{2}\int_S \bar{\mathbf{v}}\cdot (\mathbf{n}\times \mathbf{M}) dS = \frac{1}{4}\int_0^\lslot \bar{\mathbf{v}}\cdot (\mathbf{n}\times \mathbf{I}_m) ds.
\label{eq:mag_current_1}
\end{align}

For the other term with a magnetic current contribution in~\eqref{eq:efie_variational}, we model each wire as having a small but finite radius $a$, such that $\bar{\mathbf{I}}_m = 2\pi a \mathbf{M}$~\cite[Chap.~12]{balanis_2012}, where $\bar{\mathbf{I}}_m(s)=\bar{I}_m(s)\mathbf{s}$ denotes the conventional magnetic filament current, and $\mathbf{I}_m=2\bar{\mathbf{I}}_m$ due to the reflection resulting from a magnetic current in the presence of a conducting planar surface~\cite[Chap.~7]{balanis_2012}.
%due to the magnetic current in the presence of a conducting surface.
%
In our problem, where the slot is in a finite body, this reflection does not apply, but for consistency with~\cite{warne_1990,warne_1992,warne_1995,johnson_2002}, we still use this convention.
%
%The radius is obtained through a conformal mapping using the width and depth of the slot~\cite{warne_1995}.
The radius $a$ is an effective radius obtained through a conformal mapping using the width and depth of the slot~\cite{warne_1995}.
Therefore, in~\eqref{eq:efie_variational},
\begin{align}
\int_S \bar{\mathbf{v}}(\mathbf{x})\cdot  \int_{S'} \mathbf{M}(\mathbf{x}')\times\nabla' G(\mathbf{x},\mathbf{x}')dS' dS
=
\frac{1}{4\pi}\int_S \bar{\mathbf{v}}(\mathbf{x})\cdot  \int_0^\lslot \mathbf{I}_m(s')\times\int_0^{2\pi} \nabla' G(\mathbf{x},\mathbf{x}')d\phi' ds' dS,
\label{eq:mag_current_2}
\end{align}
where
\begin{align}
\nabla' G(\mathbf{x},\mathbf{x}') 
=
\frac{\partial G}{\partial R}\biggl(\frac{\partial R}{\partial \rho'}\boldsymbol{\rho}' + \frac{1}{\rho'}\frac{\partial R}{\partial \phi'}\boldsymbol{\phi}' + \frac{\partial R}{\partial s'}\mathbf{s}'\biggr),
\label{eq:grad_G}
\end{align}
and
\begin{align}
R=\sqrt{\rho^2 + \rho'^2 - 2\rho \rho'\cos(\phi-\phi')+(s-s')^2}.
\label{eq:R}
\end{align}
In~\eqref{eq:grad_G} and ~\eqref{eq:R}, $\rho$ is the radial distance from the wire axis, and $\phi$ is the azimuthal angle.  Because the source integral is evaluated on the wire, $\rho'=a$.

%The inner integral of this term is evaluated using cylindrical coordinates, with %.  For $G(\mathbf{x},\mathbf{x}')$, 
%

%
%\begin{align}
%R=\sqrt{\rho^2 + a^2 - 2\rho a\cos(\phi-\phi')+(s-s')^2},
%\label{eq:R}
%\end{align}
%
%where 

%$s$ and $s'$ are the positions along the filament of $\mathbf{x}$ and $\mathbf{x}'$, 
%$\rho$ is the radial distance of $\mathbf{x}$ from the filament center confined to a plane normal to the filament direction, and 
%$\phi$ and $\phi'$ are the angles of $\mathbf{x}$ and $\mathbf{x}'$ along that plane.  
%
%To account for axisymmetry, we set $\phi=0$ in~\eqref{eq:R}.
%
%Therefore, 
%

%

%
%Additionally, in~\eqref{eq:mag_current_2},
%%
%\begin{align*}
%\int_0^\lslot \mathbf{I}_m(s')\times\int_0^{2\pi} \nabla' G(\mathbf{x},\mathbf{x}')d\phi' ds'
%&{}= 
%\int_0^\lslot I_m(s')\int_0^{2\pi} \frac{\partial G}{\partial R}\biggl(-\frac{\rho \sin\phi'}{R}\boldsymbol{\rho}' + \frac{a-\rho\cos\phi'}{R}\boldsymbol{\phi}' \biggr)d\phi' ds'
%%\\
%%&{}= 
%%a\int_0^\lslot I_m(s')\int_0^{2\pi} \frac{\partial G}{\partial R}\frac{1}{R}(-\sin\phi'\mathbf{x}'+\cos\phi'\mathbf{y}') d\phi' ds'.
%\end{align*}

With~\eqref{eq:mag_current_1} and~\eqref{eq:mag_current_2}, \eqref{eq:efie_variational} is written as: find $\mathbf{J}\in\mathbb{V}$ and $\mathbf{I}_m\in\mathbb{V}^m$, such that
\begin{align}
j\omega\mu\int_S \bar{\mathbf{v}}(\mathbf{x})\cdot \int_{S'} \mathbf{J}(\mathbf{x}')G(\mathbf{x},\mathbf{x}')dS' dS 
{}-\frac{j}{\epsilon\omega}\int_S \nabla\cdot\bar{\mathbf{v}}(\mathbf{x}) \int_{S'} \nabla'\cdot \mathbf{J}(\mathbf{x}')G(\mathbf{x},\mathbf{x}')dS' dS 
& \nonumber\\ 
{}-\frac{1}{4}\int_0^\lslot \bar{\mathbf{v}}\cdot \bigl(\mathbf{n}\times \mathbf{I}_m\bigr) ds
{}+\frac{1}{4\pi}\int_S \bar{\mathbf{v}}(\mathbf{x})\cdot  \int_0^\lslot \mathbf{I}_m(s')\times\int_0^{2\pi} \nabla' G(\mathbf{x},\mathbf{x}')d\phi' ds' dS
{}+Z_s\int_S \bar{\mathbf{v}}\cdot \mathbf{J} dS
&{}=
\int_S \bar{\mathbf{v}} \cdot \mathbf{E}^\mathcal{I} dS
\label{eq:efie_variational_2}
\end{align} 
for all $\mathbf{v}\in\mathbb{V}$, where $\mathbb{V}^m$ is an appropriate space containing vector fields that are located on and tangent to the filament and vanish at $s=0$ and $s=\lslot$.
%
%For the inner surface, in the absence of sources, $\mathbf{E}^\mathcal{I}=\mathbf{0}$ in~\eqref{eq:efie_variational_2}.
%
We can write~\eqref{eq:efie_variational_2} more succinctly as
\begin{align}
\aee(\mathbf{J},\mathbf{v}) + \aem(\mathbf{I}_m,\mathbf{v}) = \be\bigl(\mathbf{E}^\mathcal{I}, \mathbf{v}\bigr),
\label{eq:efie_sesquilinear}
\end{align}
where the sesquilinear forms and inner product are defined by
\begin{align}
\aee(\mathbf{u},\mathbf{v}) ={}& j\omega\mu \int_S \bar{\mathbf{v}}(\mathbf{x})\cdot\int_{S'} \mathbf{u}(\mathbf{x}')G(\mathbf{x},\mathbf{x}')dS'dS -\frac{j}{\epsilon\omega} \int_S \nabla\cdot\bar{\mathbf{v}}(\mathbf{x})\int_{S'} \nabla'\cdot\mathbf{u}(\mathbf{x}')G(\mathbf{x},\mathbf{x}')dS' dS \nonumber \\ &+ Z_s\int_S \bar{\mathbf{v}}(\mathbf{x})\cdot \mathbf{u}(\mathbf{x}) dS, \label{eq:aee}
\\
%a_1^\mathcal{M}(\mathbf{u},\mathbf{v}) ={}& -\frac{1}{4}\int_0^\lslot \bar{\mathbf{v}}(\mathbf{x})\cdot \bigl(\mathbf{n}(\mathbf{x})\times \mathbf{u}(\mathbf{x})\bigr) ds,
%\\
%a_2^\mathcal{M}(\mathbf{u},\mathbf{v}) ={}& {\color{red}\int_S \bar{\mathbf{v}}(\mathbf{x})\cdot  \int_0^\lslot \mathbf{u}(s')\times\frac{1}{2\pi}\int_0^{2\pi} \nabla' G(\mathbf{x},\mathbf{x}')d\phi' ds' dS},
%\\
\aem(\mathbf{u},\mathbf{v}) ={}& -\frac{1}{4}\int_0^\lslot \bar{\mathbf{v}}(\mathbf{x})\cdot \bigl[\mathbf{n}(\mathbf{x})\times \mathbf{u}(s)\bigr] ds + \frac{1}{4\pi}\int_S \bar{\mathbf{v}}(\mathbf{x})\cdot  \int_0^\lslot \mathbf{u}(s')\times\int_0^{2\pi} \nabla' G(\mathbf{x},\mathbf{x}')d\phi' ds' dS,
\label{eq:aem}
\\
\be(\mathbf{u},\mathbf{v})  ={}& \int_S \bar{\mathbf{v}}(\mathbf{x})\cdot\mathbf{u}(\mathbf{x})  dS. \nonumber%\label{eq:efie_b}
\end{align}

%===============================================================================
\subsection{The Thick Slot Model} %=============================================
%===============================================================================
% Warne_2016: [4] is Warne_1992, [5] is Warne_1990
% Johnson_2002: [2] is Warne_1990, [12] is Warne_1992
% Warne_1992: [2] is Warne_1990

%\Delta \tilde{Y}_L
%\Delta \tilde{Y}_C
Letting $\mathbf{H}$ denote the total magnetic field, the magnetic current along each wire is modeled using transmission line theory~\cite{warne_1990,warne_1992,johnson_2002}:
\begin{align}
\mathbf{s}\cdot\biggl[\mathbf{H} + \frac{1}{4}\biggl(Y_L \frac{d^2}{ds^2} - Y_C\biggr)\mathbf{I}_m\biggr] = 0,
\label{eq:thick_slot}
\end{align}
where $\mathbf{I}_m(0)=\mathbf{I}_m(\lslot)=\mathbf{0}$.
The transmission line parameters are defined by~\cite{warne_1992,johnson_2002}
\begin{align*}
Y_L &{}= \tilde{Y} + \frac{1}{j\omega L_0}, \\
Y_C &{}= j\omega C_0,
\end{align*}
where
\begin{align*}
\tilde{Y} = \frac{2 Z_s}{\omega L_0 (\omega L_0 d - 2j Z_s)}
\end{align*}
represents the effect of the finite conductivity of the metallic slot walls~\cite{johnson_2002}.
$L_0 = \mu_0 w/d$ is the interior inductance per unit length,
$C_0 = \epsilon_0 d/w$ is the interior capacitance per unit length,
$Z_s = (1 + j)R_s$ is the resistive surface impedance of the walls, 
$R_s = \sqrt{\omega\mu/(2\sigma)}$ is the surface resistance,
$\sigma$ is the wall electric conductivity,
$\mu$ is the wall magnetic permeability, and
$\mu_0$ and $\epsilon_0$ are the permeability and permittivity of free space~\cite{warne_1990,warne_1992,johnson_2002}.

Noting that $\mathbf{J}=\mathbf{n}\times\mathbf{H}$~\cite[Chap.~1]{balanis_2012} and using~\eqref{eq:identity},~\eqref{eq:thick_slot} can be written as
\begin{align}
\mathbf{s}\cdot\biggl[\mathbf{J}\times\mathbf{n} + \frac{1}{4}\biggl(Y_L \frac{d^2}{ds^2} - Y_C\biggr)\mathbf{I}_m\biggr] = 0.
\label{eq:thick_slot_2}
\end{align}
We project~\eqref{eq:thick_slot_2} onto $\mathbb{V}^m$ and integrate by parts.  This yields the variational form of the slot equation: find $\mathbf{I}_m\in\mathbb{V}^m$ and $\mathbf{J}\in\mathbb{V}$, such that
%
%\begin{align}
%\int_0^\lslot \bar{\mathbf{v}}^m(s)\cdot\biggl[\mathbf{J}(\mathbf{x}(s))\times\mathbf{n} + \frac{1}{2}\biggl(\Delta \tilde{Y}_L \frac{d^2}{ds^2} - \Delta \tilde{Y}_C\biggr)\mathbf{I}_m(s)\biggr] ds = 0.
%\label{eq:thick_slot_variational}
%\end{align} 
%
\begin{align}
\int_0^\lslot \bar{\mathbf{v}}^m\cdot(\mathbf{J}\times\mathbf{n}) ds 
{}-{}
\frac{Y_L}{4}\int_0^\lslot \bar{\mathbf{v}}^m\strut'\cdot \mathbf{I}_m'  ds
{}-{}
\frac{Y_C}{4}\int_0^\lslot \bar{\mathbf{v}}^m\cdot \mathbf{I}_m ds
{}= 0
\label{eq:thick_slot_variational}
\end{align} 
for all $\mathbf{v}^m\in\mathbb{V}^m$.  We can write~\eqref{eq:thick_slot_variational} more succinctly as
\begin{align}
\ame(\mathbf{J},\mathbf{v}^m) + \amm(\mathbf{I}_m,\mathbf{v}^m) = 0,
\label{eq:slot_sesquilinear}
\end{align}
where the sesquilinear forms are defined by
\begin{align*}
\ame(\mathbf{u},\mathbf{v}) ={}& \int_0^\lslot \bar{\mathbf{v}}(s)\cdot\bigl[\mathbf{u}(\mathbf{x})\times\mathbf{n}(\mathbf{x})\bigr] ds,
\\
\amm(\mathbf{u},\mathbf{v})  ={}& -\frac{1}{4}\biggl(
Y_L\int_0^\lslot \bar{\mathbf{v}}'(s)\cdot \mathbf{u}'(s) ds
+
Y_C\int_0^\lslot \bar{\mathbf{v}}(s)\cdot \mathbf{u}(s) ds\biggr). %\label{eq:efie_b}
\end{align*}

%===============================================================================
\section{Discretization} %======================================================
%===============================================================================
\label{sec:discretization}

To solve \eqref{eq:efie_sesquilinear} and \eqref{eq:slot_sesquilinear}, we discretize $S$ with a mesh composed of triangular elements and approximate $\mathbf{J}$ with $\mathbf{J}_h$ using the Rao--Wilton--Glisson (RWG) basis functions $\boldsymbol{\Lambda}_{\srcidx}(\mathbf{x})$~\cite{rao_1982}:
\begin{align}
\mathbf{J}_h(\mathbf{x}) = \sum_{\srcidx=1}^{\nbasis} J_{\srcidx} \boldsymbol{\Lambda}_{\srcidx}(\mathbf{x}),
\label{eq:J_h}
\end{align}%
where $\nbasis$ is the number of RWG basis functions. 
The RWG basis functions are second-order accurate~\cite[pp.\ 155--156]{warnick_2008}, and are defined for a triangle pair by
\begin{align}
\boldsymbol{\Lambda}_{\srcidx}(\mathbf{x}) = \left\{
\begin{matrix}
\displaystyle\frac{\ell_{\srcidx}}{2A_{\srcidx}^+}\boldsymbol{\rho}_{\srcidx}^+, & \text{for }\mathbf{x}\in T_{\srcidx}^+ \\[1em]
\displaystyle\frac{\ell_{\srcidx}}{2A_{\srcidx}^-}\boldsymbol{\rho}_{\srcidx}^-, & \text{for }\mathbf{x}\in T_{\srcidx}^- \\[1em]
\mathbf{0}, & \text{otherwise}
\end{matrix}
\right.,
\label{eq:rwg}
\end{align}
where $\ell_{\srcidx}$ is the length of the edge shared by the triangle pair, and $A_{\srcidx}^+$ and $A_{\srcidx}^-$ are the areas of the triangles $T_{\srcidx}^+$ and $T_{\srcidx}^-$ associated with basis function $\srcidx$.  $\boldsymbol{\rho}_{\srcidx}^+$ denotes the vector from the vertex of $T_{\srcidx}^+$ opposite the shared edge to $\mathbf{x}$, and $\boldsymbol{\rho}_{\srcidx}^-$ denotes the vector to the vertex of $T_{\srcidx}^-$ opposite the shared edge from $\mathbf{x}$.

These basis functions ensure that $\mathbf{J}_h$ is tangent to the mesh when using planar triangular elements.
Additionally, along the shared edge of the triangle pair, the component of $\boldsymbol{\Lambda}_{\srcidx}(\mathbf{x})$ normal to that edge is unity.  Therefore, for a triangle edge shared by only two triangles, the component of $\mathbf{J}_h$ normal to that edge is $J_\srcidx$.  The solution is considered most accurate at the midpoint of the edge~\cite[pp.\ 155--156]{warnick_2008}; therefore, we measure the solution at the midpoints.

Similarly, we discretize each wire with \reviewerTwo{one-dimensional} bar elements and approximate $\mathbf{I}_m$ with $\mathbf{I}_h$ using a one-dimensional analog to the RWG basis functions $\boldsymbol{\Lambda}_{\srcidx}^m(s)$:
\begin{align}
\mathbf{I}_h(s) = \sum_{\srcidx=1}^{\nbasis^m} I_{\srcidx} \boldsymbol{\Lambda}_{\srcidx}^m(s),
\label{eq:I_h}
\end{align}%
where $\nbasis^m$ is the number of one-dimensional basis functions. 
$\boldsymbol{\Lambda}_{\srcidx}^m$ is defined for a bar element pair by
\begin{align}
\boldsymbol{\Lambda}_{\srcidx}^m(s) = \left\{
\begin{matrix}
\displaystyle\frac{s            -s_{\srcidx-1}}{|s_{\srcidx  }-s_{\srcidx-1}|}\mathbf{s}, & \text{for }s\in [s_{\srcidx-1},\,s_{\srcidx  }] \\[1em]
\displaystyle\frac{s_{\srcidx+1}-s            }{|s_{\srcidx+1}-s_{\srcidx  }|}\mathbf{s}, & \text{for }s\in [s_{\srcidx}  ,\,s_{\srcidx+1}] \\[1em]
\mathbf{0}, & \text{otherwise}
\end{matrix}
\right..
\label{eq:rwg_bar}
\end{align}
%
%where 
%%
%\begin{align*}
%s(\mathbf{x}) = |\mathbf{x}-\mathbf{x}_{\srcidx-1}| + \sum_{i=1}^{\srcidx-1} |\mathbf{x}_i-\mathbf{x}_{i-1}|, & \quad\text{for }\mathbf{x}\in [\mathbf{x}_{\srcidx-1},\,\mathbf{x}_{\srcidx}].
%\end{align*}

%\subsection{Thick Slot}

%\clearpage
Defining $\mathbb{V}_h$ to be the span of RWG basis functions~\eqref{eq:rwg} and $\mathbb{V}_h^m$ to be the span of the one-dimensional basis functions~\eqref{eq:rwg_bar}, the Galerkin approximation of~\eqref{eq:efie_sesquilinear} and~\eqref{eq:slot_sesquilinear} is now: find $\mathbf{J}_h\in\mathbb{V}_h$ and $\mathbf{I}_h\in\mathbb{V}_h^m$, such that
\begin{align}
\aee(\mathbf{J}_h,\boldsymbol{\Lambda}_{\testidx}) + 
\aem(\mathbf{I}_h,\boldsymbol{\Lambda}_{\testidx}) = \be\bigl(\mathbf{E}^\mathcal{I}, \boldsymbol{\Lambda}_{\testidx}\bigr)
\label{eq:proj_disc_efie}
\end{align}
for $i=1,\hdots,\nbasis$, and 
\begin{align}
\ame(\mathbf{J}_h,\boldsymbol{\Lambda}_{\testidx}^m) + \amm(\mathbf{I}_h,\boldsymbol{\Lambda}_{\testidx}^m) = 0
\label{eq:proj_disc_slot}
\end{align}
for $i=1,\hdots,\nbasis^m$.

Equation~\eqref{eq:proj_disc_efie} is evaluated on the exterior and interior surfaces of the scatterer, such that there are $\nbasis^\text{ext}+\nbasis^\text{int}$ unknowns for $\mathbf{J}_h$.  Similarly,~\eqref{eq:proj_disc_slot} is evaluated for the wires on the exterior and interior surfaces.  However, for the thick slot model, $\mathbf{I}_m$ is modeled as equal and opposite at the corresponding locations on the interior and exterior surface wires, reducing the number of unknowns for $\mathbf{I}_h$ to $\nbasis^m$.  Physically, this equality is due to the assumed invariance of the voltage along the small electrical depth of the slot. The opposite direction is due to the assumption that $\mathbf{n}^\text{ext}=-\mathbf{n}^\text{int}$ and the fact that $\mathbf{M}=\mathbf{E}\times\mathbf{n}$~\cite[Chap.~1]{balanis_2012}.

The discretized system of equations can be written in matrix--vector form as
\begin{align}
\mathbf{Z}\solvec = \mathbf{V}.
\label{eq:system}
\end{align}
The impedance matrix $\mathbf{Z}$ is given by
\begin{align*}
\mathbf{Z} = \left[\begin{array}{@{} c @{} l r @{} c @{} l c @{} l @{}}
\mathbf{A}&^\text{ext} &  &\mathbf{0} &            & \phantom{-}\mathbf{B} &^\text{ext} \\ 
\mathbf{0}&            &  &\mathbf{A} &^\text{int} & -\mathbf{B} &^\text{int} \\ 
\mathbf{C}&^\text{ext} & -&\mathbf{C} &^\text{int} & \phantom{-}\mathbf{D} \end{array}\right]\in\mathbb{C}^{(\nbasis+\nbasis^m)\times(\nbasis+\nbasis^m)},%\label{eq:Z}
\end{align*}
where 
\begin{alignat*}{9}
A_{\testidx,\srcidx}&{}={}&\pz\aee(\boldsymbol{\Lambda}_{\srcidx}^{\phantom{m}},\boldsymbol{\Lambda}_{\testidx}^{\phantom{m}}),  &&\qquad
\mathbf{A}^\text{ext}\in\mathbb{C}&^{\nbasis^\text{ext}} &&^{\times\nbasis^\text{ext}}&&, \qquad &
\mathbf{A}^\text{int}\in\mathbb{C}&^{\nbasis^\text{int}} &&^{\times\nbasis^\text{int}}&&,
\\
B_{\testidx,\srcidx}&{}={}&\pz\aem(\boldsymbol{\Lambda}_{\srcidx}^m ,\boldsymbol{\Lambda}_{\testidx}^{\phantom{m}}), &&\qquad
\mathbf{B}^\text{ext}\in\mathbb{C}&^{\nbasis^\text{ext}} &&^{\times\nbasis^m}&&,  \qquad &
\mathbf{B}^\text{int}\in\mathbb{C}&^{\nbasis^\text{int}} &&^{\times\nbasis^m}&&,
\\
C_{\testidx,\srcidx}&{}={}&\pz\ame(\boldsymbol{\Lambda}_{\srcidx}^{\phantom{m}},\boldsymbol{\Lambda}_{\testidx}^m ),  &&\qquad
\mathbf{C}^\text{ext}\in\mathbb{R}&^{\nbasis^m} &&^{\times\nbasis^\text{ext}}&&,  \qquad &
\mathbf{C}^\text{int}\in\mathbb{R}&^{\nbasis^m} &&^{\times\nbasis^\text{int}}&&,
\\
D_{\testidx,\srcidx}&{}={}&2  \amm(\boldsymbol{\Lambda}_{\srcidx}^m ,\boldsymbol{\Lambda}_{\testidx}^m ),  &&\qquad
\mathbf{D}\phantom{^\text{ext}}\in\mathbb{C}&^{\nbasis^m} &&^{\times\nbasis^m}&&.
\end{alignat*}
$\mathbf{Z}$ can be written more compactly as
\begin{align}
\mathbf{Z} = \left[\begin{matrix}
\mathbf{A} & \mathbf{B} \\ 
\mathbf{C} & \mathbf{D} \end{matrix}\right],\label{eq:Z}
\end{align}
where
\begin{align*}
\mathbf{A} = \left[\begin{array}{@{} c @{} l c @{} l @{}}
\mathbf{A}&^\text{ext} & \mathbf{0} \\ 
\mathbf{0}&            & \mathbf{A} &^\text{int} \end{array}\right] \in\mathbb{C}^{\nbasis\times\nbasis},
\qquad
\mathbf{B} = \left[\begin{array}{@{} r @{} l @{}}
 \mathbf{B} &^\text{ext} \\ 
-\mathbf{B} &^\text{int} \end{array}\right]\in\mathbb{C}^{\nbasis\times\nbasis^m},
\qquad
\mathbf{C} = \left[\begin{array}{@{} c @{} l c @{} l @{}}
\mathbf{C}&^\text{ext} & -\mathbf{C} &^\text{int}\end{array}\right]\in\mathbb{R}^{\nbasis^m\times\nbasis},
\end{align*}
and $\nbasis=\nbasis^\text{ext}+\nbasis^\text{int}$.
The solution vector $\solvec$, which contains the coefficients used to construct $\mathbf{J}_h$~\eqref{eq:J_h} and $\mathbf{I}_h$~\eqref{eq:I_h}, is given by
\begin{align*}
\solvec = \left\{\begin{array}{@{} r @{} l @{}}
\mathbf{J}&^h\strut^\text{ext} \\ 
\mathbf{J}&^h\strut^\text{int} \\ 
\mathbf{I}&^h\end{array}\right\}\in\mathbb{C}^{\nbasis+\nbasis^m},
\end{align*}
where
\begin{alignat*}{9}
J&_{\srcidx}^h &&{}={} &J_{\srcidx},  &&\qquad \mathbf{J}&^h\strut^\text{ext}&&{}\in\mathbb{C}^{\nbasis^\text{ext}}&&, \qquad & {\mathbf{J}^h}^\text{int}&{}\in\mathbb{C}^{\nbasis^\text{int}},
\\
I&_{\srcidx}^h &&{}={} &I_{\srcidx},  &&\qquad \mathbf{I}&^h               &&{}\in\mathbb{C}^{\nbasis^m}         &&.
\end{alignat*}
$\solvec$ can be written more compactly as
\begin{align*}
\solvec = \left\{\begin{array}{@{} r @{} l @{}}
\mathbf{J}&^h \\ 
\mathbf{I}&^h\end{array}\right\},
\end{align*}
where
\begin{align*}
\mathbf{J}^h = \left\{\begin{array}{@{} r @{} l @{}}
\mathbf{J}&^h\strut^\text{ext} \\ 
\mathbf{J}&^h\strut^\text{int}\end{array}\right\}\in\mathbb{C}^{\nbasis}.
\end{align*}
Finally, the excitation vector $\mathbf{V}$ is given by
\begin{align*}
\mathbf{V} = \left\{\begin{array}{@{} c @{} l @{}}
\mathbf{V}& ^\mathcal{E}\strut^\text{ext} \\ 
\mathbf{V}& ^\mathcal{E}\strut^\text{int} \\ 
\mathbf{0}& \end{array}\right\}\in\mathbb{C}^{\nbasis+\nbasis^m},
\end{align*}
where
\begin{alignat*}{7}
V^\mathcal{E}_{\srcidx} &{}= \be\bigl(\mathbf{E}^\mathcal{I} , \boldsymbol{\Lambda}_{\testidx}\bigr),  \qquad &
{\mathbf{V}^\mathcal{E}}^\text{ext}&{}\in\mathbb{C}^{\nbasis^\text{ext}}, \qquad &
{\mathbf{V}^\mathcal{E}}^\text{int}&{}\in\mathbb{C}^{\nbasis^\text{int}}.
\end{alignat*}
$\mathbf{V}$ can be written more compactly as
\begin{align*}
\mathbf{V} = \left\{\begin{array}{@{} c @{} l @{}}
\mathbf{V}& ^\mathcal{E} \\ 
\mathbf{0}& \end{array}\right\},
\end{align*}
where
\begin{align*}
\mathbf{V}^\mathcal{E} = \left\{\begin{array}{@{} c @{} l @{}}
\mathbf{V}& ^\mathcal{E}\strut^\text{ext} \\ 
\mathbf{V}& ^\mathcal{E}\strut^\text{int}\end{array}\right\}\in\mathbb{C}^{\nbasis}.
\end{align*}

%===============================================================================
\section{Manufactured Solutions} %==============================================
%===============================================================================
\label{sec:mms}

We define residual functionals for the surfaces and wires as
\begin{alignat}{7}
r_{\mathcal{E}_{\testidx}}(\mathbf{u},\mathbf{v}) &{}={}& \aee(\mathbf{u},\boldsymbol{\Lambda}_{\testidx}^{\phantom{m}}) &{}+{}& \aem(\mathbf{v},\boldsymbol{\Lambda}_{\testidx}^{\phantom{m}}) &{}-{}&\be\bigl(\mathbf{E}^\mathcal{I}, \boldsymbol{\Lambda}_{\testidx}\bigr),
\label{eq:res_func_efie}
\\
r_{\mathcal{M}_{\testidx}}(\mathbf{u},\mathbf{v}) &{}={}& \ame(\mathbf{u},\boldsymbol{\Lambda}_{\testidx}^m) &{}+{}& \amm(\mathbf{v},\boldsymbol{\Lambda}_{\testidx}^m)&.
\label{eq:res_func_slot}
\end{alignat}
%
%For each test basis function, 
We can write the variational forms from~\eqref{eq:efie_sesquilinear} and~\eqref{eq:slot_sesquilinear} in terms of~\eqref{eq:res_func_efie} and~\eqref{eq:res_func_slot} as
\begin{alignat}{7}
r_{\mathcal{E}_{\testidx}}(\mathbf{J},\mathbf{I}_m) &{}={}& \aee(\mathbf{J},\boldsymbol{\Lambda}_{\testidx}^{\phantom{m}}) &{}+{}& \aem(\mathbf{I}_m,\boldsymbol{\Lambda}_{\testidx}^{\phantom{m}}) &{}- \be\bigl(\mathbf{E}^\mathcal{I}, \boldsymbol{\Lambda}_{\testidx}\bigr)=0,
\label{eq:res_efie}
\\
r_{\mathcal{M}_{\testidx}}(\mathbf{J},\mathbf{I}_m) &{}={}& \ame(\mathbf{J},\boldsymbol{\Lambda}_{\testidx}^m) &{}+{}& \amm(\mathbf{I}_m,\boldsymbol{\Lambda}_{\testidx}^m)&{}=0.
\label{eq:res_slot}
\end{alignat}
Similarly, we can write the discretized problems in~\eqref{eq:proj_disc_efie} and~\eqref{eq:proj_disc_slot} in terms of~\eqref{eq:res_func_efie} and~\eqref{eq:res_func_slot} as
\begin{alignat}{7}
r_{\mathcal{E}_{\testidx}}(\mathbf{J}_h,\mathbf{I}_h) &{}={}& \aee(\mathbf{J}_h,\boldsymbol{\Lambda}_{\testidx}^{\phantom{m}}) &{}+{}& \aem(\mathbf{I}_h,\boldsymbol{\Lambda}_{\testidx}^{\phantom{m}}) &{}- \be\bigl(\mathbf{E}^\mathcal{I}, \boldsymbol{\Lambda}_{\testidx}\bigr) = 0,
\label{eq:res_disc_efie}
\\
r_{\mathcal{M}_{\testidx}}(\mathbf{J}_h,\mathbf{I}_h) &{}={}& \ame(\mathbf{J}_h,\boldsymbol{\Lambda}_{\testidx}^m) &{}+{}& \amm(\mathbf{I}_h,\boldsymbol{\Lambda}_{\testidx}^m)&{}=0.
\label{eq:res_disc_slot}
\end{alignat}
%
%where $\mathbf{r}_h(\mathbf{J}_h) = \mathbf{Z}\mathbf{J}^h - \mathbf{V}$.

The method of manufactured solutions modifies~\eqref{eq:res_disc_efie} and~\eqref{eq:res_disc_slot} to be
\begin{alignat}{7}
r_{\mathcal{E}_{\testidx}}(\mathbf{J}_h,\mathbf{I}_h) &{}={}& r_{\mathcal{E}_{\testidx}}(\mathbf{J}_\text{MS},\mathbf{I}_\text{MS}),
\label{eq:mms_efie}
\\
r_{\mathcal{M}_{\testidx}}(\mathbf{J}_h,\mathbf{I}_h) &{}={}& r_{\mathcal{M}_{\testidx}}(\mathbf{J}_\text{MS},\mathbf{I}_\text{MS}),
\label{eq:mms_slot}
\end{alignat}
where $\mathbf{J}_\text{MS}$ and $\mathbf{I}_\text{MS}$ are the manufactured solutions, and $\mathbf{r}_\mathcal{E}(\mathbf{J}_\text{MS},\mathbf{I}_\text{MS})$ and $\mathbf{r}_\mathcal{M}(\mathbf{J}_\text{MS},\mathbf{I}_\text{MS})$ are computed exactly.

Inserting~\eqref{eq:res_efie} and~\eqref{eq:res_disc_efie} into~\eqref{eq:mms_efie} yields
\begin{align}
\aee(\mathbf{J}_h,\boldsymbol{\Lambda}_{\testidx}) + \aem(\mathbf{I}_h,\boldsymbol{\Lambda}_{\testidx})= \aee(\mathbf{J}_\text{MS},\boldsymbol{\Lambda}_{\testidx}) + \aem(\mathbf{I}_\text{MS},\boldsymbol{\Lambda}_{\testidx}).
\label{eq:proj_disc_mms_efie}
\end{align}
%
%
%%
%\begin{align}
%a(\mathbf{J}_h,\boldsymbol{\Lambda}_{\testidx}) = a(\mathbf{J},\boldsymbol{\Lambda}_{\testidx}) 
%%\label{eq:zjv_mmsp}\\
%%
%\mathbf{Z}\mathbf{J}^h = \mathbf{V}_{\text{MS}}(\mathbf{J}_\text{MS}), \nonumber
%\end{align}
%%
%where 
%%
%\begin{align}
%V_{\text{MS}_{\testidx}}(\mathbf{u})= a(\mathbf{u},\boldsymbol{\Lambda}_{\testidx}).
%\label{eq:V_mms}
%\end{align}
%%
However, instead of solving~\eqref{eq:proj_disc_mms_efie}, we can equivalently solve~\eqref{eq:proj_disc_efie} by setting %$\mathbf{V}=\mathbf{V}_{\text{MS}}(\mathbf{J}_\text{MS})$, such that, from~\eqref{eq:V} and~\eqref{eq:V_mms},
\begin{align}
\be\bigl(\mathbf{E}^\mathcal{I}, \boldsymbol{\Lambda}_{\testidx}\bigr) = \aee(\mathbf{J}_\text{MS},\boldsymbol{\Lambda}_{\testidx})+\aem(\mathbf{I}_\text{MS},\boldsymbol{\Lambda}_{\testidx}).
\label{eq:E_mms_1}
\end{align}
Equation~\eqref{eq:E_mms_1} is satisfied by
\begin{align}
\mathbf{E}^\mathcal{I}(\mathbf{x}) ={}& \frac{j}{\epsilon\omega} \int_{S'}\bigl[ k^2\mathbf{J}_\text{MS} (\mathbf{x}')G(\mathbf{x},\mathbf{x}') +\nabla'\cdot\mathbf{J}_\text{MS}(\mathbf{x}')\nabla G(\mathbf{x},\mathbf{x}')\bigr]dS' \nonumber
\\
&-\frac{1}{4}\bigl(\mathbf{n}(\mathbf{x})\times \mathbf{I}_\text{MS}(\mathbf{x})\bigr)\delta_\text{slot}(\mathbf{x})
{}+\frac{1}{4\pi}\int_0^\lslot \mathbf{I}_\text{MS}(s')\times\int_0^{2\pi} \nabla' G(\mathbf{x},\mathbf{x}')d\phi' ds' + Z_s \mathbf{J}_\text{MS}(\mathbf{x}),
\label{eq:E_i}
\end{align}
where $\delta_\text{slot}$ is defined such that
\begin{align}
\be\bigl((\mathbf{n}\times \mathbf{I}_\text{MS})\delta_\text{slot},\boldsymbol{\Lambda}_{\testidx}\bigr)
=
\int_S \boldsymbol{\Lambda}_{\testidx}\cdot (\mathbf{n}\times \mathbf{I}_\text{MS})\delta_\text{slot} dS
=
\int_0^\lslot \boldsymbol{\Lambda}_{\testidx}\cdot (\mathbf{n}\times \mathbf{I}_\text{MS}) ds.
\label{eq:delta_slot}
\end{align}

Inserting~\eqref{eq:res_slot} and~\eqref{eq:res_disc_slot} into~\eqref{eq:mms_slot} yields
\begin{align}
\ame(\mathbf{J}_h,\boldsymbol{\Lambda}_{\testidx}^m) + \amm(\mathbf{I}_h,\boldsymbol{\Lambda}_{\testidx}^m)= \ame(\mathbf{J}_\text{MS},\boldsymbol{\Lambda}_{\testidx}^m) + \amm(\mathbf{I}_\text{MS},\boldsymbol{\Lambda}_{\testidx}^m).
\label{eq:proj_disc_mms_slot}
\end{align}
As an alternative to solving~\eqref{eq:proj_disc_mms_slot}, we can solve~\eqref{eq:proj_disc_slot} by choosing $\mathbf{I}_\text{MS}$, such that, for a given $\mathbf{J}_\text{MS}$,
\begin{align}
\ame(\mathbf{J}_\text{MS},\boldsymbol{\Lambda}_{\testidx}^m) + \amm(\mathbf{I}_\text{MS},\boldsymbol{\Lambda}_{\testidx}^m)=0.
\label{eq:no_mms_source}
\end{align}
With these approaches, the manufactured source term for the EFIE is incorporated through the incident electric field, and the slot equation does not require a manufactured source term.

%===============================================================================
\subsection{Solution-Discretization Error} %====================================
%===============================================================================
\label{sec:sde}

If the integrals are evaluated exactly in~\eqref{eq:proj_disc_efie} and~\eqref{eq:proj_disc_slot}, the only contribution to the discretization error is the solution-discretization error.  Solving for $\mathbf{J}^h$ and $\mathbf{I}^h$ enables us to compute the discretization errors 
\begin{alignat}{7}
&\mathbf{e}_\mathbf{J} &&{}={}& \mathbf{J}^h &{}-{}& \mathbf{J}&_n&,
\label{eq:solution_error_J}
\\
&\mathbf{e}_\mathbf{I} &&{}={}& \mathbf{I}^h &{}-{}& \mathbf{I}&_s&,
\label{eq:solution_error_I}
\end{alignat}
where $J_{n_\srcidx}$ denotes the component of $\mathbf{J}_\text{MS}$ flowing from $T_\srcidx^+$ to $T_\srcidx^-$ and $I_{s_\srcidx}$ denotes the component of $\mathbf{I}_\text{MS}$ flowing along $\mathbf{s}$ at position $s_\srcidx$.  The norms of~\eqref{eq:solution_error_J} and~\eqref{eq:solution_error_I} have the properties $\|\mathbf{e}_\mathbf{J}\|\le C_\mathbf{J} h^{p_\mathbf{J}}$ and $\|\mathbf{e}_\mathbf{I}\|\le C_\mathbf{I} h^{p_\mathbf{I}}$, where
$C_\mathbf{J}$ and $C_\mathbf{I}$ are functions of the solution derivatives, $h$ is representative of the mesh size, and $p_\mathbf{J}$ and $p_\mathbf{I}$ are the orders of accuracy.  By performing a mesh-convergence study of the norms of the discretization errors, we can assess whether the expected orders of accuracy are obtained.  For $\boldsymbol{\Lambda}_{\srcidx}(\mathbf{x})$~\eqref{eq:rwg}, the expectation is second-order accuracy $(p_\mathbf{J}=2)$ when the error is evaluated at the edge centers~\cite{warnick_2008}.  For $\boldsymbol{\Lambda}_{\srcidx}^m(s)$~\eqref{eq:rwg_bar}, the expectation is second-order accuracy $(p_\mathbf{I}=2)$.

These expected orders of accuracy are based on the assumption of smoothness in the equations and their solutions.  For the EFIE, the first term in $\aem(\mathbf{u},\mathbf{v})$~\eqref{eq:aem} introduces a discontinuity on the surface where the wire is located, which is characterized by $\delta_\text{slot}$, as described in~\eqref{eq:delta_slot}.  For the manufactured solutions, this implication is additionally present in $\mathbf{E}^\mathcal{I}$~\eqref{eq:E_i}.  This discontinuity will contaminate the convergence studies used to assess the correctness of the implementation of the numerical methods, reducing the convergence rate from $\mathcal{O}(h^2)$ to $\mathcal{O}(h)$~\cite{dangelo_2012,li_2021}.

To mitigate the effects of the discontinuity, we first separate the two terms in $\aem(\mathbf{u},\mathbf{v})$:
\begin{align*}
\aem(\mathbf{u},\mathbf{v})={\aem}_1(\mathbf{u},\mathbf{v})+{\aem}_2(\mathbf{u},\mathbf{v}),
\end{align*}
where
\begin{align*}
{\aem}_1(\mathbf{u},\mathbf{v}) ={}& -\frac{1}{4}\int_0^\lslot \bar{\mathbf{v}}\cdot (\mathbf{n}\times \mathbf{u}) ds,
\\
{\aem}_2(\mathbf{u},\mathbf{v}) ={}& \frac{1}{4\pi}\int_S \bar{\mathbf{v}}(\mathbf{x})\cdot  \int_0^\lslot \mathbf{u}(s')\times\int_0^{2\pi} \nabla' G(\mathbf{x},\mathbf{x}')d\phi' ds' dS,
\end{align*}
and ${\aem}_1(\mathbf{u},\mathbf{v})$ is the term that introduces the singularity.
We can write $\mathbf{Z}$~\eqref{eq:Z} as
%
%\begin{align}
%\mathbf{Z} = \left[\begin{array}{@{} c @{} l c @{} l r @{} c @{} l @{} l @{}}
%\mathbf{A}&^\text{ext} & \mathbf{0} &            & (\mathbf{B}_1&{}+{}&\mathbf{B}_2) &^\text{ext} \\ 
%\mathbf{0}&            & \mathbf{A} &^\text{int} & (\mathbf{B}_1&{}+{}&\mathbf{B}_2) &^\text{int} \\ 
%\mathbf{C}&^\text{ext} & \mathbf{C} &^\text{int} & &\mathbf{D}& 
%\end{array}\right],\label{eq:Z2}
%\end{align}
%%
%\begin{align}
%\mathbf{Z} = \left[\begin{array}{@{} c r @{} c @{} l @{}}
%\mathbf{A}& (\mathbf{B}_1&{}+{}&\mathbf{B}_2) \\ 
%\mathbf{C}&  &\mathbf{D} 
%\end{array}\right],\label{eq:Z2}
%\end{align}
%
\begin{align}
\mathbf{Z} = \left[\begin{matrix}
\mathbf{A}& (\mathbf{B}_1+\mathbf{B}_2) \\ 
\mathbf{C}&  \mathbf{D} 
\end{matrix}\right],\label{eq:Z2}
\end{align}
where $B_{1_{\testidx,\srcidx}}={\aem}_1(\boldsymbol{\Lambda}_{\srcidx}^m,\boldsymbol{\Lambda}_{\testidx})\in\mathbb{R}$ and $B_{2_{\testidx,\srcidx}}={\aem}_2(\boldsymbol{\Lambda}_{\srcidx}^m,\boldsymbol{\Lambda}_{\testidx})\in\mathbb{C}$.
Noting that ${\aem}_1(\mathbf{u},\mathbf{v})=-\frac{1}{4}{\ame}(\bar{\mathbf{v}},\bar{\mathbf{u}})$, $\mathbf{B}_1=-\frac{1}{4}\mathbf{C}^T$, such that~\eqref{eq:Z2} can be written as
%
%\begin{align}
%\mathbf{Z} = \left[\begin{array}{@{} c @{} l c @{} l l @{} c @{} l @{} }
%\mathbf{A}&^\text{ext} & \mathbf{0} &            & -{\mathbf{C}^\text{ext}}^T/4&{}+{}&\mathbf{B}_2^\text{ext} \\
%\mathbf{0}&            & \mathbf{A} &^\text{int} & -{\mathbf{C}^\text{int}}^T/4&{}+{}&\mathbf{B}_2^\text{int} \\
%\mathbf{C}&^\text{ext} & \mathbf{C} &^\text{int} & &\mathbf{D} \end{array}\right].\label{eq:Z3}
%\end{align}
%
\begin{align*}
\mathbf{Z} = \left[\begin{array}{@{} c   l @{} c @{} l @{} }
\mathbf{A}  & \bigl(-\frac{1}{4}\mathbf{C}^T&{}+{}&\mathbf{B}_2\bigr) \\
\mathbf{C}  & &\mathbf{D} \end{array}\right].%\label{eq:Z3}
\end{align*}
%
%\begin{align}
%\mathbf{Z} = \left[\begin{matrix}
%\mathbf{A}  & -\frac{1}{4}\mathbf{C}^T+\mathbf{B}_2 \\
%\mathbf{C}  & \mathbf{D} \end{matrix}\right].\label{eq:Z3}
%\end{align}
%
Taking the transpose of $\mathbf{C}$, dividing it by four, and adding it to $\mathbf{B}$, we can solve a modified problem, where $\mathbf{Z}$ is modified to be
%
%\begin{align}
%\mathbf{Z} = \left[\begin{array}{@{} c @{} l c @{} l c @{} l @{}}
%\mathbf{A}&^\text{ext} & \mathbf{0} &            & \mathbf{B} &_2^\text{ext} \\ 
%\mathbf{0}&            & \mathbf{A} &^\text{int} & \mathbf{B} &_2^\text{int} \\ 
%\mathbf{C}&^\text{ext} & \mathbf{C} &^\text{int} & \mathbf{D} \end{array}\right],\label{eq:Z4}
%\end{align}
%
\begin{align}
\mathbf{Z} = \left[\begin{array}{@{} c  c @{} l @{}}
\mathbf{A} & \mathbf{B} &_2 \\ 
\mathbf{C} & \mathbf{D} \end{array}\right],\label{eq:Z4}
\end{align}
and $\mathbf{E}^\mathcal{I}$~\eqref{eq:E_i} is modified to be
\begin{align}
\mathbf{E}^\mathcal{I}(\mathbf{x}) ={}& \frac{j}{\epsilon\omega} \int_{S'}\bigl[ k^2\mathbf{J}_\text{MS} (\mathbf{x}')G(\mathbf{x},\mathbf{x}') +\nabla'\cdot\mathbf{J}_\text{MS}(\mathbf{x}')\nabla G(\mathbf{x},\mathbf{x}')\bigr]dS' \nonumber
\\
&+\frac{1}{4\pi}\int_0^\lslot \mathbf{I}_\text{MS}(s')\times\int_0^{2\pi} \nabla' G(\mathbf{x},\mathbf{x}')d\phi' ds' + Z_s \mathbf{J}_\text{MS}(\mathbf{x}).
\label{eq:E_i2}
\end{align}
With the modifications in~\eqref{eq:Z4} and~\eqref{eq:E_i2}, the discontinuity is removed. The correctness of the implementation of $\mathbf{B}_1$ is assessed by its successful removal using $\mathbf{C}$, and the correctness of the implementation of $\mathbf{C}$ is assessed through the aforementioned mesh-convergence study.

%===============================================================================
\subsection{Numerical-Integration Error} %======================================
%===============================================================================
\label{sec:nie}

In practice, the integrals in~\eqref{eq:proj_disc_efie} and~\eqref{eq:proj_disc_slot} are evaluated numerically by integrating over each triangular or bar element using quadrature.  These evaluations are generally approximations, which incur a numerical-integration error.  Therefore, it is important to measure the numerical-integration error without contamination from the solution-discretization error.

In~\cite{freno_mfie_2022}, approaches are presented to isolate the numerical-integration error by canceling or eliminating the solution-discretization error.  In this paper, we cancel the solution-discretization error and measure the numerical-integration error from
\begin{alignat}{7}
&e_a&{}={}&
%a^q(\mathbf{J}_{h_\text{MS}},\mathbf{I}_{h_\text{MS}}) - 
%a  (\mathbf{J}_{h_\text{MS}},\mathbf{I}_{h_\text{MS}}) = 
\boldsymbol{\mathcal{J}}^H(&\mathbf{Z}^q{}-{}&\mathbf{Z}&&)\boldsymbol{\mathcal{J}},  \label{eq:a_error_cancel}
\\
&e_b&{}={}&
%\be^q\bigl(\mathbf{E}^\mathcal{I}_\text{MS},\mathbf{J}_{h_\text{MS}}\bigr) {}-{} \be\bigl(\mathbf{E}^\mathcal{I}_\text{MS},\mathbf{J}_{h_\text{MS}}\bigr) = 
\boldsymbol{\mathcal{J}}^H(&\mathbf{V}^q{}-{}&\mathbf{V}&&), \label{eq:b_error_cancel}
\end{alignat}
where
%
%\begin{align*}
%\mathbf{J}_\text{MS}^h = \left\{\begin{array}{@{} r @{} l @{}}
%\mathbf{J}_n&\strut^{\!\!\!\text{ext}} \\ 
%\mathbf{J}_n&\strut^{\!\!\!\text{int}} \\ 
%\mathbf{I}_n&\end{array}\right\}.
%\end{align*}
\begin{align*}
\boldsymbol{\mathcal{J}} = \left\{\begin{array}{@{} r @{} l @{}}
\mathbf{J}&_n \\ 
\mathbf{I}&_s\end{array}\right\}.
\end{align*}
%
%\begin{align*}
%a(\mathbf{u},\mathbf{v}) &{}= 
%\aee(\mathbf{u},\mathbf{u}) + 
%\aem(\mathbf{v},\mathbf{u}) +
%\ame(\mathbf{u},\mathbf{v}) + 
%\amm(\mathbf{v},\mathbf{v}), 
%\end{align*}
% 
%
Equations~\eqref{eq:a_error_cancel} and~\eqref{eq:b_error_cancel} have the properties $|e_a| \le C_a h^{p_a}$ and $|e_b| \le C_b h^{p_b}$, where $C_a$ and $C_b$ are functions of the integrand derivatives, and $p_a$ and $p_b$ depend on the quadrature accuracy.
Unlike the solution-discretization error, the numerical-integration error is not contaminated by the discontinuity.  Therefore, we use $\mathbf{Z}$~\eqref{eq:Z} and $\mathbf{E}^\mathcal{I}$~\eqref{eq:E_i} without applying the modifications presented in Section~\ref{sec:sde}.

Reference~\cite{freno_mfie_2022} shows that $e_a$~\eqref{eq:a_error_cancel} and $e_b$~\eqref{eq:b_error_cancel} are proportional to their influence on the solution-discretization error.

%===============================================================================
\subsection{Manufactured Green's Function} %====================================
%===============================================================================
\label{sec:g_ms}
Integrals containing the Green's function~\eqref{eq:G} or its derivatives, such as those appearing in $\aee(\cdot,\cdot)$~\eqref{eq:aee}, $\aem(\cdot,\cdot)$~\eqref{eq:aem}, and $\mathbf{E}^\mathcal{I}$~\eqref{eq:E_i}, cannot be computed analytically. Additionally, the singularity when $R\to 0$ complicates their accurate approximation, potentially contaminating convergence studies.  Therefore, as is done in~\cite{freno_em_mms_2020,freno_mfie_2022}, we manufacture the Green's function, using the form
\begin{align}
G_\text{MS}(\mathbf{x},\mathbf{x}') = G_q(\mathbf{x},\mathbf{x}') = G_0\biggl(1 - \frac{R^2}{R_m^2}\biggr)^q,
\label{eq:G_mms}
\end{align}
where $G_0=1$~m$^{-1}$, $q\in\mathbb{N}$, and $R_m=\max_{\mathbf{x},\mathbf{x}'\in S} R$ is the maximum possible distance between two points on the domain.  The even powers of $R$ permit the aforementioned integrals to be computed analytically for the basis functions, as well as for many choices of $\mathbf{J}_\text{MS}$ and $\mathbf{I}_\text{MS}$, avoiding contamination from numerical-integration error.

%==============================================================================
\section{Numerical Examples} %=================================================
%==============================================================================
\label{sec:results}

\begin{figure}%[!t]
\centering
\includegraphics[scale=.28,clip=true,trim=0in 0in 0in 0in]{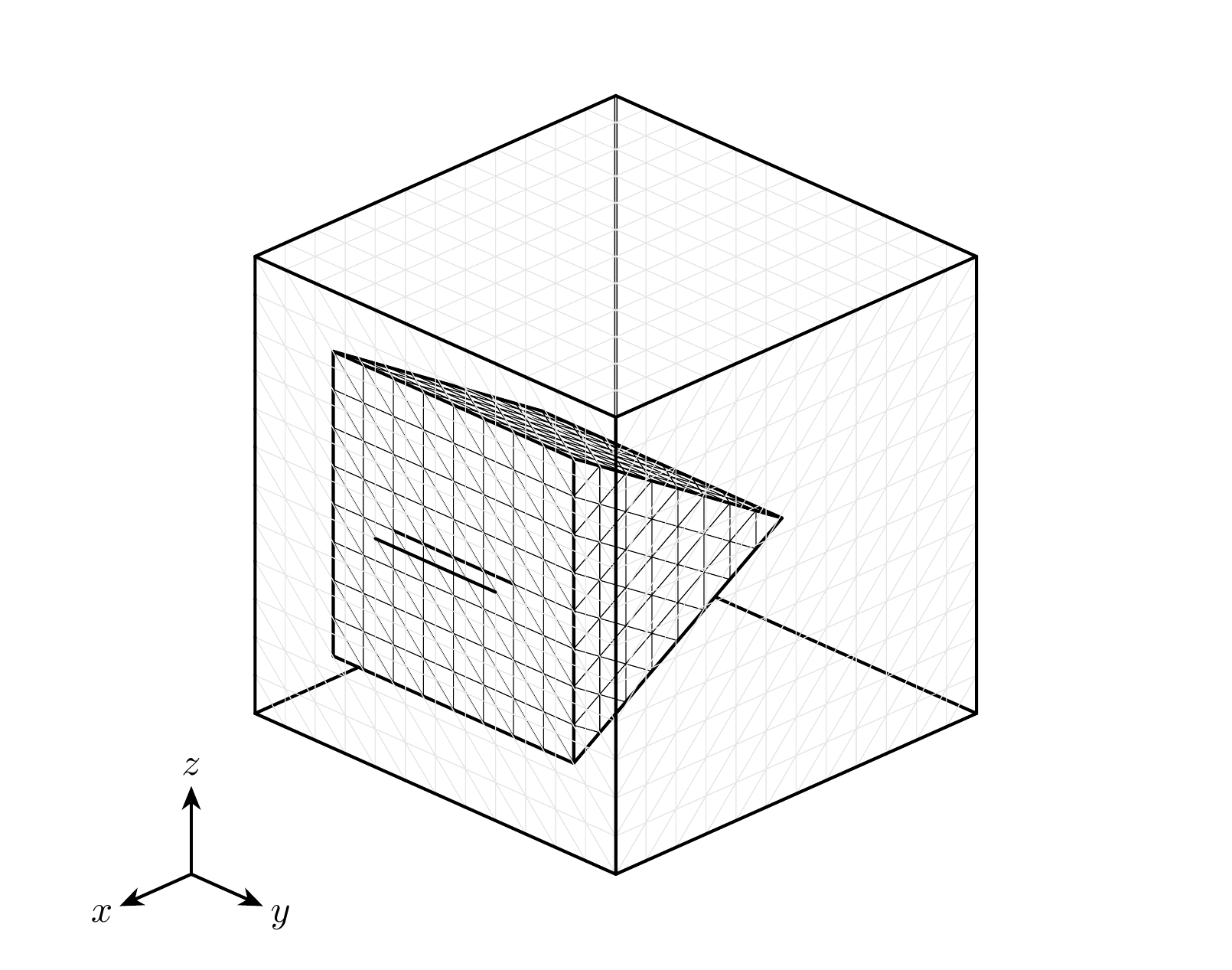}
\includegraphics[scale=.28,clip=true,trim=0in 0in 0in 0in]{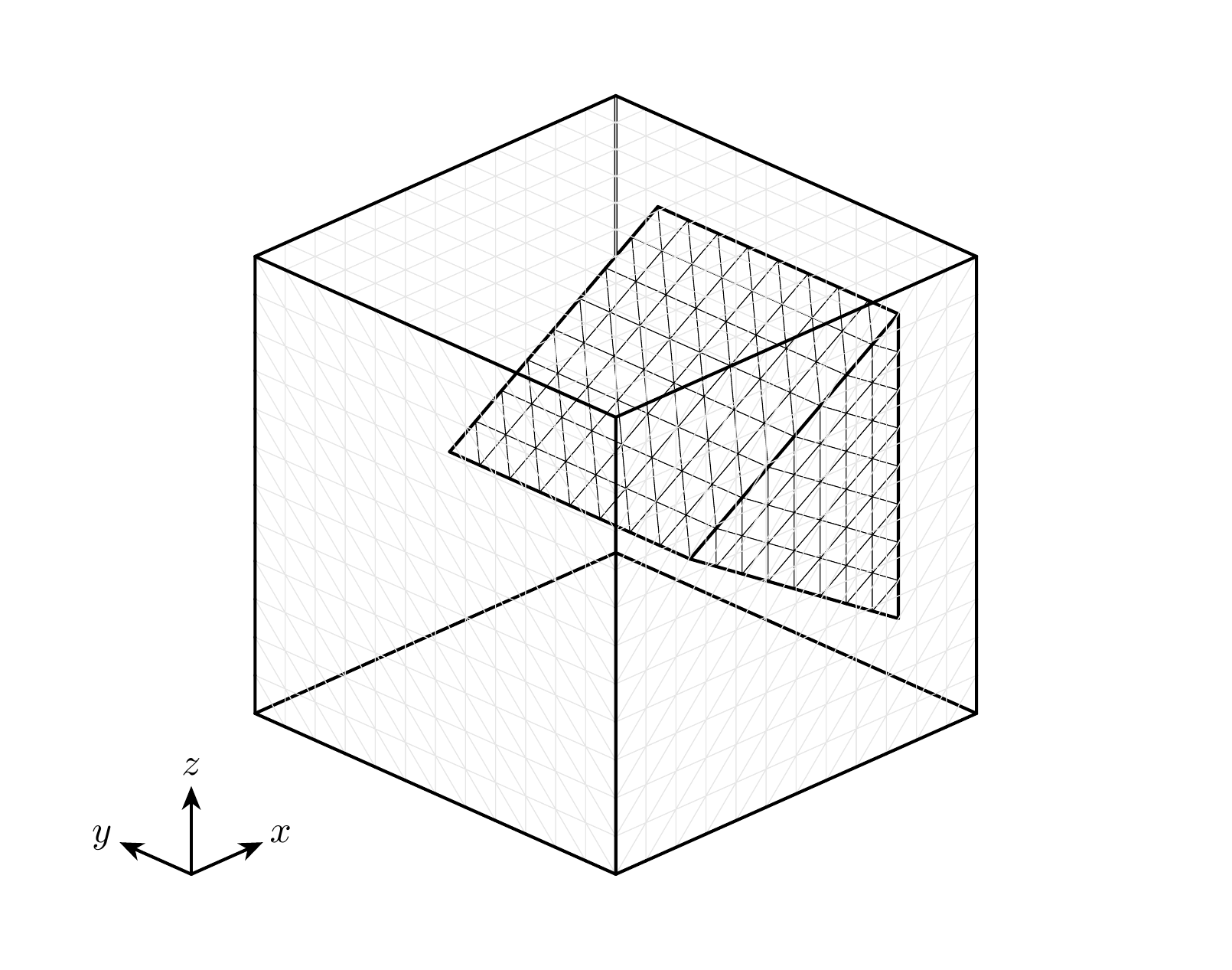}
\caption{Meshed domain with $\ntriangles=2240$ for $d=\lext/20$.}
\vskip-\dp\strutbox
\label{fig:whole_mesh_12}
\end{figure}

In this section, we demonstrate the approaches described in Section~\ref{sec:mms} by isolating and measuring the solution-discretization error (Section~\ref{sec:sde}) and the numerical-integration error (Section~\ref{sec:nie}).

%==============================================================================
\subsection{Domain and Coordinate Systems} %===================================
%==============================================================================
\label{sec:domain}

\begin{figure}%[!t]
\centering
\input{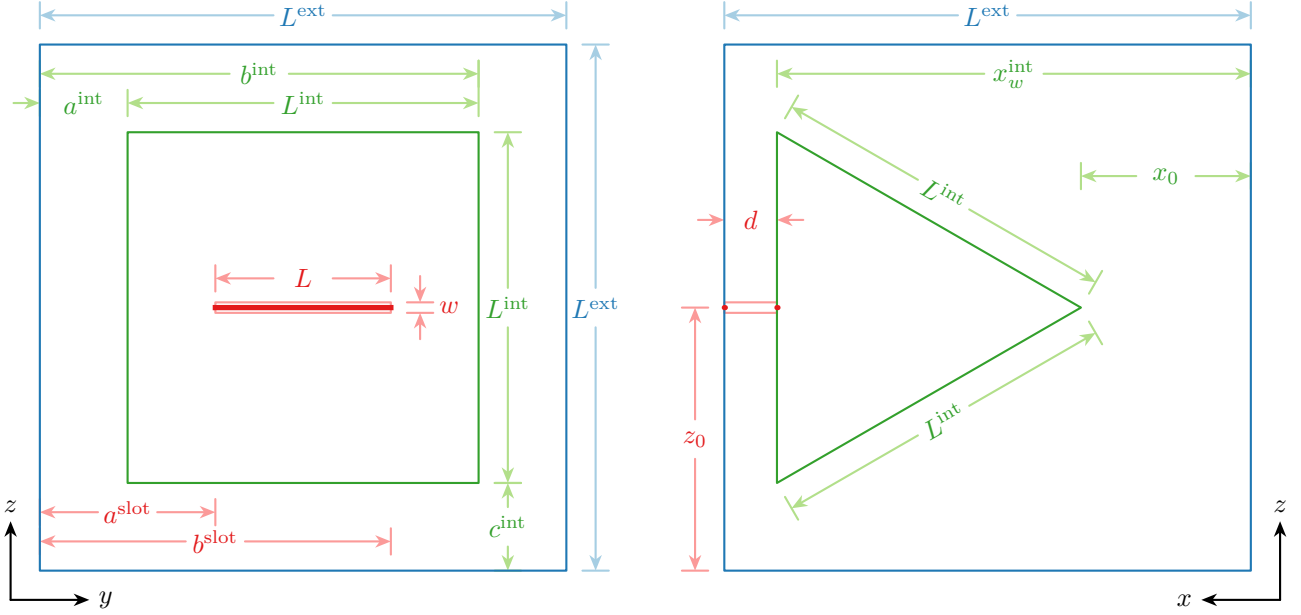}
\caption{Dimensions of the domain.}
\vskip-\dp\strutbox
\label{fig:dimensions}
\end{figure}

We consider the case of a cubic scatterer with a triangularly prismatic cavity.  There is a rectangularly prismatic slot that connects the exterior of the scatterer to the interior cavity.  The slot is modeled by two wires at the apertures.   The domain is shown in Figure~\ref{fig:whole_mesh_12}.  The dimensions of the domain are shown in Figure~\ref{fig:dimensions}, where $\lext=1$~m, and
\begin{align*}
\lint = \frac{2}{3}\lext, \quad
\lslot=\frac{\lext}{3},   \quad
w=\frac{\lext}{50},       \quad
\aint=\frac{\lext}{6},    \quad
\cint=\frac{\lext}{6},    \quad
\aslot=\frac{\lext}{3},   \quad
z_0=\frac{\lext}{2}.
\end{align*}
Additionally, we consider the presence of $\mathbf{B}_1$ and $\mathbf{B}_2$ together and separately in~\eqref{eq:Z2}, with the corresponding terms in $\be\bigl(\mathbf{E}^\mathcal{I}, \boldsymbol{\Lambda}_{\testidx}\bigr)$ adjusted accordingly; two manufactured Green's functions~\eqref{eq:G_mms}: $G_1$ and $G_2$; and three depths: $d_1=\lext/10$, $d_2=\lext/100$, and $d_3=\lext/1000$.  These three depths, along with the choice of width, test the two conformal mapping approaches, which are chosen based on the depth-to-width ratio~\cite{warne_1995}.  We set the permeability and permittivity of the surrounding medium to those of free space: $\mu =\mu_0$ and $\epsilon = \epsilon_0$, we set the wavenumber to $k=2\pi$~m$^{-1}$, and we set the electric conductivity $\sigma$ to that of aluminum.
An example discretization is shown in Figure~\ref{fig:whole_mesh_12} with $\ntriangles=2240$ total triangles for the exterior and interior surfaces and four bar elements for each of the two wires.

%{\color{gray}
%We consider the case of a cubic scatterer with edges of length $\lext = 1$~m, such that $x\in[0,\,\lext]$, $y\in[0,\,\lext]$, and $z\in[0,\,\lext]$.  
%Inside the scatterer is a triangularly prismatic cavity with each edge of length $\lint=2\lext/3$, such that $x\in[\xslot-\sqrt{3}\lint/2,\,\xslot]$, $y\in[\aint,\,\bint]$, and $z\in(\lext\pm\lint)/2\pm\sqrt{3}(x-\xslot)/2$, where $\xslot=\lext-d$, $\aint=\lext/6$, and $\bint=\aint+\lint$.  

%The cube and triangular prism each have a square face with a normal in the $\pm x$-direction and edges parallel to the $y$- and $z$-directions.  These faces are offset by a depth $d$ in the $x$-direction.  For the triangular prism, the edges of this face are offset from the $y$- and $z$-axes by $d_y=(\lext-\lint)/2$ and $d_z=(\lext-\lint)/2$.} 

%There is a slot of length $\lslot=\lext/3$, width $w=\lext/50$, and depth $d$ that connects the cube and triangular prism, with $x\in[\lext-d,\,\lext]$, $y\in[\aslot,\,\bslot]$, and $z\in\lext/2\pm w/2$, where $\aslot=\lext/3$ and $\bslot=\aslot+\lslot$.  The slot is modeled by two wires of length $\lslot$ at the apertures, with $x=\{\lext-d,\,\lext\}$, $y\in[\aslot,\,\bslot]$, and $z=\lext/2$.  The domain is shown in Figure~\ref{fig:whole_mesh_12} with $\ntriangles=2240$ total triangles for the exterior and interior surfaces and four bar elements for each of the two wires.} 

To manufacture the surface current, we introduce coordinate systems that wrap around the lateral surfaces of the exterior and interior domains.  
We use $\boldsymbol{\xi}_\theta$, which is described in Table~\ref{tab:xi_theta_cube} for the cube and Table~\ref{tab:xi_theta_triangular_prism} for the triangular prism.  For this coordinate system, $\eta=y$ and $\xi$ is perpendicular to $y$, wrapping clockwise around $y$ along the surfaces for which $\mathbf{n}\cdot\mathbf{e}_y=0$.
For the cube, $\eta\in[0,\,1]\lext$, and $\xi\in[0  ,\,  4] \lext$, beginning at $x=0$ and $z=\lext$.
For the triangular prism, $\eta\in[\aint,\,\bint]$, and $\xi\in\xi_0+[0,\,3]\lint$, where $\xi_0=3(\lext-\lint)/2$, beginning at $x=x_0$ and $z=z_0$.   For both the cube and the triangular prism, the wires are aligned with $\xi_w=3\lext/2$ for $\eta\in[\aslot,\bslot]$.
% {\color{gray}\xslot-\sqrt{3}\lint/2}, {\color{gray}\lext/2}
For the cube, we additionally use $\boldsymbol{\xi}_\phi$, which is described in Table~\ref{tab:xi_phi_cube}.  For this coordinate system, $\eta=x$ and $\xi$ is perpendicular to $x$, wrapping clockwise around $x$ along the surfaces for which $\mathbf{n}\cdot\mathbf{e}_x=0$.
Additionally, $\eta\in[0,\,1]\lext$, and $\xi\in[0  ,\,  4] \lext$, beginning at $y=\lext$ and $z=0$.

%==============================================================================
\subsection{Manufactured Surface Current} %====================================
%==============================================================================

To manufacture compatible surface currents on the exterior and interior surfaces, we use~\eqref{eq:thick_slot_2}.
For the exterior wire,
\begin{align}
\mathbf{s}\cdot\biggl[\mathbf{J}^\text{ext}\times\mathbf{n}^\text{ext} + \frac{1}{4}\biggl(Y_L \frac{d^2}{ds^2} - Y_C\biggr)\mathbf{I}_m^\text{ext}\biggr] = 0,
\label{eq:thick_slot_2_ext}
\end{align}
and, for the interior wire,
\begin{align}
\mathbf{s}\cdot\biggl[\mathbf{J}^\text{int}\times\mathbf{n}^\text{int} + \frac{1}{4}\biggl(Y_L \frac{d^2}{ds^2} - Y_C\biggr)\mathbf{I}_m^\text{int}\biggr] = 0.
\label{eq:thick_slot_2_int}
\end{align}
Since $\mathbf{n}^\text{ext}=-\mathbf{n}^\text{int}$ and $\mathbf{I}_m^\text{ext}=-\mathbf{I}_m^\text{int}$, \eqref{eq:thick_slot_2_ext} and~\eqref{eq:thick_slot_2_int} are combined to yield
\begin{align}
\mathbf{s}\cdot\bigl[(\mathbf{J}^\text{ext}-\mathbf{J}^\text{int})\times\mathbf{n}^\text{ext} \bigr]=0.\label{eq:current_match_0}
\end{align}
In the $\boldsymbol{\xi}_\theta$-coordinate system, $\mathbf{s}=\mathbf{e}_{\eta_\theta}$ and $\mathbf{n}^\text{ext}=\mathbf{e}_{\zeta_\theta}$, such that~\eqref{eq:current_match_0} requires that
\begin{align}
\mathbf{J}^\text{ext}\cdot\mathbf{e}_{\xi_\theta}=\mathbf{J}^\text{int}\cdot\mathbf{e}_{\xi_\theta}.
\label{eq:current_match}
\end{align} 

%===============================================================================
\begin{table}%[!t]
\centering
\begin{tabular}{c c c c c}
\toprule
$j$ & $\mathbf{n}_j$ & $\boldsymbol{\xi}_{\theta_j}$ & $[\xi_{a_j},\,\xi_{b_j}]$ & $\mathbf{x}_{\theta_j}(\boldsymbol{\xi})$ \\ \midrule

1   & 
$\left\{\begin{matrix}\pn0 \\ \pn0 \\ \pn1\end{matrix}\right\}$ & 
$\left\{\begin{matrix}\pn1 \\ \pn0 \\ \pn0\end{matrix}\right\}$ & 
$[0,\,1] \lext$ &
$\displaystyle \left[\begin{matrix}\pn1 & \pn0 & \pn0 \\ \pn0 & \pn1 & \pn0 \\ \pn0 & \pn0 & \pn1\end{matrix}\right]\boldsymbol{\xi} + \left\{\begin{matrix}0 \\ 0 \\ 1\end{matrix}\right\} \lext$ 
\\[1.5em]

2   & 
$\left\{\begin{matrix}\pn0 \\ \pn0 \\   -1\end{matrix}\right\}$ &
$\left\{\begin{matrix}  -1 \\ \pn0 \\ \pn0\end{matrix}\right\}$ &
$[2,\,3] \lext$ &
$\displaystyle \left[\begin{matrix} -1 & \pn0 & \pn0 \\ \pn0 & \pn1 & \pn0 \\ \pn0 & \pn0 & -1\end{matrix}\right]\boldsymbol{\xi} + \left\{\begin{matrix}3 \\ 0 \\ 0\end{matrix}\right\} \lext$ 
\\[1.5em]

5   & 
$\left\{\begin{matrix}\pn1 \\ \pn0 \\ \pn0\end{matrix}\right\}$ & 
$\left\{\begin{matrix}\pn0 \\ \pn0 \\   -1\end{matrix}\right\}$ & 
$[1,\,2] \lext$ &
$\displaystyle \left[\begin{matrix} \pn0 & \pn0 & \pn1 \\ \pn0 & \pn1 & \pn0 \\ -1 & \pn0 & \pn0\end{matrix}\right]\boldsymbol{\xi} + \left\{\begin{matrix}1 \\ 0 \\ 2\end{matrix}\right\} \lext$ 
\\[1.5em]

6   & 
$\left\{\begin{matrix}  -1 \\ \pn0 \\ \pn0\end{matrix}\right\}$ & 
$\left\{\begin{matrix}\pn0 \\ \pn0 \\ \pn1\end{matrix}\right\}$ & 
$[3,\,4] \lext$ &
$\displaystyle \left[\begin{matrix} \pn0 & \pn0 & -1 \\ \pn0 & \pn1 & \pn0 \\ \pn1 & \pn0 & \pn0\end{matrix}\right]\boldsymbol{\xi} - \left\{\begin{matrix}0 \\ 0 \\ 3\end{matrix}\right\} \lext$ 
\\
\bottomrule
\end{tabular}
\caption{Transformation between $\boldsymbol{\xi}_\theta$ and $\mathbf{x}$ for Face $j$ of the cube.  The normal $\mathbf{n}$ points outward ($+\zeta_\theta$).}
\label{tab:xi_theta_cube}
\end{table}

%===============================================================================

\begin{table}%[!h]
\centering
\begin{tabular}{c c c c c}
\toprule
$j$ & $\mathbf{n}_j$ & $\boldsymbol{\xi}_{\theta_j}$ & $[\xi_{a_j},\,\xi_{b_j}]$ & $\mathbf{x}_{\theta_j}(\boldsymbol{\xi})$ \\ \midrule

1   & 
$\left\{\begin{array}{@{} r @{} c @{}}  &\beta  \\ &0 \\ \pn&\alpha \end{array}\right\}$ & 
$\left\{\begin{array}{@{} r @{} c @{}} -&\alpha \\ &0 \\    &\beta  \end{array}\right\}$ & 
$\displaystyle\xi_0+[2,\,3]\lint$ &
\multicolumn{1}{l}{%
$\displaystyle\left[\begin{array}{@{} r @{} c c r @{} c @{}} 
-&\alpha \pn&0 & -&\beta \\ 
 & 0     \pn&1 &  &0     \\ 
 &\beta  \pn&0 & -&\alpha 
\end{array}\right]\boldsymbol{\xi} + \left\{\begin{matrix}\xslot-\alpha(\lint-\xi_1) \\ 0 \\ \beta(\lext-\xi_1)\end{matrix}\right\}$ 
}
\\[1.5em]

2   & 
$\left\{\begin{array}{@{} r @{} c @{}}    &\beta  \\ &0 \\ -&\alpha\end{array}\right\}$ & 
$\left\{\begin{array}{@{} r @{} c @{}} \pn&\alpha \\ &0 \\  &\beta \end{array}\right\}$ &
$\displaystyle\xi_0+[0,\,1]\lint$ &
\multicolumn{1}{l}{%
$\displaystyle\left[\begin{array}{@{} r @{} c c r @{} c @{}} 
\pn &\alpha \pn&0 & -&\beta \\ 
    & 0     \pn&1 &  &0     \\ 
    &\beta  \pn&0 &  &\alpha 
\end{array}\right]\boldsymbol{\xi} + \left\{\begin{matrix}\xslot -\alpha(\lint+\xi_0) \\ 0 \\ \beta(\lext-\xi_0)\end{matrix}\right\}$ 
}
\\[1.5em]

3   & 
$\left\{\begin{array}{@{} r @{} c @{}} -&\SetToWidest{1} \\ &\SetToWidest{0} \\  &\SetToWidest{0}\end{array}\right\}$ & 
$\left\{\begin{array}{@{} r @{} c @{}}  &\SetToWidest{0} \\ &\SetToWidest{0} \\ -&\SetToWidest{1}\end{array}\right\}$ & 
$\displaystyle\xi_0+[1,\,2]\lint$ &
\multicolumn{1}{l}{%
$\displaystyle\left[\begin{array}{@{} r @{} c c r @{} c @{}} 
 & \SetToWidest{0} & \pn0 &  \pn& \SetToWidest{1} \\ 
 & \SetToWidest{0} & \pn1 &  \pn& \SetToWidest{0} \\ 
-& \SetToWidest{1} & \pn0 &  \pn& \SetToWidest{0} 
\end{array}\right]\boldsymbol{\xi} + \left\{\begin{matrix}\xslot \\ \SetToWidestB{0} \\ \lext/2+\xi_w\end{matrix}\right\}$ 
}
\\
\bottomrule
\end{tabular}
\caption{Transformation between $\boldsymbol{\xi}_\theta$ and $\mathbf{x}$ for Face $j$ of the triangular prism.  The normal $\mathbf{n}$ points inward ($-\zeta_\theta$).  $\alpha=\sqrt{3}/2$, $\beta=1/2$, $\xi_0=(3\lext-\lint)/2$, $\xi_1=\xi_0+3\lint$, $\xi_w=3\lext/2$.}
\label{tab:xi_theta_triangular_prism}
\end{table}

%===============================================================================

\begin{table}%[!b]
\centering
\begin{tabular}{c c c c c}
\toprule
$j$ & $\mathbf{n}_j$ & $\boldsymbol{\xi}_{\phi_j}$ & $[\xi_{a_j},\,\xi_{b_j}]$ & $\mathbf{x}_{\phi_j}(\boldsymbol{\xi})$ \\ \midrule

1   & 
$\left\{\begin{matrix}\pn0 \\ \pn0 \\ \pn1\end{matrix}\right\}$ &
$\left\{\begin{matrix}\pn0 \\   -1 \\ \pn0\end{matrix}\right\}$ &
$[1,\,2] \lext$ &
$\displaystyle\left[\begin{matrix} \pn0 & \pn1 & \pn0 \\ -1 & \pn0 & \pn0 \\ \pn0 & \pn0 & \pn1\end{matrix}\right]\boldsymbol{\xi} + \left\{\begin{matrix}0 \\ 2 \\ 1\end{matrix}\right\} \lext$
\\[1.5em]

2   & 
$\left\{\begin{matrix}\pn0 \\ \pn0 \\   -1\end{matrix}\right\}$  & 
$\left\{\begin{matrix}\pn0 \\ \pn1 \\ \pn0\end{matrix}\right\}$  & 
$[3,\,4] \lext$ &
$\displaystyle\left[\begin{matrix} \pn0 & \pn1 & \pn0 \\ \pn1 & \pn0 & \pn0 \\ \pn0 & \pn0 & -1\end{matrix}\right]\boldsymbol{\xi} - \left\{\begin{matrix}0 \\ 3 \\ 0\end{matrix}\right\} \lext$
\\[1.5em]

3   & 
$\left\{\begin{matrix}\pn0 \\ \pn1 \\ \pn0\end{matrix}\right\}$ & 
$\left\{\begin{matrix}\pn0 \\ \pn0 \\ \pn1\end{matrix}\right\}$ & 
$[0,\,1] \lext$ &
$\displaystyle\left[\begin{matrix} \pn0 & \pn1 & \pn0 \\ \pn0 & \pn0 & \pn1 \\ \pn1 & \pn0 & \pn0\end{matrix}\right]\boldsymbol{\xi} + \left\{\begin{matrix}0 \\ 1 \\ 0\end{matrix}\right\} \lext$
\\[1.5em]

4   & 
$\left\{\begin{matrix}\pn0 \\   -1 \\ \pn0\end{matrix}\right\}$ & 
$\left\{\begin{matrix}\pn0 \\ \pn0 \\   -1\end{matrix}\right\}$ & 
$[2,\,3] \lext$ &
$\displaystyle\left[\begin{matrix} \pn0 & \pn1 & \pn0 \\ \pn0 & \pn0 & -1 \\ -1 & \pn0 & \pn0\end{matrix}\right]\boldsymbol{\xi} + \left\{\begin{matrix}0 \\ 0 \\ 3\end{matrix}\right\} \lext$
\\
\bottomrule
\end{tabular}
\caption{Transformation between $\boldsymbol{\xi}_\phi$ and $\mathbf{x}$ for Face $j$ of the cube.  The normal $\mathbf{n}$ points outward ($+\zeta_\phi$).}
\label{tab:xi_phi_cube}
\end{table}

We manufacture surface current densities for the cube and triangular prism using the aforementioned coordinate systems.  For the cube, 
\begin{align}
\mathbf{J}_\text{MS}(\mathbf{x}) = J_{\xi_\theta}(\boldsymbol{\xi}_\theta)\mathbf{e}_{\xi_\theta} + J_{\xi_\phi}(\boldsymbol{\xi}_\phi)\mathbf{e}_{\xi_\phi}.
\label{eq:J_MS_cube}
\end{align}
For the triangular prism, 
\begin{align}
\mathbf{J}_\text{MS}(\mathbf{x}) = J_{\xi_\theta}(\boldsymbol{\xi}_\theta)\mathbf{e}_{\xi_\theta}.
\label{eq:J_MS_triangular_prism}
\end{align}
In~\eqref{eq:J_MS_cube} and~\eqref{eq:J_MS_triangular_prism}, $\mathbf{e}_{\xi_\theta}=(\partial\mathbf{x}/\partial\xi)_{\theta_j}$ and $\mathbf{e}_{\xi_\phi}=(\partial\mathbf{x}/\partial\xi)_{\phi_j}$ in the $\mathbf{x}$-coordinate system.  Additionally,
\begin{align}
J_{\xi_\theta}(\boldsymbol{\xi}) &{}= J_0 f_{\xi_\theta}(\xi)g_{\eta_\theta}(\eta), \label{eq:j_xi_theta}
\\
J_{\xi_\phi}  (\boldsymbol{\xi}) &{}= J_0 f_{\xi_\phi}  (\xi)g_{\eta_\phi}(\eta), \label{eq:j_xi_phi}
\end{align}
where $J_0=1$~A/m, and
\begin{align*}
f_{\xi_\theta}(\xi)&{}=\sin(\gamma(\xi-\bar{\xi}_1)), \\
f_{\xi_\phi}(\xi)&{}=\sin(\gamma(\xi-\bar{\xi}_2)), \\
g_{\eta_\theta}(\eta)&{}=\left\{\begin{matrix}\displaystyle\sin^3\biggl(\pi\frac{\eta-\aint}{\lint}\biggr), & \text{for }\eta\in[\aint,\,\bint]
\\[1em]
0, & \text{otherwise}
\end{matrix}\right., \\
g_{\eta_\phi}(\eta)&{}=\sin^3\biggl(\frac{\pi\eta}{\lext}\biggr).
\end{align*}
For the cube, $\gamma=\pi/(2\lext)$, $\bar{\xi}_1=0$, and $\bar{\xi}_2=\lext/2$; for the triangular prism, $\gamma=2\pi/(3\lint)$ and $\bar{\xi}_1=5\lext/4$.  At the wire locations, where $\xi_w=3\lext/2$, $J_{\xi_\theta}^\text{ext}=J_{\xi_\theta}^\text{int}$, such that~\eqref{eq:current_match} is satisfied.

\begin{figure}[!t]
\centering
\includegraphics[scale=.28,clip=true,trim=0in 0in 0in 0in]{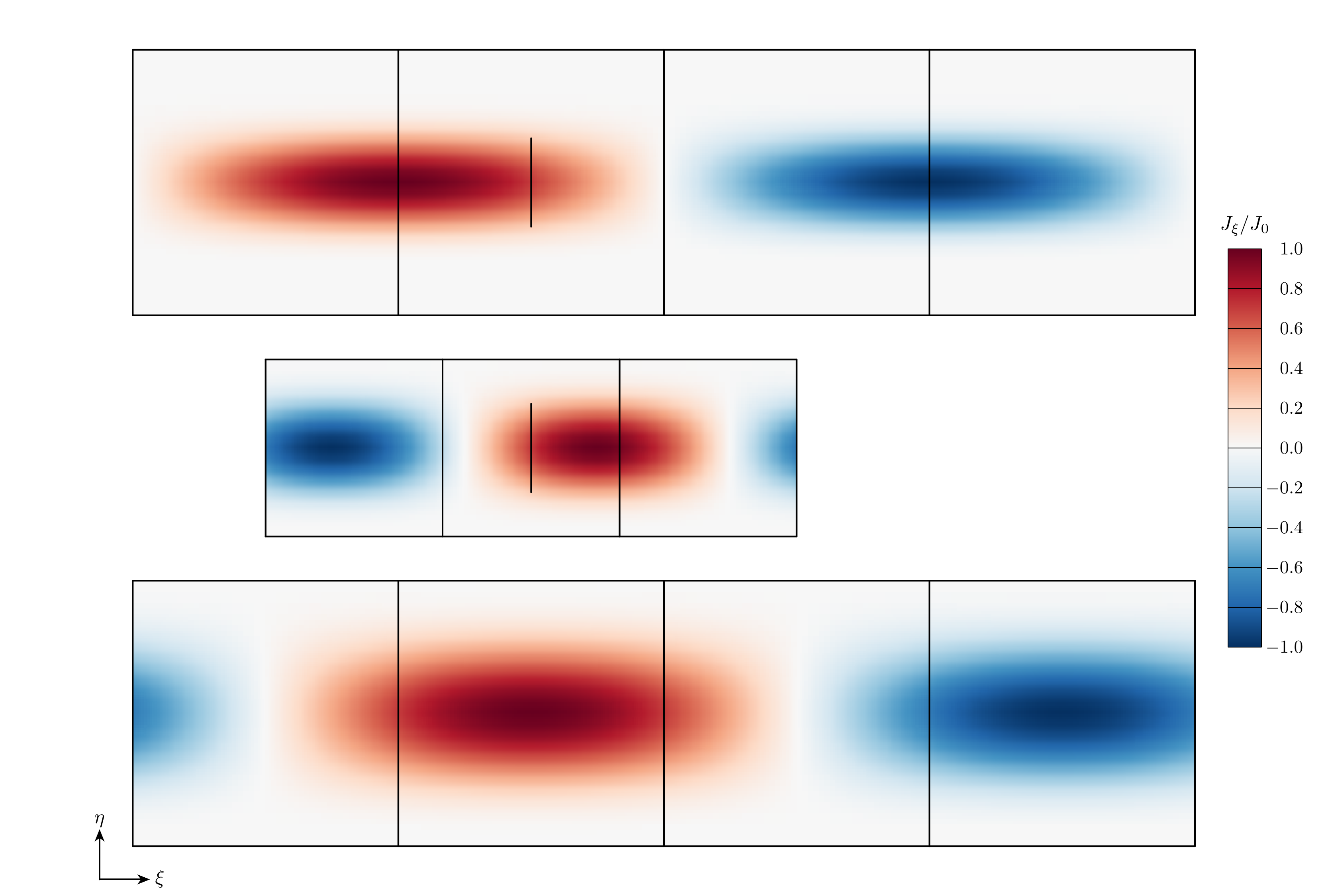}
\caption{\strut Manufactured surface current density $\mathbf{J}_\text{MS}$: $J_{\xi_\theta}$~\eqref{eq:j_xi_theta} for the cube (top) and triangular prism (middle), and $J_{\xi_\phi}$~\eqref{eq:j_xi_phi} for the cube (bottom).}
\vskip-\dp\strutbox
\label{fig:2d_solutions}
\end{figure}

\begin{figure}%[!t]
\centering
\includegraphics[scale=.28,clip=true,trim=0in 0in 0in 0in]{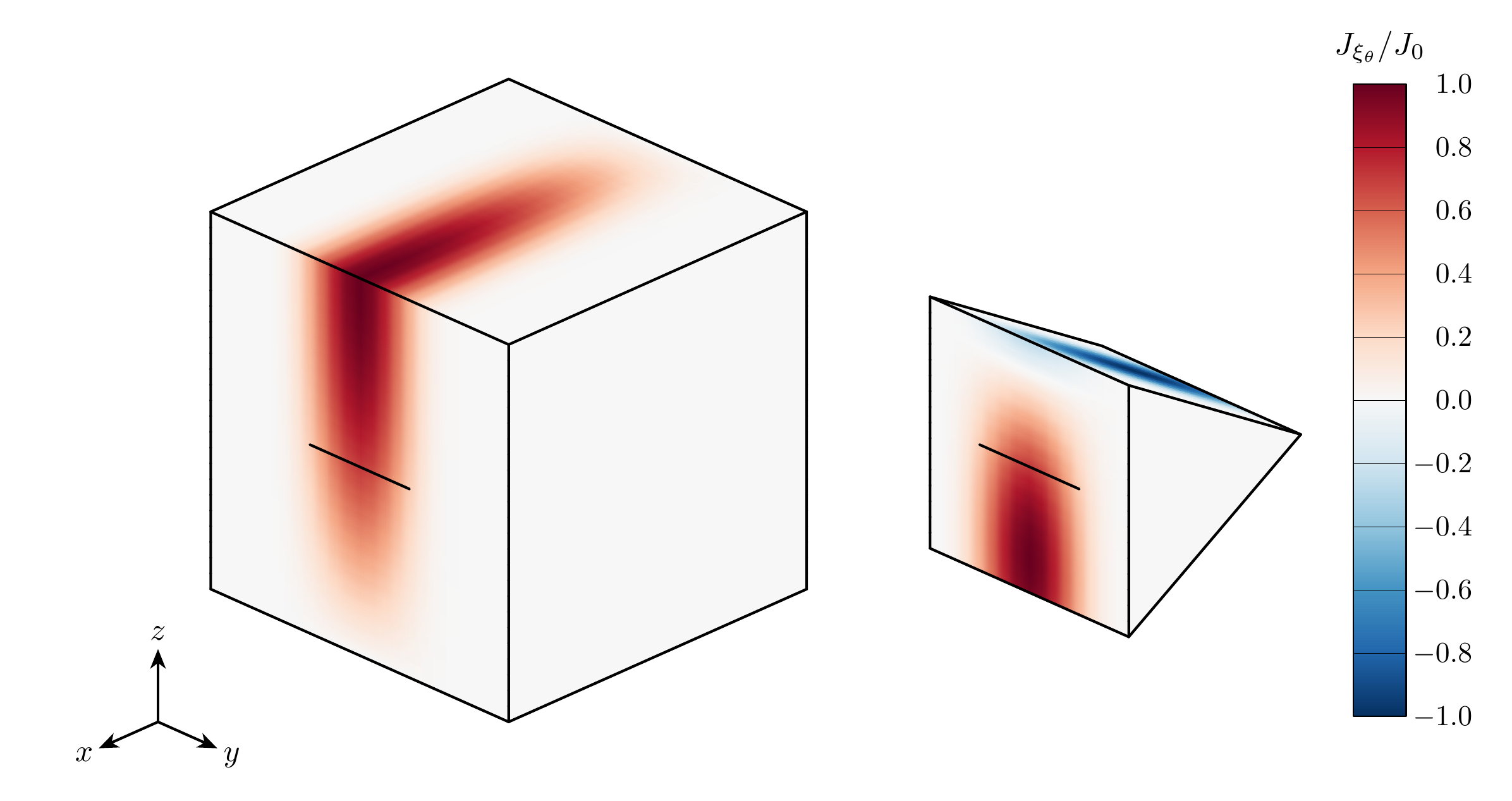}
\includegraphics[scale=.28,clip=true,trim=0in 0in 0in 0in]{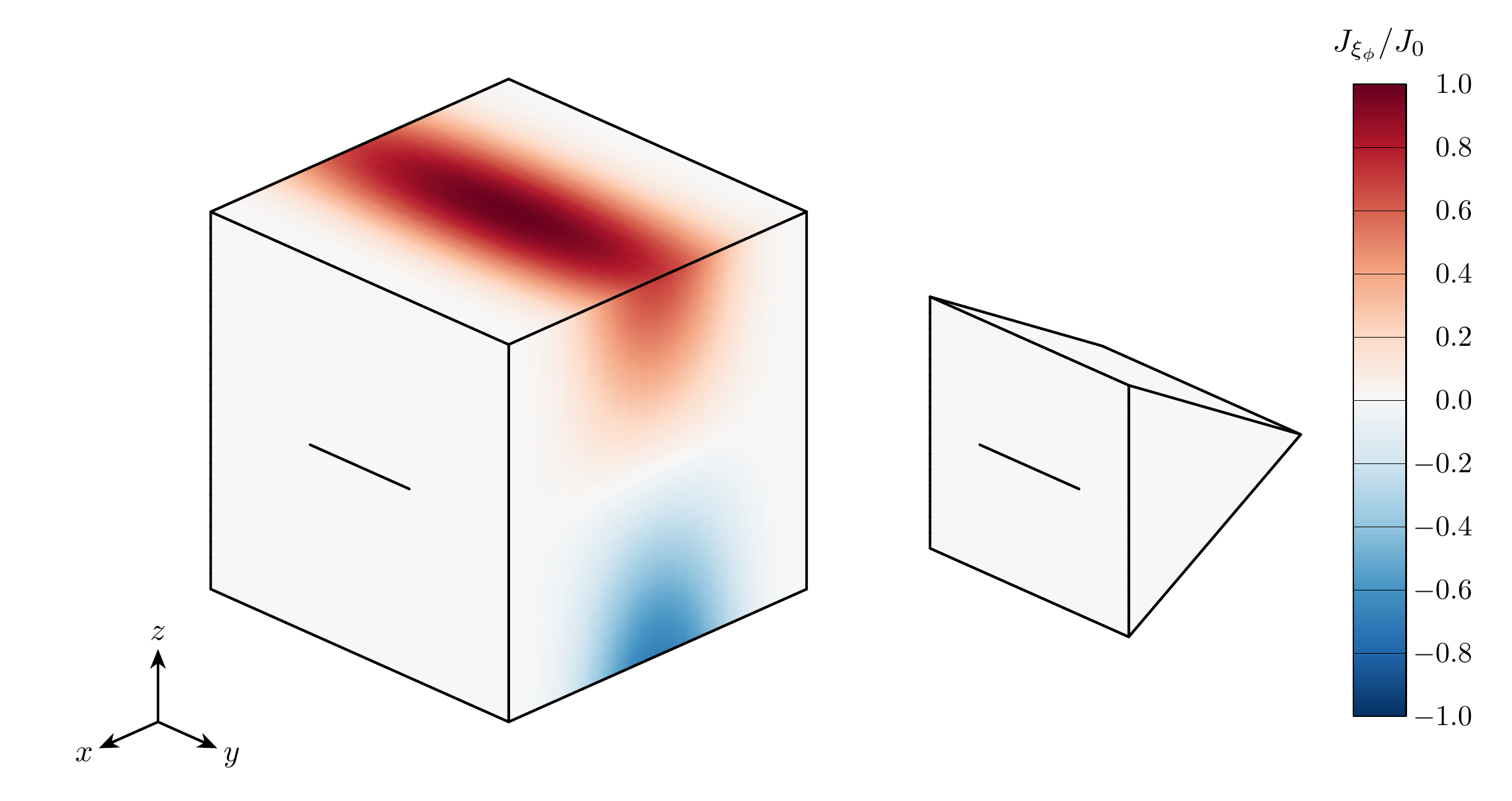}
\includegraphics[scale=.28,clip=true,trim=0in 0in 0in 0in]{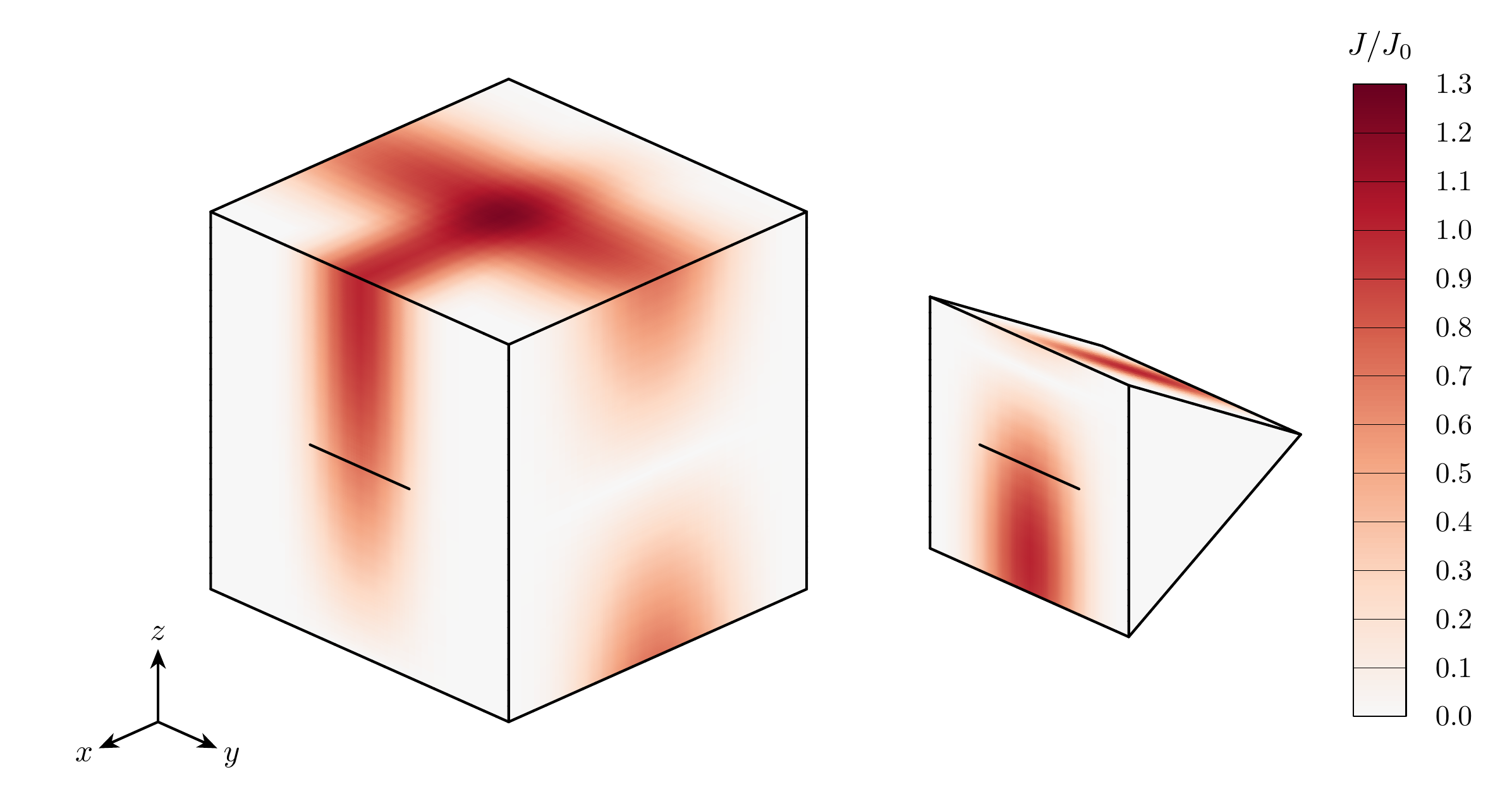}
\caption{\strut Manufactured surface current density $\mathbf{J}_\text{MS}$: $J_{\xi_\theta}$~\eqref{eq:j_xi_theta} (top), $J_{\xi_\phi}$~\eqref{eq:j_xi_phi} (middle), and $J=|\mathbf{J}_\text{MS}|$~\eqref{eq:J_MS_cube} and~\eqref{eq:J_MS_triangular_prism} (bottom).}
\vskip-\dp\strutbox
\label{fig:3d_solutions}
\end{figure}

These equations are chosen because $g_{\eta_\theta}(\eta)$ and $g_{\eta_\phi}(\eta)$ are of class $C^2$, and $f_{\xi_\theta}(\xi)$ and $f_{\xi_\phi}(\xi)$ are periodic with minimal oscillations, such that finer meshes are not required for mesh-convergence studies.
In Figures~\ref{fig:2d_solutions} and~\ref{fig:3d_solutions}, \eqref{eq:j_xi_theta} is plotted for the cube and triangular prism, and~\eqref{eq:j_xi_phi} is plotted for the cube.

%==============================================================================
\subsection{Magnetic Current} %================================================
%==============================================================================

Next, instead of arbitrarily manufacturing $\mathbf{I}_\text{MS}$, we choose $\mathbf{I}_\text{MS}$ to satisfy~\eqref{eq:no_mms_source}, given our choice of $\mathbf{J}_\text{MS}$.  With $\mathbf{I}_m(s)=I_m(s)\mathbf{s}$ and using the $\boldsymbol{\xi}_\theta$-coordinate system, \eqref{eq:thick_slot_2} for the external wire becomes
\begin{align}
-J_{\xi_\theta}(\boldsymbol{\xi}) + \frac{1}{4}\biggl(Y_L \frac{d^2}{ds^2} - Y_C\biggr)I_m(s) = 0,
\label{eq:ji_de}
\end{align}
where $s=\eta-\aslot$, and the boundary conditions are $I_m(0)=I_m(\lslot)=0$.  Solving~\eqref{eq:ji_de} yields
\begin{align*}
I_m(s) ={}
C_0
\biggl[ C_1 \cosh\biggl(\frac{s}{Z}\biggr) + C_2 \sinh\biggl(\frac{s}{Z}\biggr)
  +C_3 \sin\biggl(\frac{ \pi (s+\Delta a)}{\lint}\biggr) + C_4\sin\biggl(\frac{ 3\pi (s+\Delta a)}{\lint}\biggr)
\biggr],
\end{align*}
where 
\begin{alignat*}{7}
C_0      &{}= \frac{J_0 f_{\xi_\theta}(\xi_w) {\lint}^2 }{ Y_1 Y_2}, &\quad
C_1      &{}=3 Y_1 \sin\biggl(\frac{ \pi \Delta a}{\lint}\biggr) - Y_2 \sin\biggl(\frac{ 3\pi \Delta a}{\lint}\biggr), &\quad
C_2      &{} = -C_1\coth\biggl(\frac{L}{Z}\biggr) + C_5\csch\biggl(\frac{L}{Z}\biggr), \\
C_3      &{}= -3 Y_1, &
C_4      &{}= Y_2, &
C_5      &{}=3 Y_1 \sin\biggl(\frac{ \pi \Delta b}{\lint}\biggr) - Y_2 \sin\biggl(\frac{ 3\pi \Delta b}{\lint}\biggr), \\
\Delta a &{}= \aslot - \aint, &
\Delta b &{}= \bslot - \aint, &
Z        &{}= \sqrt{Y_L/Y_C}, \\
Y_1      &{}= {\lint}^2Y_C + 9\pi^2Y_L, &
Y_2      &{}= {\lint}^2Y_C +  \pi^2Y_L.
\end{alignat*}

\begin{table}[!t]
\centering
\begin{tabular}{c c c c}
\toprule
Maximum          & Number of       & Number of  & Convergence        \\
integrand degree & triangle points & bar points & rate               \\
\midrule
1                & 1               & 1          & $\mathcal{O}(h^2)$ \\
2                & 3               & ---        & $\mathcal{O}(h^4)$ \\
3                & 4               & 2          & $\mathcal{O}(h^4)$ \\
4                & 6               & ---        & $\mathcal{O}(h^6)$ \\
5                & 7               & 3          & $\mathcal{O}(h^6)$ \\
\bottomrule
\end{tabular}
\caption{Polynomial quadrature rule properties.}
\vskip-\dp\strutbox
\label{tab:quadrature_properties}
\end{table}

\begin{figure}%[!t]
\centering
\begin{subfigure}[b]{.49\textwidth}
\includegraphics[scale=.64,clip=true,trim=2.3in 0in 2.8in 0in]{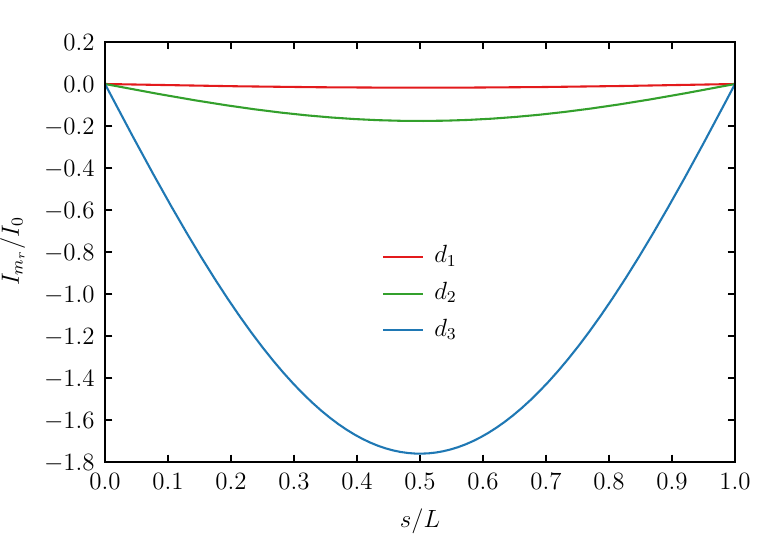}
\caption{$I_{m_r}$\vpad}
\end{subfigure}
\hspace{0.25em}
\begin{subfigure}[b]{.49\textwidth}
\includegraphics[scale=.64,clip=true,trim=2.3in 0in 2.8in 0in]{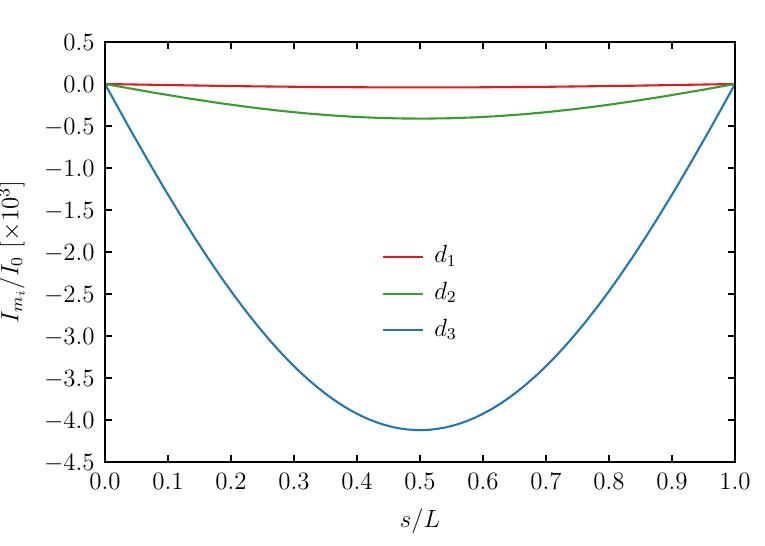}
\caption{$I_{m_i}$\vpad}
\end{subfigure}

\caption{Real and imaginary components of $I_m=I_{m_r}+jI_{m_i}$ for 3 depths.}
\vskip-\dp\strutbox
\label{fig:I_m}
\end{figure}

For $d\in\{d_1,d_2,d_3\}$, Figure~\ref{fig:I_m} shows the real and imaginary components of $I_m(s)$, normalized by $I_0=f_{\xi_\theta}(\xi_w)$~V.

%% I_m =======================================================================

%==============================================================================
\subsection{Numerical Integration} %===========================================
%==============================================================================
When solving~\eqref{eq:proj_disc_efie} and~\eqref{eq:proj_disc_slot}, numerical integration is performed using two-dimensional polynomial quadrature rules for triangles and one-dimensional polynomial quadrature rules for bars.  For multiple quadrature point amounts, Table~\ref{tab:quadrature_properties} lists the maximum polynomial degree of the integrand the points can integrate exactly in two dimensions~\cite{lyness_1975,dunavant_1985} and one dimension~\cite[Chap.~5]{kahaner_1989}, as well as the convergence rates of the errors for inexact integrations of nonsingular integrands.  The properties listed assume optimal point locations and weights.  %Figure~\ref{fig:quad_optimal_6} shows the optimal 6-point quadrature rule, which can exactly integrate polynomials up to degree 4.

When integrating the left-hand sides of~\eqref{eq:proj_disc_efie} and~\eqref{eq:proj_disc_slot}, we note that, in~\eqref{eq:proj_disc_efie}, the maximum polynomial degree of the two-dimensional test and source integrands of $\aee(\mathbf{J}_h,\boldsymbol{\Lambda}_{\testidx})$ is $2q+1$.  The maximum polynomial degree of the one-dimensional integrand of ${\aem}_1(\mathbf{I}_h,\boldsymbol{\Lambda}_{\testidx})$ is $1$.  For ${\aem}_2(\mathbf{I}_h,\boldsymbol{\Lambda}_{\testidx})$, the one-dimensional integral with respect to $\phi'$ is precomputed.  The maximum polynomial degree of the one-dimensional integrand with respect to $s'$ is $2q-1$.  The maximum polynomial degree of the two-dimensional test integrand is $2q$.
In~\eqref{eq:proj_disc_slot}, the maximum polynomial degrees of the one-dimensional integrands are $1$ for $\ame(\mathbf{J}_h,\boldsymbol{\Lambda}_{\testidx}^m)$ and $2$ for $\amm(\mathbf{I}_h,\boldsymbol{\Lambda}_{\testidx}^m)$.  
Therefore, for $G_1$, four quadrature points integrate exactly for triangular elements and two points integrate exactly for bar elements.  For $G_2$, seven quadrature points integrate exactly for triangular elements and two points integrate exactly for bar elements.

When integrating the right-hand side of~\eqref{eq:proj_disc_efie}, we note that the terms in $\mathbf{E}^\mathcal{I}$~\eqref{eq:E_i}, excluding $Z_s \mathbf{J}_\text{MS}$ and $(\mathbf{n}\times \mathbf{I}_\text{MS})\delta_\text{slot}$, are polynomials that do not exceed degree $2q$.  The corresponding terms in $\be\bigl(\mathbf{E}^\mathcal{I}, \boldsymbol{\Lambda}_{\testidx}\bigr)$ do not exceed degree $2q+1$ and can be integrated exactly for triangular elements using four points for $G_1$ and seven points for $G_2$.  
The contributions to $\be\bigl(\mathbf{E}^\mathcal{I}, \boldsymbol{\Lambda}_{\testidx}\bigr)$ from $\be(Z_s \mathbf{J}_\text{MS}, \boldsymbol{\Lambda}_{\testidx})$~\eqref{eq:E_i} and $\be\bigl((\mathbf{n}\times \mathbf{I}_\text{MS})\delta_\text{slot},\boldsymbol{\Lambda}_{\testidx}\bigr)$~\eqref{eq:delta_slot} are computed analytically.

%==============================================================================
\subsection{Solution-Discretization Error} %===================================
%==============================================================================

% Part 1 =======================================================================
\begin{figure}%[!t]
\centering
\begin{subfigure}[b]{.49\textwidth}
\includegraphics[scale=.64,clip=true,trim=2.3in 0in 2.8in 0in]{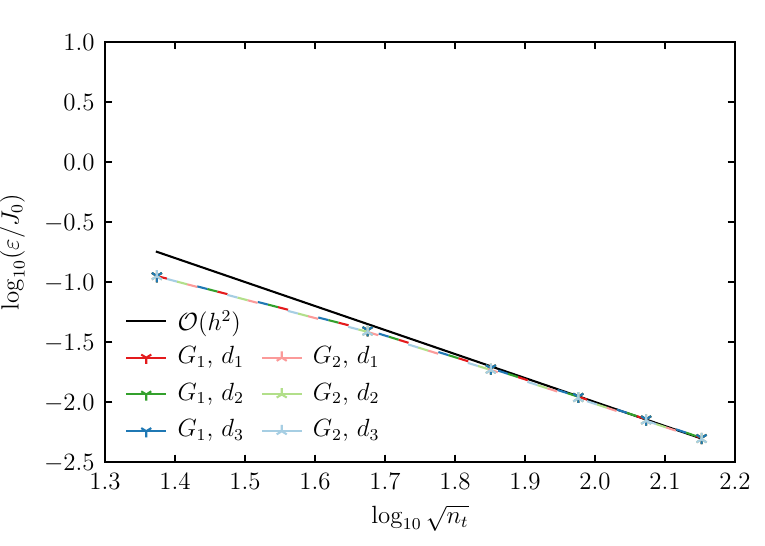}
\caption{$\mathbf{B}_1\ne\mathbf{0}$, $\mathbf{B}_2=\mathbf{0}$, $\mathbf{e}=\mathbf{e}_\mathbf{J}$~\eqref{eq:solution_error_J}\vpad}
\label{fig:p1_error_1J}
\end{subfigure}
\hspace{0.25em}
\begin{subfigure}[b]{.49\textwidth}
\includegraphics[scale=.64,clip=true,trim=2.3in 0in 2.8in 0in]{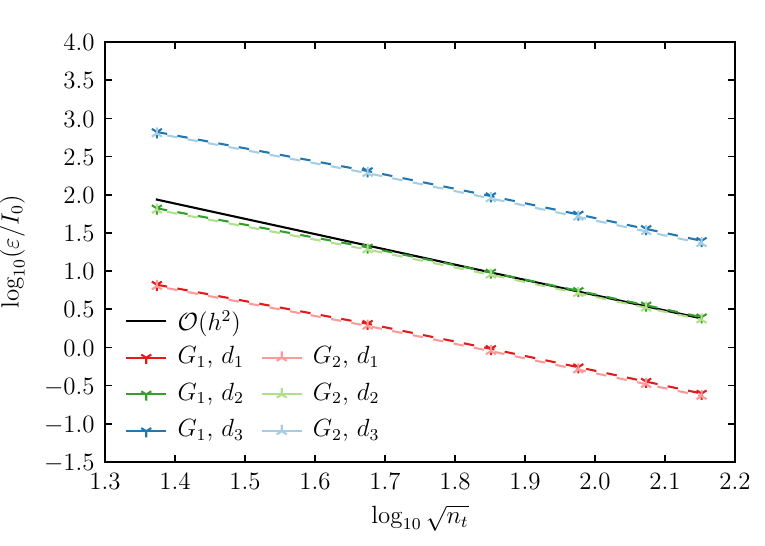}
\caption{$\mathbf{B}_1\ne\mathbf{0}$, $\mathbf{B}_2=\mathbf{0}$, $\mathbf{e}=\mathbf{e}_\mathbf{I}$~\eqref{eq:solution_error_I}\vpad}
\label{fig:p1_error_1I}
\end{subfigure}
\\
\begin{subfigure}[b]{.49\textwidth}
\includegraphics[scale=.64,clip=true,trim=2.3in 0in 2.8in 0in]{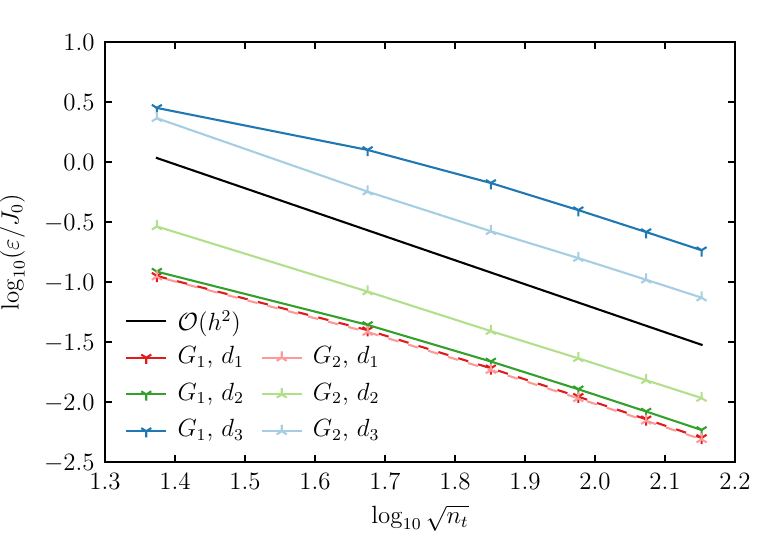}
\caption{$\mathbf{B}_1=\mathbf{0}$, $\mathbf{B}_2\ne\mathbf{0}$, $\mathbf{e}=\mathbf{e}_\mathbf{J}$~\eqref{eq:solution_error_J}\vpad}
\label{fig:p1_error_2J}
\end{subfigure}
\hspace{0.25em}
\begin{subfigure}[b]{.49\textwidth}
\includegraphics[scale=.64,clip=true,trim=2.3in 0in 2.8in 0in]{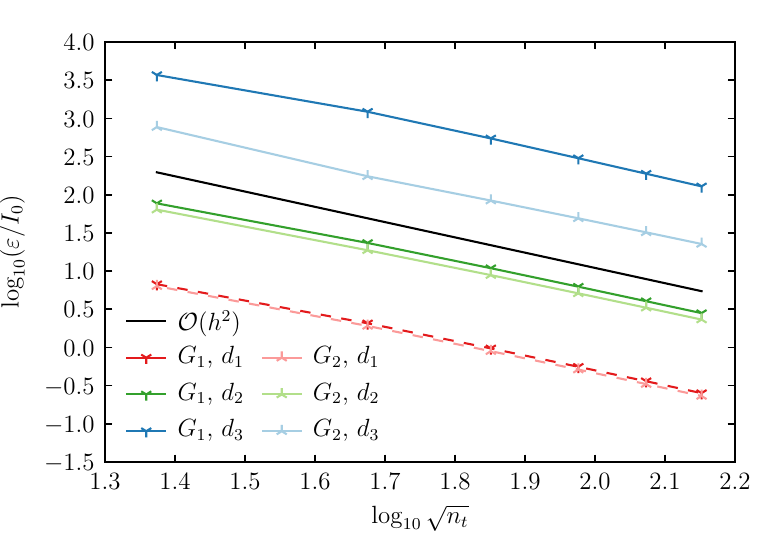}
\caption{$\mathbf{B}_1=\mathbf{0}$, $\mathbf{B}_2\ne\mathbf{0}$, $\mathbf{e}=\mathbf{e}_\mathbf{I}$~\eqref{eq:solution_error_I}\vpad}
\label{fig:p1_error_2I}
\end{subfigure}
\\
\begin{subfigure}[b]{.49\textwidth}
\includegraphics[scale=.64,clip=true,trim=2.3in 0in 2.8in 0in]{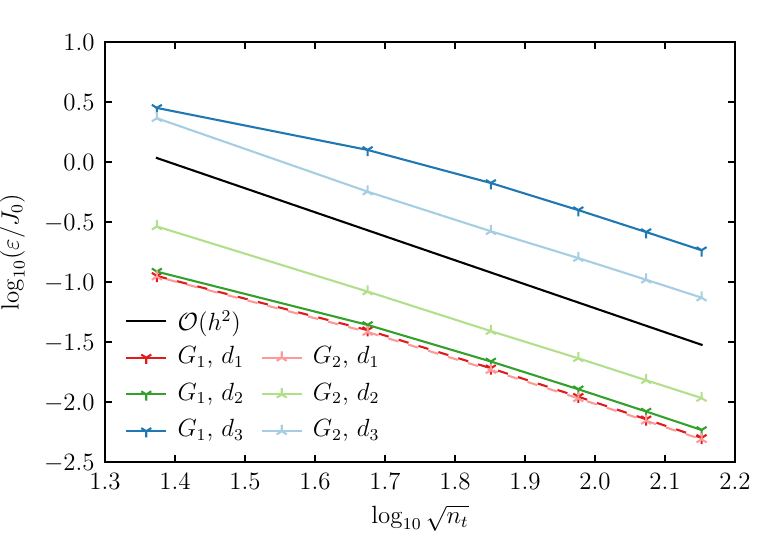}
\caption{$\mathbf{B}_1\ne\mathbf{0}$, $\mathbf{B}_2\ne\mathbf{0}$, $\mathbf{e}=\mathbf{e}_\mathbf{J}$~\eqref{eq:solution_error_J}\vpad}
\label{fig:p1_error_3J}
\end{subfigure}
\hspace{0.25em}
\begin{subfigure}[b]{.49\textwidth}
\includegraphics[scale=.64,clip=true,trim=2.3in 0in 2.8in 0in]{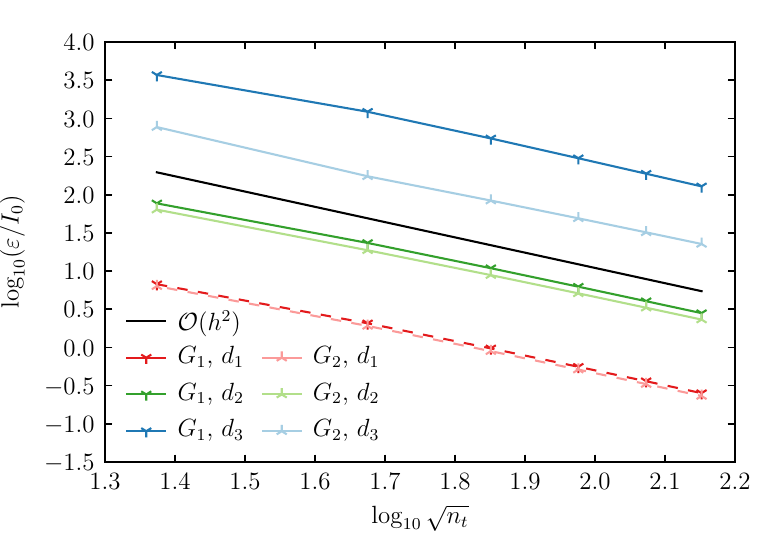}
\caption{$\mathbf{B}_1\ne\mathbf{0}$, $\mathbf{B}_2\ne\mathbf{0}$, $\mathbf{e}=\mathbf{e}_\mathbf{I}$~\eqref{eq:solution_error_I}\vpad}
\label{fig:p1_error_3I}
\end{subfigure}
\caption{Solution-discretization error: $\varepsilon={\|\mathbf{e}\|}_\infty$ %for 2 Green's functions~\eqref{eq:G_mms} and 3 depths 
with the discontinuity removed when $\mathbf{B}_1\ne\mathbf{0}$.}
\vskip-\dp\strutbox
\label{fig:p1_error}
\end{figure}

To isolate and measure the solution-discretization error, we proceed with the assessment described in Section~\ref{sec:sde} to remove the discontinuity.  \reviewerTwo{As mentioned in Section~\ref{sec:domain}, we consider three cases: 
\begin{itemize}
\item $\mathbf{B}_1\ne \mathbf{0}$ and $\mathbf{B}_2 = \mathbf{0}$,
\item $\mathbf{B}_1 = \mathbf{0}$ and $\mathbf{B}_2\ne \mathbf{0}$,
\item $\mathbf{B}_1\ne \mathbf{0}$ and $\mathbf{B}_2\ne \mathbf{0}$.
\end{itemize}
When $\mathbf{B}_1\ne \mathbf{0}$, we remove the discontinuity; when $\mathbf{B}_1=\mathbf{0}$, there is no discontinuity to remove.}
The integrals on both sides of~\eqref{eq:proj_disc_efie} and~\eqref{eq:proj_disc_slot} are computed exactly. 

Figure~\ref{fig:p1_error} shows the $L^\infty$-norm of the discretization errors in~\eqref{eq:solution_error_J} and~\eqref{eq:solution_error_I}: ${\|\mathbf{e}_\mathbf{J}\|}_\infty$ and ${\|\mathbf{e}_\mathbf{I}\|}_\infty$, which arise from only the solution-discretization error.  Error norms are shown for $G_\text{MS}\in\{G_1,\,G_2\}$~\eqref{eq:G_mms} and $d\in\{d_1,\,d_2,\,d_3\}$.  Removing the discontinuity for the case with $\mathbf{B}_1\ne\mathbf{0}$ and $\mathbf{B}_2\ne\mathbf{0}$ in Figures~\ref{fig:p1_error_3J} and~\ref{fig:p1_error_3I} yields the same errors as the case with $\mathbf{B}_1=\mathbf{0}$ and $\mathbf{B}_2\ne\mathbf{0}$ in Figures~\ref{fig:p1_error_2J} and~\ref{fig:p1_error_2I}.  The convergence rates for all of these cases are $\mathcal{O}(h^2)$, as expected.

\begin{figure}%[!t]
\centering
\begin{subfigure}[b]{.49\textwidth}
\includegraphics[scale=.64,clip=true,trim=2.3in 0in 2.8in 0in]{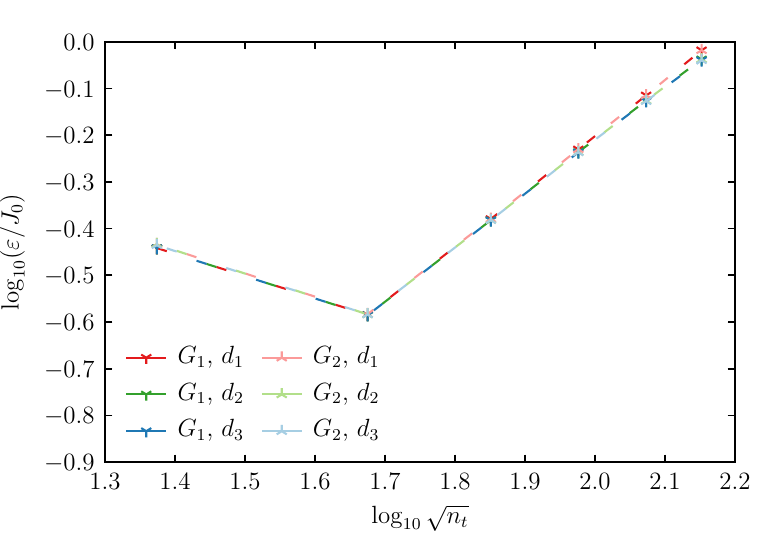}
\caption{$\mathbf{e}=\mathbf{e}_\mathbf{J}$~\eqref{eq:solution_error_J}, $\mathbf{e}_\mathbf{J}\blra \mathbf{e}_\mathbf{I}$~\eqref{eq:system2}\vpad}
\label{fig:p1_error_b4J}
\end{subfigure}
\hspace{0.25em}
\begin{subfigure}[b]{.49\textwidth}
\includegraphics[scale=.64,clip=true,trim=2.3in 0in 2.8in 0in]{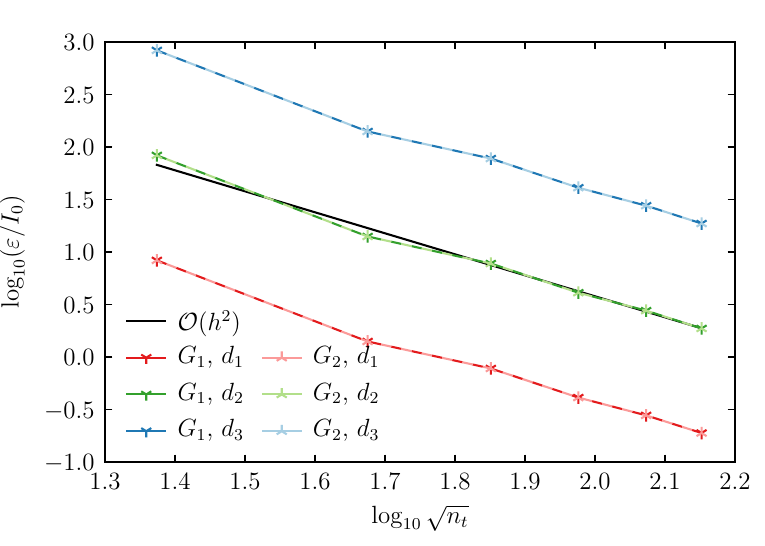}
\caption{$\mathbf{e}=\mathbf{e}_\mathbf{I}$~\eqref{eq:solution_error_I}, $\mathbf{e}_\mathbf{J}\blra \mathbf{e}_\mathbf{I}$~\eqref{eq:system2}\vpad}
\label{fig:p1_error_b4I}
\end{subfigure}

\caption{Solution-discretization error: $\varepsilon={\|\mathbf{e}\|}_\infty$ with the discontinuity.}
\vskip-\dp\strutbox
\label{fig:p1_error_b4}
\end{figure}

To demonstrate the consequence of not using the approach described in Section~\ref{sec:sde}, Figure~\ref{fig:p1_error_b4} shows the convergence rates for $\mathbf{B}_1\ne\mathbf{0}$ and $\mathbf{B}_2\ne\mathbf{0}$ when the discontinuity is not removed.  \reviewerTwo{When the discontinuity is not removed, $\mathbf{Z}$ is defined by~\eqref{eq:Z} instead of~\eqref{eq:Z4}, and $\mathbf{E}^\mathcal{I}$ is defined by~\eqref{eq:E_i} instead of~\eqref{eq:E_i2}.}  For the meshes considered, asymptotic convergence is not demonstrated.  Though ${\|\mathbf{e}_\mathbf{I}\|}_\infty$ is approximately $\mathcal{O}(h^2)$ in Figure~\ref{fig:p1_error_b4I}, ${\|\mathbf{e}_\mathbf{J}\|}_\infty$ does not decrease with refinement in Figure~\ref{fig:p1_error_b4J}.

In light of the results in Figure~\ref{fig:p1_error_b4}, a natural concern is whether~\eqref{eq:E_i} has been correctly implemented.  To assess this, we reconsider the system of equations~\eqref{eq:system}:
\begin{align}
\left[\begin{matrix}
\mathbf{A} & \mathbf{B} \\ 
\mathbf{C} & \mathbf{D} \end{matrix}\right]
\left\{\begin{array}{@{} r @{} l @{}}
\mathbf{J}&^h \\ 
\mathbf{I}&^h\end{array}\right\}
{}={}
\left\{\begin{array}{@{} c @{} l @{}}
\mathbf{V}& ^\mathcal{E} \\ 
\mathbf{0}& \end{array}\right\}.
\label{eq:system2}
\end{align}

\begin{figure}%[!t]
\centering
\begin{subfigure}[b]{.49\textwidth}
\includegraphics[scale=.64,clip=true,trim=2.3in 0in 2.8in 0in]{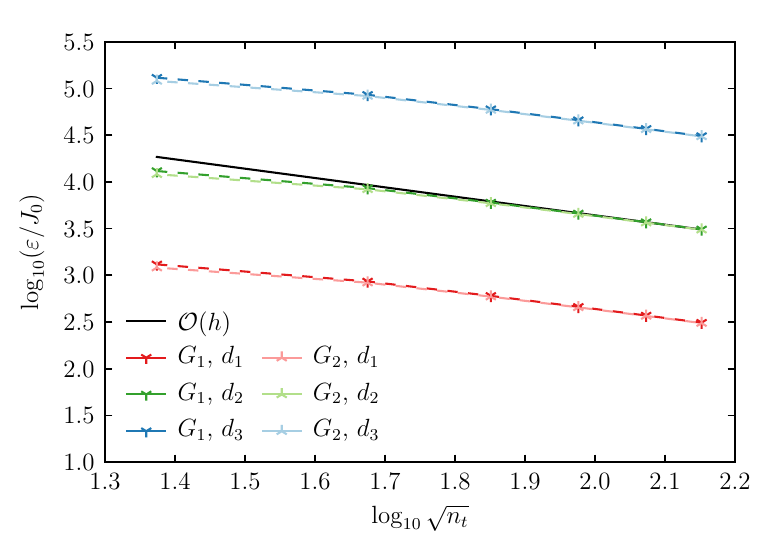}
\caption{$\mathbf{e}=\mathbf{e}_\mathbf{J}$~\eqref{eq:solution_error_J}, $\mathbf{e}_\mathbf{J}\bnlra \mathbf{e}_\mathbf{I}$~\eqref{eq:system3}\vpad}
\label{fig:p1_error_b5J}
\end{subfigure}
\hspace{0.25em}
\begin{subfigure}[b]{.49\textwidth}
\includegraphics[scale=.64,clip=true,trim=2.3in 0in 2.8in 0in]{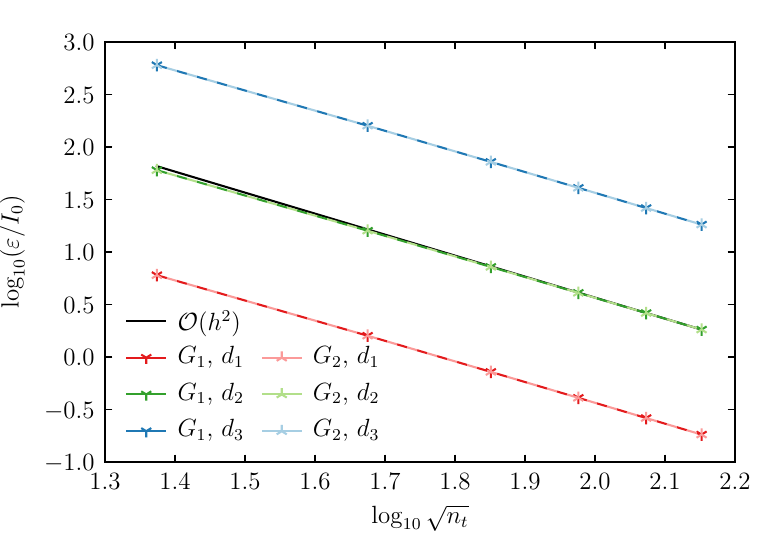}
\caption{$\mathbf{e}=\mathbf{e}_\mathbf{I}$~\eqref{eq:solution_error_I}, $\mathbf{e}_\mathbf{J}\bnlra \mathbf{e}_\mathbf{I}$~\eqref{eq:system3}\vpad}
\label{fig:p1_error_b5I}
\end{subfigure}
\\
\begin{subfigure}[b]{.49\textwidth}
\includegraphics[scale=.64,clip=true,trim=2.3in 0in 2.8in 0in]{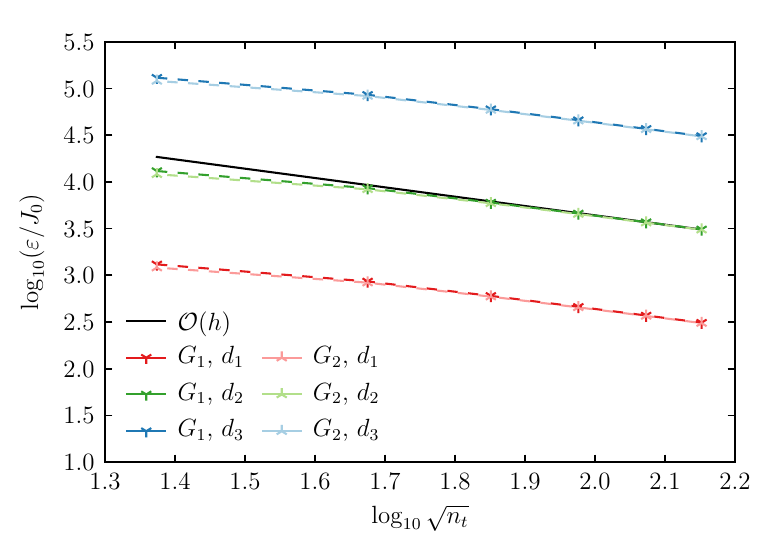}
\caption{$\mathbf{e}=\mathbf{e}_\mathbf{J}$~\eqref{eq:solution_error_J}, $\mathbf{e}_\mathbf{J}\bra \mathbf{e}_\mathbf{I}$~\eqref{eq:system4}\vpad}
\label{fig:p1_error_b6J}
\end{subfigure}
\hspace{0.25em}
\begin{subfigure}[b]{.49\textwidth}
\includegraphics[scale=.64,clip=true,trim=2.3in 0in 2.8in 0in]{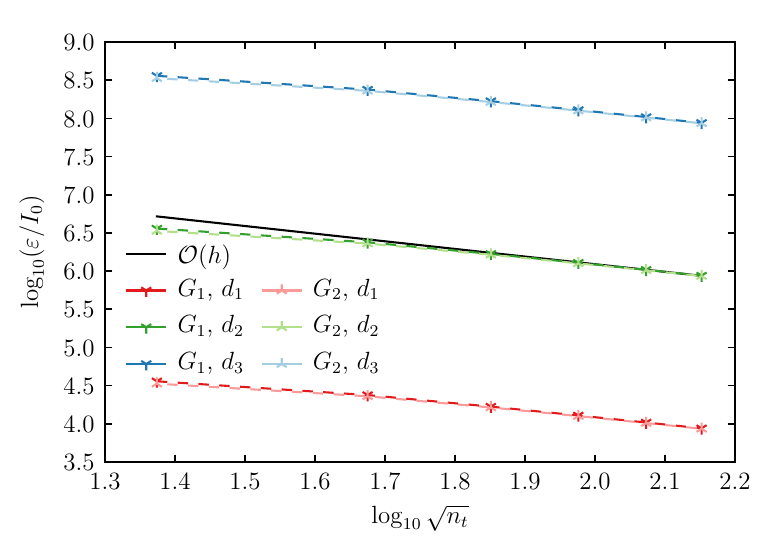}
\caption{$\mathbf{e}=\mathbf{e}_\mathbf{I}$~\eqref{eq:solution_error_I}, $\mathbf{e}_\mathbf{J}\bra \mathbf{e}_\mathbf{I}$~\eqref{eq:system4}\vpad}
\label{fig:p1_error_b6I}
\end{subfigure}
\\
\begin{subfigure}[b]{.49\textwidth}
\includegraphics[scale=.64,clip=true,trim=2.3in 0in 2.8in 0in]{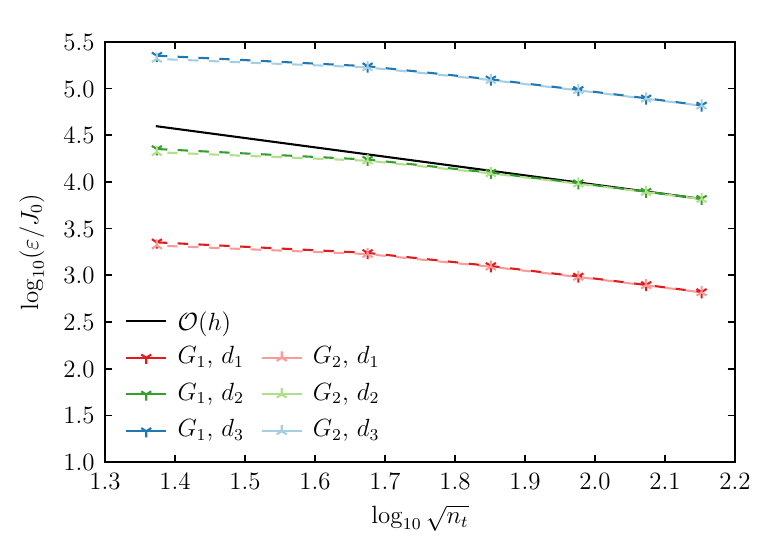}
\caption{$\mathbf{e}=\mathbf{e}_\mathbf{J}$~\eqref{eq:solution_error_J}, $\mathbf{e}_\mathbf{J}\bla \mathbf{e}_\mathbf{I}$~\eqref{eq:system5}\vpad}
\label{fig:p1_error_b7J}
\end{subfigure}
\hspace{0.25em}
\begin{subfigure}[b]{.49\textwidth}
\includegraphics[scale=.64,clip=true,trim=2.3in 0in 2.8in 0in]{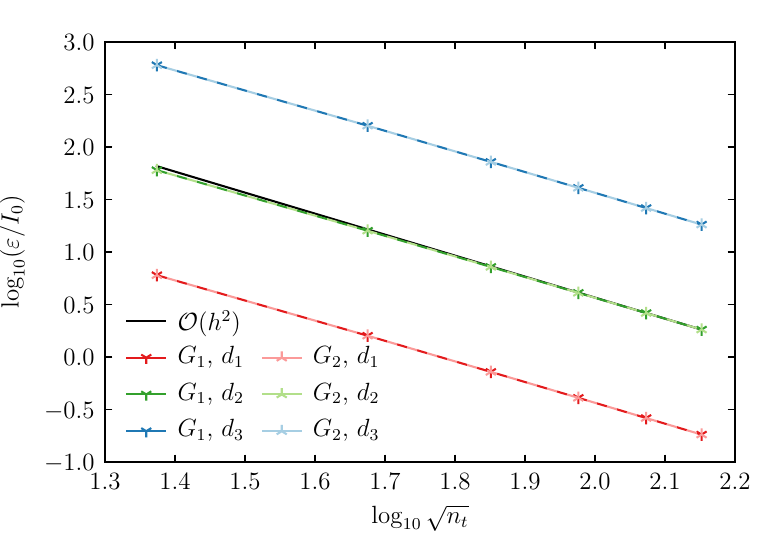}
\caption{$\mathbf{e}=\mathbf{e}_\mathbf{I}$~\eqref{eq:solution_error_I}, $\mathbf{e}_\mathbf{J}\bla \mathbf{e}_\mathbf{I}$~\eqref{eq:system5}\vpad}
\label{fig:p1_error_b7I}
\end{subfigure}
\caption{Solution-discretization error: $\varepsilon={\|\mathbf{e}\|}_\infty$ with the discontinuity for different discretization error interactions.}
\vskip-\dp\strutbox
\label{fig:p1_error_b567}
\end{figure}

We first decouple the interaction of the discretization errors $\mathbf{e}_\mathbf{J}$ and $\mathbf{e}_\mathbf{I}$ by modifying~\eqref{eq:system2} to be
\begin{align}
\left[\begin{matrix}
\mathbf{A} & \mathbf{0} \\ 
\mathbf{0} & \mathbf{D} \end{matrix}\right]
\left\{\begin{array}{@{} r @{} l @{}}
\mathbf{J}&^h \\ 
\mathbf{I}&^h\end{array}\right\}
{}={}
\left\{\begin{matrix}
\mathbf{V}^\mathcal{E} - \mathbf{B}\mathbf{I}_s \\ 
-\mathbf{C}\mathbf{J}_n \end{matrix}\right\},
\label{eq:system3}
\end{align}
where $\mathbf{J}_n$~\eqref{eq:solution_error_J} and $\mathbf{I}_s$~\eqref{eq:solution_error_I} are the exact solutions.
In~\eqref{eq:system3}, $\mathbf{e}_\mathbf{J}$ and $\mathbf{e}_\mathbf{I}$ are independent of each other ($\mathbf{e}_\mathbf{J}\nleftrightarrow \mathbf{e}_\mathbf{I}$), but still depend on both $\mathbf{J}_\text{MS}$ and $\mathbf{I}_\text{MS}$.  Solving~\eqref{eq:system3}, Figures~\ref{fig:p1_error_b5J} and~\ref{fig:p1_error_b5I} show that, by decoupling $\mathbf{e}_\mathbf{J}$ and $\mathbf{e}_\mathbf{I}$, ${\|\mathbf{e}_\mathbf{J}\|}_\infty$ is $\mathcal{O}(h)$ in Figure~\ref{fig:p1_error_b5J} and ${\|\mathbf{e}_\mathbf{I}\|}_\infty$ is $\mathcal{O}(h^2)$ in Figure~\ref{fig:p1_error_b5I}, both as expected.

Next, instead of fully decoupling the discretization errors, we remove the influence of $\mathbf{e}_\mathbf{I}$ on $\mathbf{e}_\mathbf{J}$, but we preserve the influence of $\mathbf{e}_\mathbf{J}$ on $\mathbf{e}_\mathbf{I}$ ($\mathbf{e}_\mathbf{J}\rightarrow \mathbf{e}_\mathbf{I}$).  The modification to~\eqref{eq:system2} is
\begin{align}
\left[\begin{matrix}
\mathbf{A} & \mathbf{0} \\ 
\mathbf{C} & \mathbf{D} \end{matrix}\right]
\left\{\begin{array}{@{} r @{} l @{}}
\mathbf{J}&^h \\ 
\mathbf{I}&^h\end{array}\right\}
{}={}
\left\{\begin{matrix}
\mathbf{V}^\mathcal{E} - \mathbf{B}\mathbf{I}_s \\ 
\mathbf{0} \end{matrix}\right\}.
\label{eq:system4}
\end{align}
Solving~\eqref{eq:system4}, Figures~\ref{fig:p1_error_b6J} and~\ref{fig:p1_error_b6I} show that ${\|\mathbf{e}_\mathbf{J}\|}_\infty$ is $\mathcal{O}(h)$ in Figure~\ref{fig:p1_error_b6J}, which, in turn, causes ${\|\mathbf{e}_\mathbf{I}\|}_\infty$ to be $\mathcal{O}(h)$ in Figure~\ref{fig:p1_error_b6I}.

Finally, we remove the influence of $\mathbf{e}_\mathbf{J}$ on $\mathbf{e}_\mathbf{I}$, but we preserve the influence of $\mathbf{e}_\mathbf{I}$ on $\mathbf{e}_\mathbf{J}$ ($\mathbf{e}_\mathbf{J}\leftarrow \mathbf{e}_\mathbf{I}$).  The modification to~\eqref{eq:system2} is
\begin{align}
\left[\begin{matrix}
\mathbf{A} & \mathbf{B}\\ 
\mathbf{0} & \mathbf{D} \end{matrix}\right]
\left\{\begin{array}{@{} r @{} l @{}}
\mathbf{J}&^h \\ 
\mathbf{I}&^h\end{array}\right\}
{}={}
\left\{\begin{matrix}
\mathbf{V}^\mathcal{E} \\ 
-\mathbf{C}\mathbf{J}_n \end{matrix}\right\}.
\label{eq:system5}
\end{align}
Solving~\eqref{eq:system5}, Figures~\ref{fig:p1_error_b7J} and~\ref{fig:p1_error_b7I} show that ${\|\mathbf{e}_\mathbf{I}\|}_\infty$ is $\mathcal{O}(h^2)$ in Figure~\ref{fig:p1_error_b7I} and ${\|\mathbf{e}_\mathbf{J}\|}_\infty$ is $\mathcal{O}(h)$ in Figure~\ref{fig:p1_error_b7J}, both as expected.

It is worth noting that, while the expected convergence rates are obtained in Figure~\ref{fig:p1_error_b567} from solving~\eqref{eq:system3}--\eqref{eq:system5}, ${\|\mathbf{e}_\mathbf{J}\|}_\infty$ is much greater in Figure~\ref{fig:p1_error_b567} than in Figure~\ref{fig:p1_error_b4} from solving~\eqref{eq:system2}.  However, the lack of convergence of ${\|\mathbf{e}_\mathbf{J}\|}_\infty$ from solving~\eqref{eq:system2} renders traditional convergence studies ineffective.  This issue is mitigated by removing the discontinuity, as described in Section~\ref{sec:sde} and shown in Figures~\ref{fig:p1_error_3J} and~\ref{fig:p1_error_3I}.

% Part 1, B4 ===================================================================

% Part 1, B5--B7 ===============================================================

%===============================================================================
\subsection{Numerical-Integration Error} %======================================
%===============================================================================

% Part 5 =======================================================================
\begin{figure}%[!t]
\centering
\begin{subfigure}[b]{.49\textwidth}
\includegraphics[scale=.64,clip=true,trim=2.3in 0in 2.8in 0in]{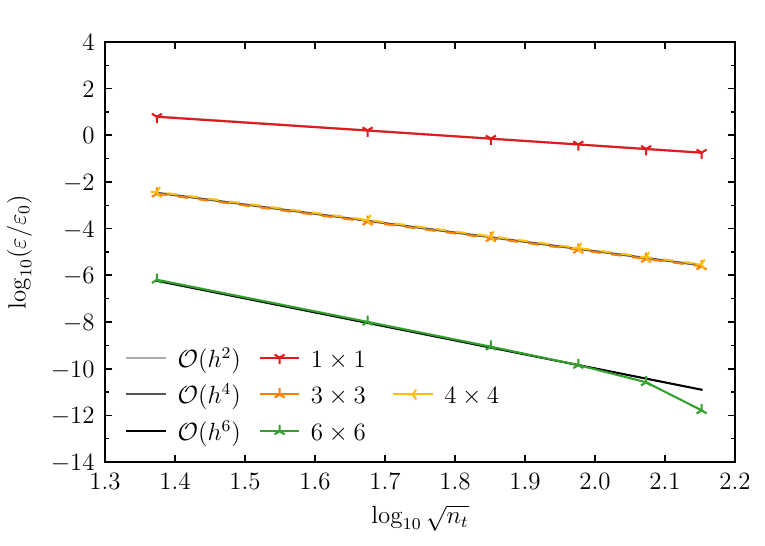}
\caption{$\mathbf{B}_1\ne\mathbf{0}$, $\mathbf{B}_2=\mathbf{0}$, $n_q^b=\bar{n}_q^b$\vpad}
\label{fig:part5a_1}
\end{subfigure}
\hspace{0.25em}
\begin{subfigure}[b]{.49\textwidth}
\includegraphics[scale=.64,clip=true,trim=2.3in 0in 2.8in 0in]{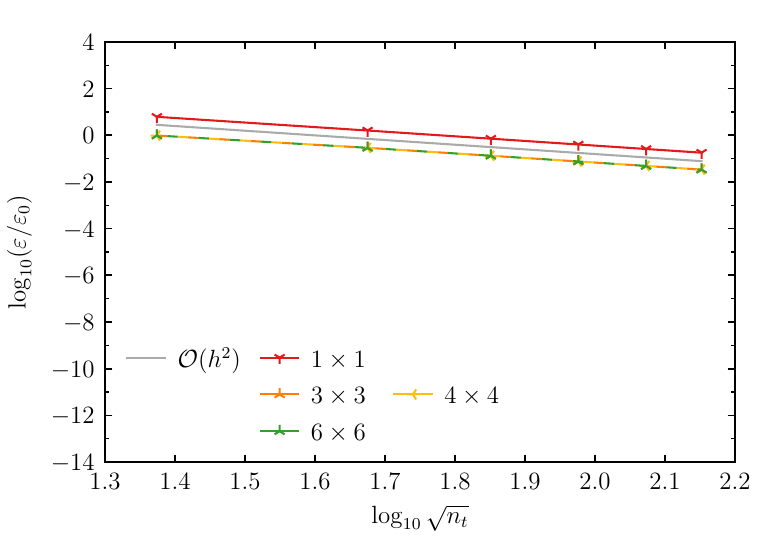}
\caption{$\mathbf{B}_1\ne\mathbf{0}$, $\mathbf{B}_2=\mathbf{0}$, $n_q^b=1$ \vpad}
\label{fig:part5a_1b}
\end{subfigure}
\\
\begin{subfigure}[b]{.49\textwidth}
\includegraphics[scale=.64,clip=true,trim=2.3in 0in 2.8in 0in]{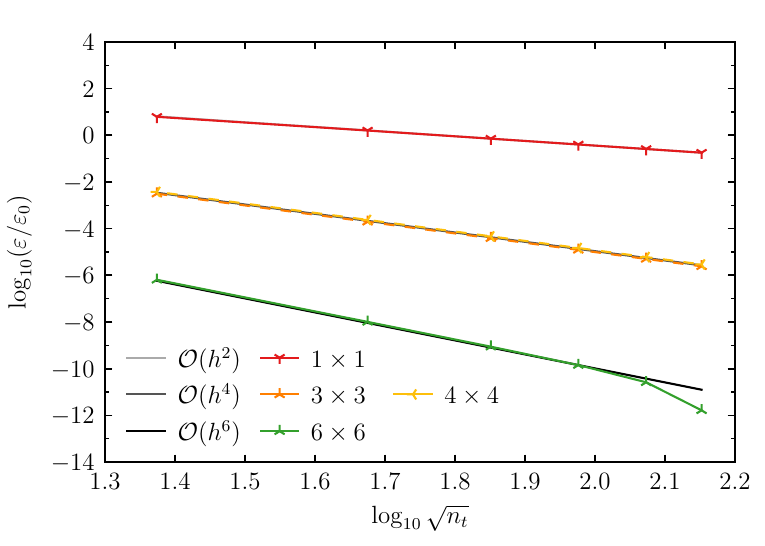}
\caption{$\mathbf{B}_1=\mathbf{0}$, $\mathbf{B}_2\ne\mathbf{0}$, $n_q^b=\bar{n}_q^b$\vpad}
\label{fig:part5a_2}
\end{subfigure}
\hspace{0.25em}
\begin{subfigure}[b]{.49\textwidth}
\includegraphics[scale=.64,clip=true,trim=2.3in 0in 2.8in 0in]{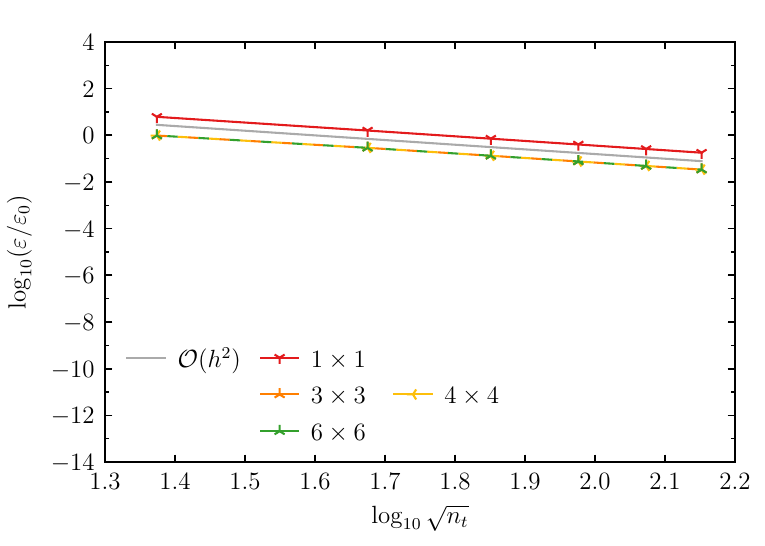}
\caption{$\mathbf{B}_1=\mathbf{0}$, $\mathbf{B}_2\ne\mathbf{0}$, $n_q^b=1$\vpad}
\label{fig:part5a_2b}
\end{subfigure}
\\
\begin{subfigure}[b]{.49\textwidth}
\includegraphics[scale=.64,clip=true,trim=2.3in 0in 2.8in 0in]{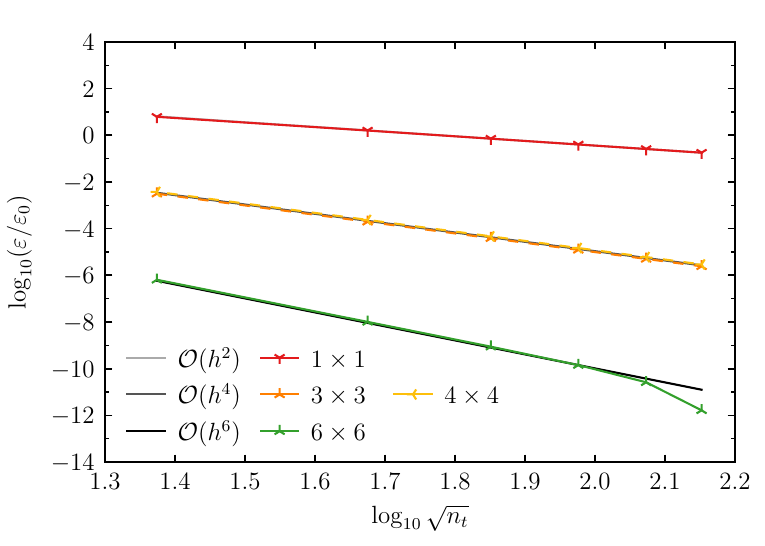}
\caption{$\mathbf{B}_1\ne\mathbf{0}$, $\mathbf{B}_2\ne\mathbf{0}$, $n_q^b=\bar{n}_q^b$\vpad}
\label{fig:part5a_3}
\end{subfigure}
\hspace{0.25em}
\begin{subfigure}[b]{.49\textwidth}
\includegraphics[scale=.64,clip=true,trim=2.3in 0in 2.8in 0in]{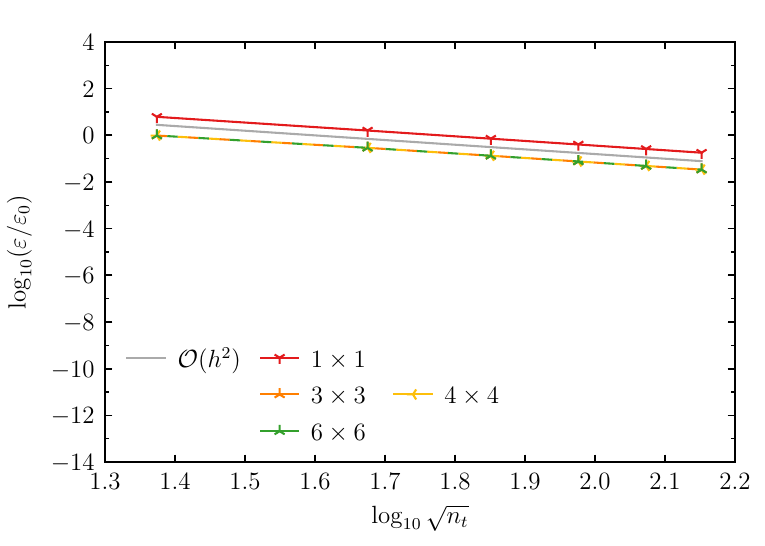}
\caption{$\mathbf{B}_1\ne\mathbf{0}$, $\mathbf{B}_2\ne\mathbf{0}$, $n_q^b=1$\vpad}
\label{fig:part5a_3b}
\end{subfigure}
\caption{Numerical-integration error: $\varepsilon=|e_a|$~\eqref{eq:a_error_cancel} for $G_2$ and $d_1$ with different amounts of quadrature points.}
\vskip-\dp\strutbox
\label{fig:part5a}
\end{figure}

\begin{figure}%[!t]
\centering
\begin{subfigure}[b]{.49\textwidth}
\includegraphics[scale=.64,clip=true,trim=2.3in 0in 2.8in 0in]{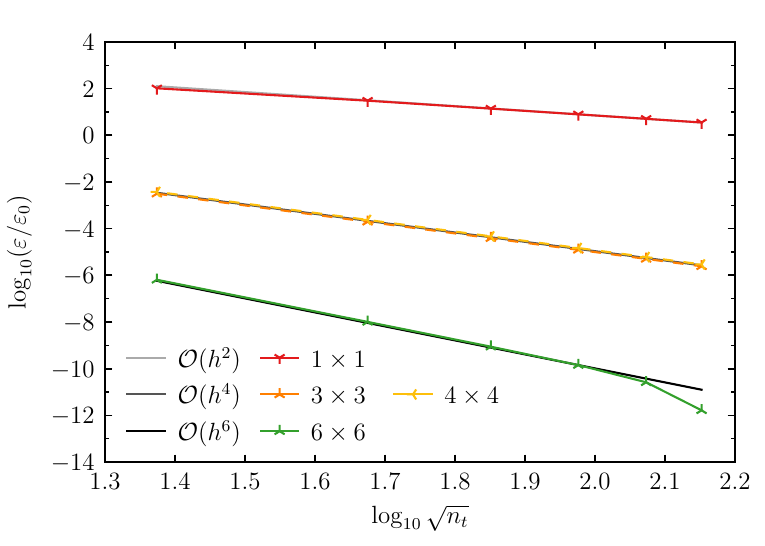}
\caption{$\mathbf{B}_1\ne\mathbf{0}$, $\mathbf{B}_2=\mathbf{0}$, $n_q^b=\bar{n}_q^b$\vpad}
\label{fig:part5b_1}
\end{subfigure}
\hspace{0.25em}
\begin{subfigure}[b]{.49\textwidth}
\includegraphics[scale=.64,clip=true,trim=2.3in 0in 2.8in 0in]{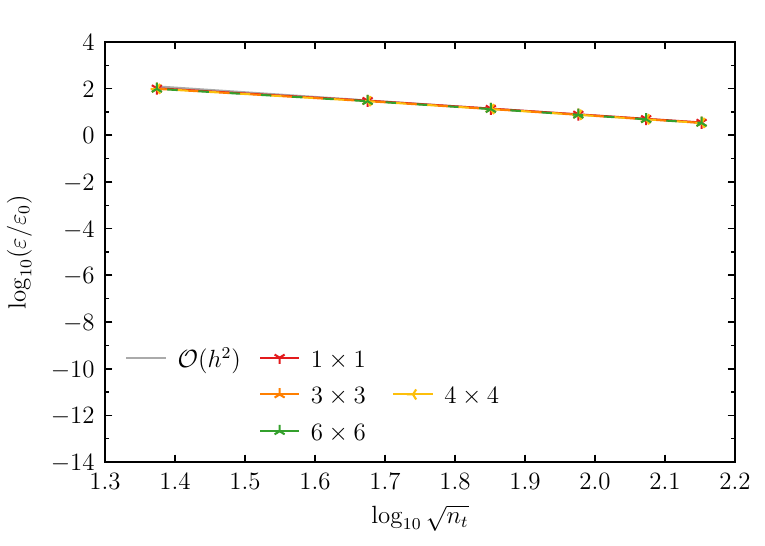}
\caption{$\mathbf{B}_1\ne\mathbf{0}$, $\mathbf{B}_2=\mathbf{0}$, $n_q^b=1$ \vpad}
\label{fig:part5b_1b}
\end{subfigure}
\\
\begin{subfigure}[b]{.49\textwidth}
\includegraphics[scale=.64,clip=true,trim=2.3in 0in 2.8in 0in]{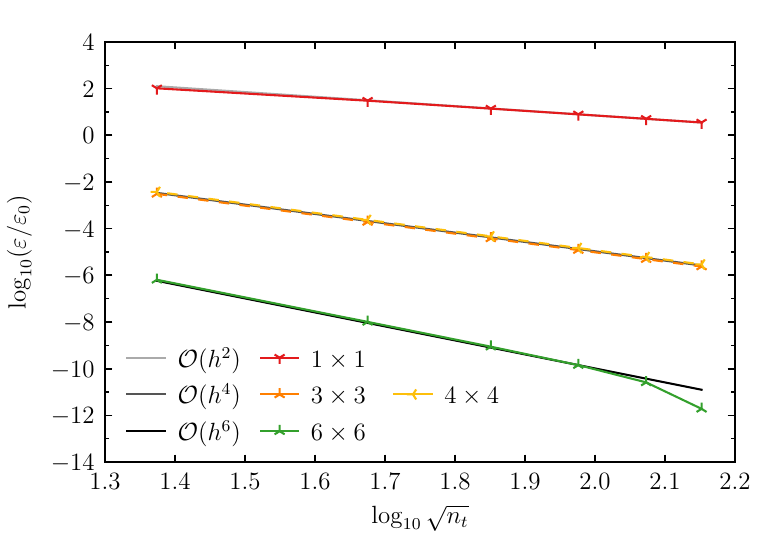}
\caption{$\mathbf{B}_1=\mathbf{0}$, $\mathbf{B}_2\ne\mathbf{0}$, $n_q^b=\bar{n}_q^b$\vpad}
\label{fig:part5b_2}
\end{subfigure}
\hspace{0.25em}
\begin{subfigure}[b]{.49\textwidth}
\includegraphics[scale=.64,clip=true,trim=2.3in 0in 2.8in 0in]{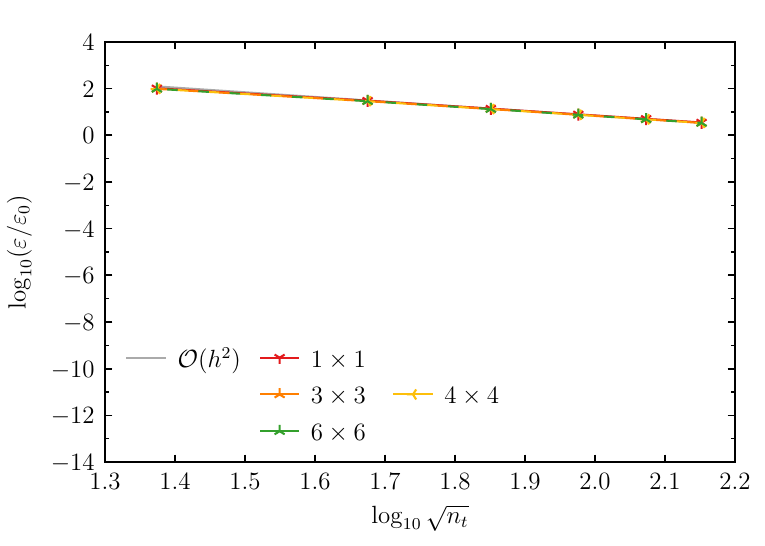}
\caption{$\mathbf{B}_1=\mathbf{0}$, $\mathbf{B}_2\ne\mathbf{0}$, $n_q^b=1$\vpad}
\label{fig:part5b_2b}
\end{subfigure}
\\
\begin{subfigure}[b]{.49\textwidth}
\includegraphics[scale=.64,clip=true,trim=2.3in 0in 2.8in 0in]{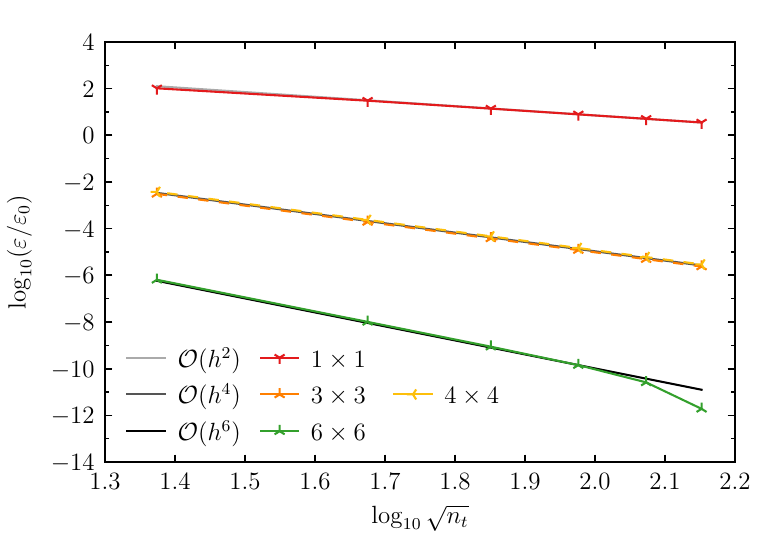}
\caption{$\mathbf{B}_1\ne\mathbf{0}$, $\mathbf{B}_2\ne\mathbf{0}$, $n_q^b=\bar{n}_q^b$\vpad}
\label{fig:part5b_3}
\end{subfigure}
\hspace{0.25em}
\begin{subfigure}[b]{.49\textwidth}
\includegraphics[scale=.64,clip=true,trim=2.3in 0in 2.8in 0in]{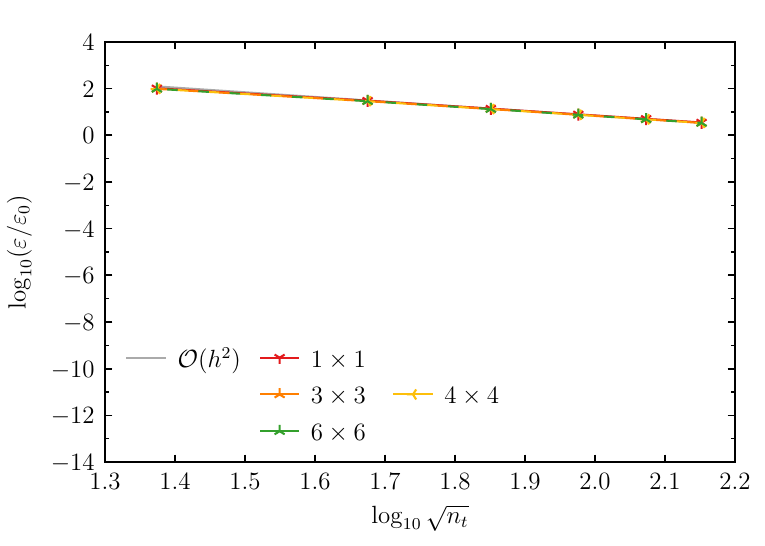}
\caption{$\mathbf{B}_1\ne\mathbf{0}$, $\mathbf{B}_2\ne\mathbf{0}$, $n_q^b=1$\vpad}
\label{fig:part5b_3b}
\end{subfigure}
\caption{Numerical-integration error: $\varepsilon=|e_a|$~\eqref{eq:a_error_cancel} for $G_2$ and $d_3$ with different amounts of quadrature points.}
\vskip-\dp\strutbox
\label{fig:part5b}
\end{figure}

To isolate and measure the numerical-integration error, we perform the assessments described in Section~\ref{sec:nie}.  For $G_2$, Figures~\ref{fig:part5a} and~\ref{fig:part5b} show the numerical-integration error $e_a$~\eqref{eq:a_error_cancel} for $d_1$ and $d_3$.  
\reviewerTwo{We consider different amounts of triangle quadrature points for each simulation.  The simulation entries in the legends take the form $n_q^t\times n_q^s$, where $n_q^t$ and $n_q^s$ respectively denote the amounts of quadrature points used to evaluate the test and source integrals.}
%In the legend entries, the first number is the amount of triangle quadrature points used to evaluate the test integrals, whereas the second is the amount used to evaluate the source integrals.  
The numerical-integration error is nondimensionalized by the constant $\varepsilon_0=1$~A$\cdot$V.  
For the subfigures in the left columns of Figures~\ref{fig:part5a} and~\ref{fig:part5b}, the number of bar quadrature points is chosen to match the convergence rates of the triangle quadrature points ($n_q^b=\bar{n}_q^b$).  \reviewerTwo{The entries in the left column of the legends are for reference convergence rates.  The simulation entries in a given row are expected to have the same convergence rates as the reference rate, as listed in Table~\ref{tab:quadrature_properties}.}
Because $\mathbf{B}_1$ is exactly evaluated with $n_q^b=1$, the errors for $\mathbf{B}_2\ne\mathbf{0}$ are the same when $\mathbf{B}_1=\mathbf{0}$ and $\mathbf{B}_1\ne\mathbf{0}$.
Each of the solutions in the left columns of Figures~\ref{fig:part5a} and~\ref{fig:part5b} converges at the expected rate.  For the finest meshes considered, the round-off error arising from the double-precision calculations exceeds the numerical-integration error.
To test the ability to detect a coding error, we set $n_q^b=1$ for all of the cases for the subfigures in the right columns of Figures~\ref{fig:part5a} and~\ref{fig:part5b}.  The cases with the coding error all have convergence rates that are $\mathcal{O}(h^2)$.  Therefore, this approach detects the coding error.

\begin{figure}%[!t]
\centering
\begin{subfigure}[b]{.49\textwidth}
\includegraphics[scale=.64,clip=true,trim=2.3in 0in 2.8in 0in]{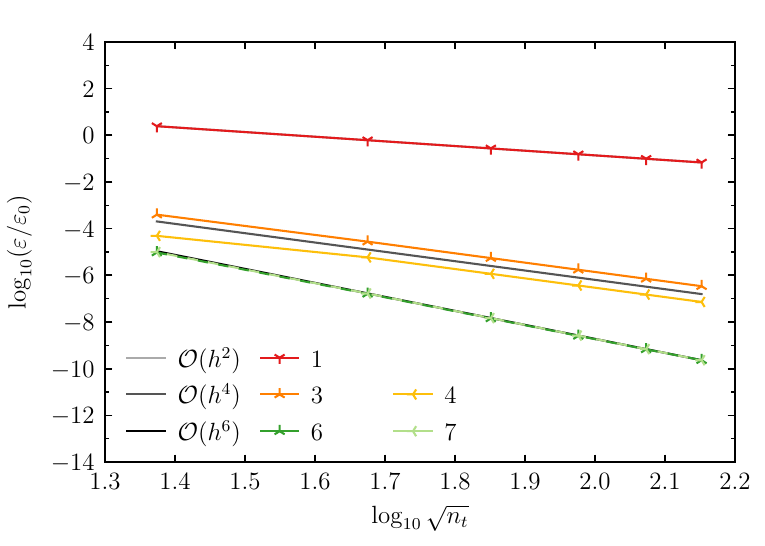}
\caption{$\mathbf{B}_1\ne\mathbf{0}$, $\mathbf{B}_2=\mathbf{0}$, $n_q^b=\bar{n}_q^b$\vpad}
\label{fig:part6a_1}
\end{subfigure}
\hspace{0.25em}
\begin{subfigure}[b]{.49\textwidth}
\includegraphics[scale=.64,clip=true,trim=2.3in 0in 2.8in 0in]{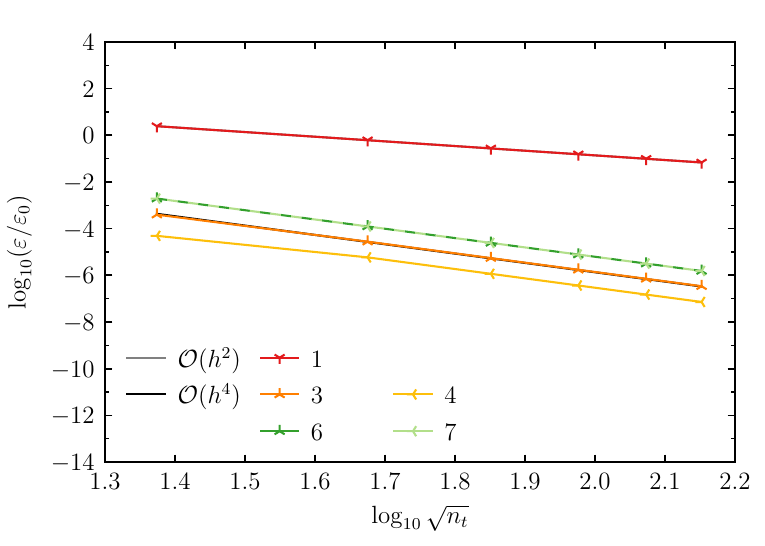}
\caption{$\mathbf{B}_1\ne\mathbf{0}$, $\mathbf{B}_2=\mathbf{0}$, $n_q^b=1$ for $1\times 1$, $n_q^b=2$ otherwise\vpad}
\label{fig:part6a_1b}
\end{subfigure}
\\
\begin{subfigure}[b]{.49\textwidth}
\includegraphics[scale=.64,clip=true,trim=2.3in 0in 2.8in 0in]{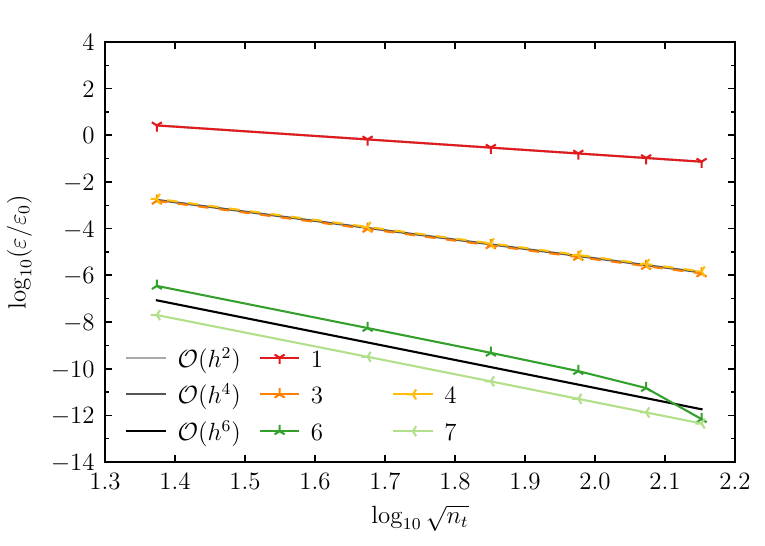}
\caption{$\mathbf{B}_1=\mathbf{0}$, $\mathbf{B}_2\ne\mathbf{0}$, $n_q^b=\bar{n}_q^b$\vpad}
\label{fig:part6a_2}
\end{subfigure}
\hspace{0.25em}
\begin{subfigure}[b]{.49\textwidth}
\includegraphics[scale=.64,clip=true,trim=2.3in 0in 2.8in 0in]{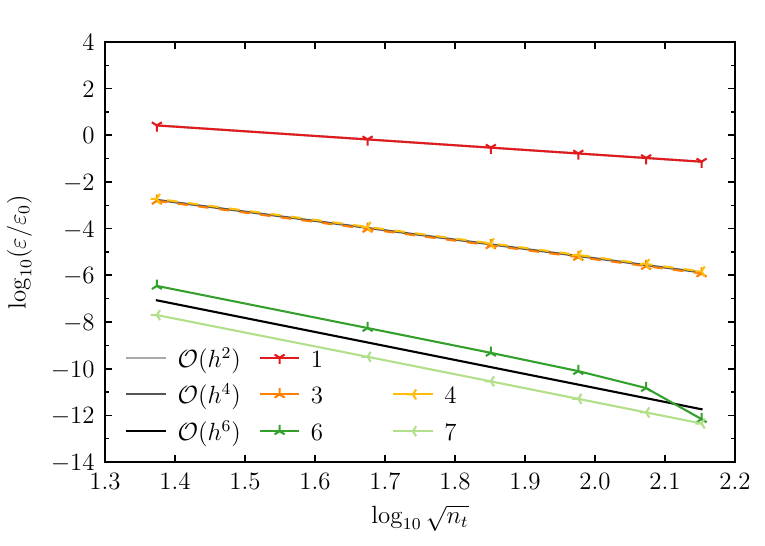}
\caption{$\mathbf{B}_1=\mathbf{0}$, $\mathbf{B}_2\ne\mathbf{0}$, $n_q^b=1$ for $1\times 1$, $n_q^b=2$ otherwise\vpad}
\label{fig:part6a_2b}
\end{subfigure}
\\
\begin{subfigure}[b]{.49\textwidth}
\includegraphics[scale=.64,clip=true,trim=2.3in 0in 2.8in 0in]{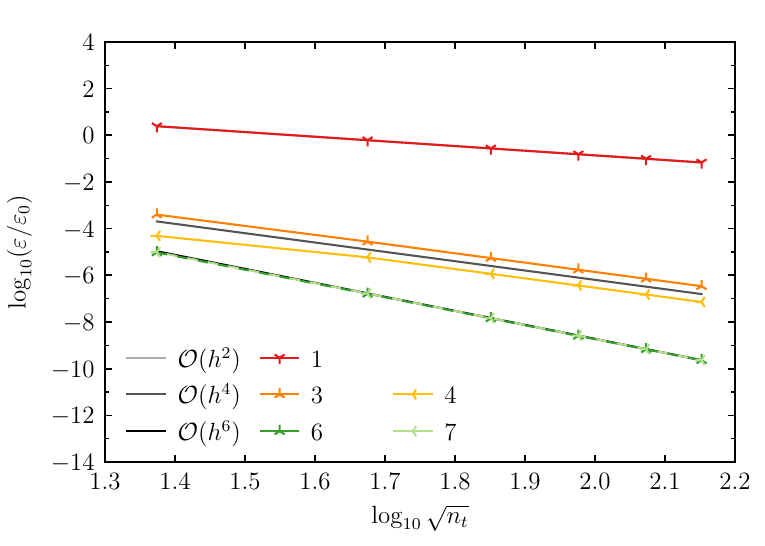}
\caption{$\mathbf{B}_1\ne\mathbf{0}$, $\mathbf{B}_2\ne\mathbf{0}$, $n_q^b=\bar{n}_q^b$\vpad}
\label{fig:part6a_3}
\end{subfigure}
\hspace{0.25em}
\begin{subfigure}[b]{.49\textwidth}
\includegraphics[scale=.64,clip=true,trim=2.3in 0in 2.8in 0in]{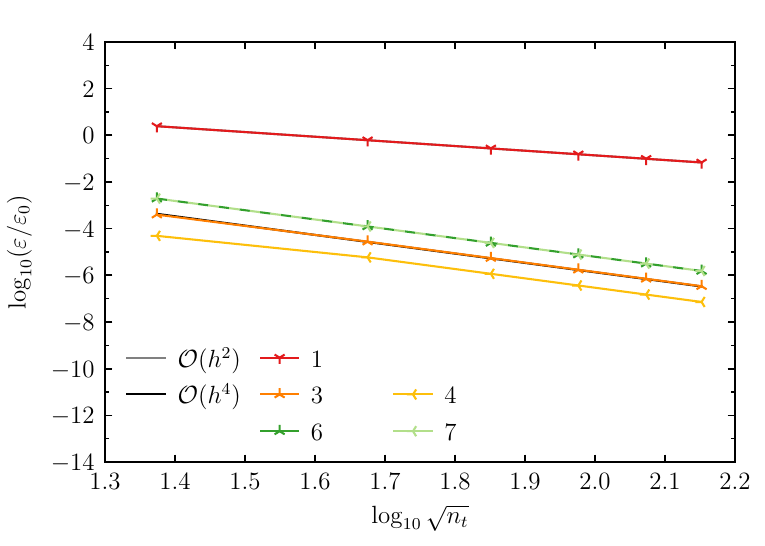}
\caption{$\mathbf{B}_1\ne\mathbf{0}$, $\mathbf{B}_2\ne\mathbf{0}$, $n_q^b=1$ for $1\times 1$, $n_q^b=2$ otherwise\vpad}
\label{fig:part6a_3b}
\end{subfigure}
\caption{Numerical-integration error: $\varepsilon=|e_b|$~\eqref{eq:b_error_cancel} for $G_2$ and $d_1$ with different amounts of quadrature points.}
\vskip-\dp\strutbox
\label{fig:part6a}
\end{figure}

\begin{figure}%[!t]
\centering
\begin{subfigure}[b]{.49\textwidth}
\includegraphics[scale=.64,clip=true,trim=2.3in 0in 2.8in 0in]{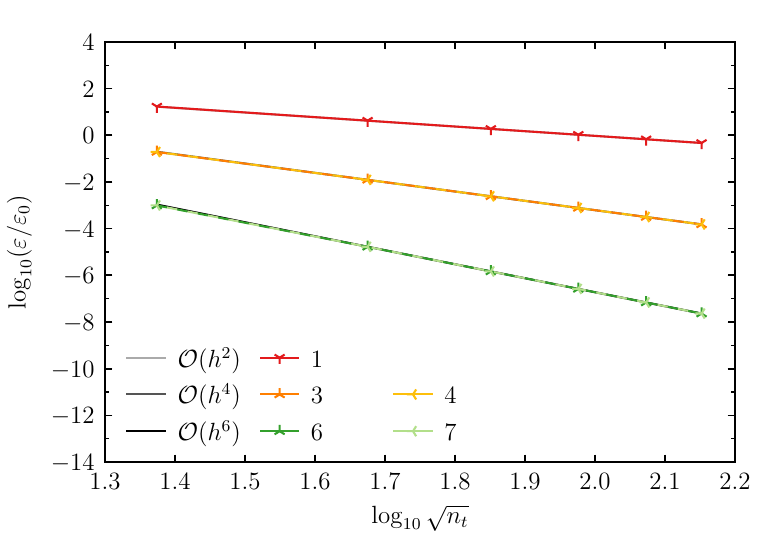}
\caption{$\mathbf{B}_1\ne\mathbf{0}$, $\mathbf{B}_2=\mathbf{0}$, $n_q^b=\bar{n}_q^b$\vpad}
\label{fig:part6b_1}
\end{subfigure}
\hspace{0.25em}
\begin{subfigure}[b]{.49\textwidth}
\includegraphics[scale=.64,clip=true,trim=2.3in 0in 2.8in 0in]{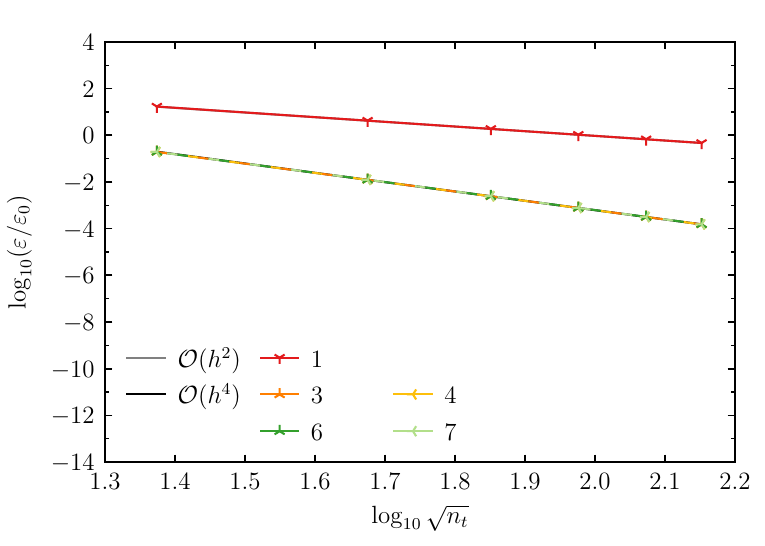}
\caption{$\mathbf{B}_1\ne\mathbf{0}$, $\mathbf{B}_2=\mathbf{0}$, $n_q^b=1$ for $1\times 1$, $n_q^b=2$ otherwise\vpad}
\label{fig:part6b_1b}
\end{subfigure}
\\
\begin{subfigure}[b]{.49\textwidth}
\includegraphics[scale=.64,clip=true,trim=2.3in 0in 2.8in 0in]{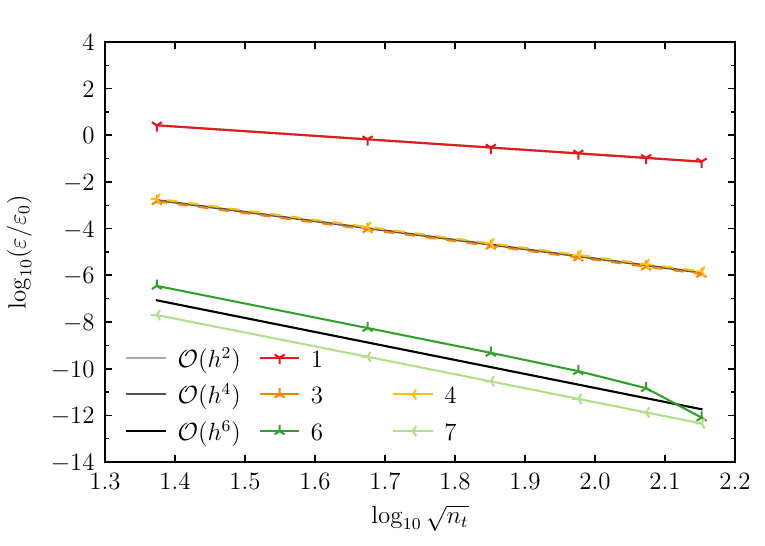}
\caption{$\mathbf{B}_1=\mathbf{0}$, $\mathbf{B}_2\ne\mathbf{0}$, $n_q^b=\bar{n}_q^b$\vpad}
\label{fig:part6b_2}
\end{subfigure}
\hspace{0.25em}
\begin{subfigure}[b]{.49\textwidth}
\includegraphics[scale=.64,clip=true,trim=2.3in 0in 2.8in 0in]{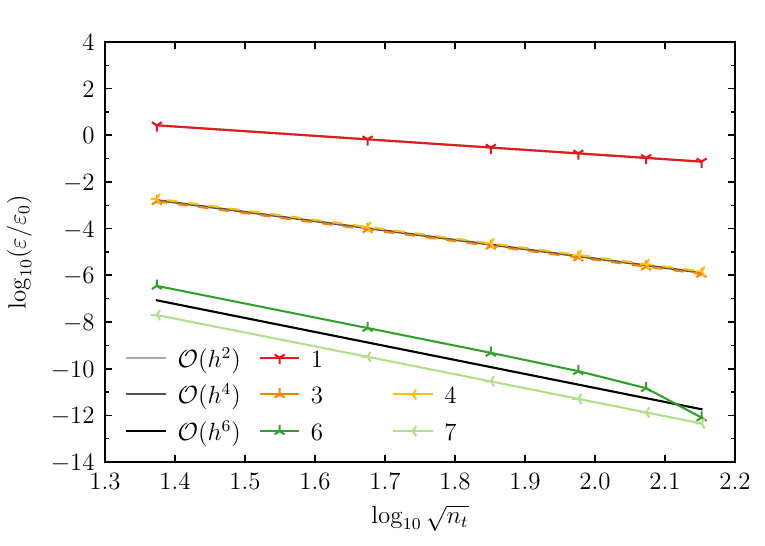}
\caption{$\mathbf{B}_1=\mathbf{0}$, $\mathbf{B}_2\ne\mathbf{0}$, $n_q^b=1$ for $1\times 1$, $n_q^b=2$ otherwise\vpad}
\label{fig:part6b_2b}
\end{subfigure}
\\
\begin{subfigure}[b]{.49\textwidth}
\includegraphics[scale=.64,clip=true,trim=2.3in 0in 2.8in 0in]{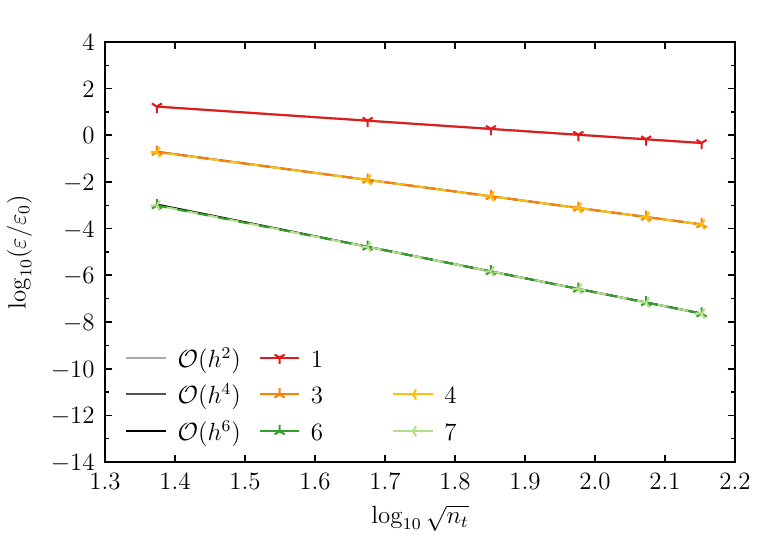}
\caption{$\mathbf{B}_1\ne\mathbf{0}$, $\mathbf{B}_2\ne\mathbf{0}$, $n_q^b=\bar{n}_q^b$\vpad}
\label{fig:part6b_3}
\end{subfigure}
\hspace{0.25em}
\begin{subfigure}[b]{.49\textwidth}
\includegraphics[scale=.64,clip=true,trim=2.3in 0in 2.8in 0in]{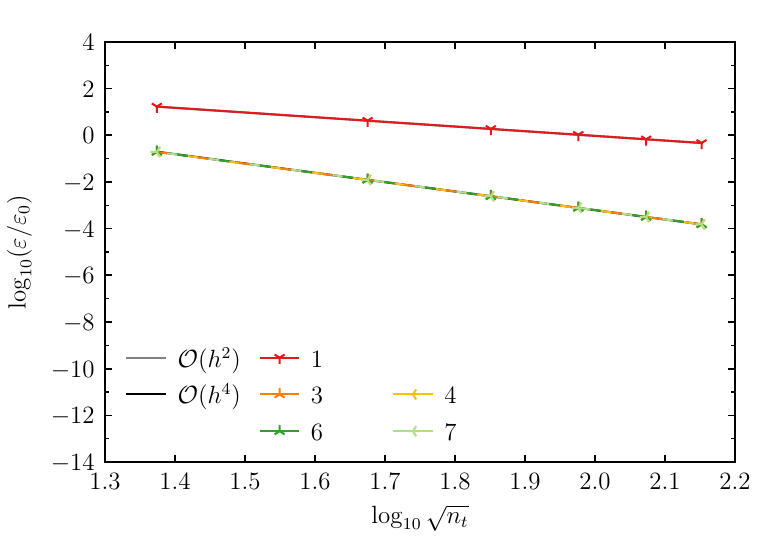}
\caption{$\mathbf{B}_1\ne\mathbf{0}$, $\mathbf{B}_2\ne\mathbf{0}$, $n_q^b=1$ for $1\times 1$, $n_q^b=2$ otherwise\vpad}
\label{fig:part6b_3b}
\end{subfigure}
\caption{Numerical-integration error: $\varepsilon=|e_b|$~\eqref{eq:b_error_cancel} for $G_2$ and $d_3$ with different amounts of quadrature points.}
\vskip-\dp\strutbox
\label{fig:part6b}
\end{figure}

For $G_2$, Figures~\ref{fig:part6a} and~\ref{fig:part6b} show the numerical-integration error $e_b$~\eqref{eq:b_error_cancel} for $d_1$ and $d_3$.  In the legend entries, the number is the amount of triangle quadrature points used to evaluate the test integrals. 
%{Additionally, in the captions and in this paragraph, the status of $\mathbf{B}_1$ and $\mathbf{B}_2$ being zero or nonzero refers to the corresponding terms in   } 
For the subfigures in the left columns of Figures~\ref{fig:part6a} and~\ref{fig:part6b}, the number of one-dimensional quadrature points is $n_q^b=\bar{n}_q^b$.  
Each of the solutions in the left column of Figures~\ref{fig:part6a} and~\ref{fig:part6b} converges at the expected rate listed in Table~\ref{tab:quadrature_properties} until the round-off error exceeds the numerical-integration error.  
To test the ability to detect a coding error, we set $n_q^b=2$ for the cases where $\bar{n}_q^b>2$ in the right columns of Figures~\ref{fig:part6a} and~\ref{fig:part6b}.  The cases with the coding error have convergence rates limited to $\mathcal{O}(h^4)$ when $\mathbf{B}_1\ne\mathbf{0}$.  When $\mathbf{B}_1=\mathbf{0}$, $n_q^b$ is not used to compute $\be\bigl(\mathbf{E}^\mathcal{I}, \boldsymbol{\Lambda}_{\testidx}\bigr)$.  Therefore, this approach detects the coding error.

%===============================================================================
\section{Conclusions} %=========================================================
%===============================================================================
\label{sec:conclusions}

In this paper, we presented code-verification approaches for the method-of-moments implementation of the electric-field integral equation and a thick slot model to isolate and measure the solution-discretization error and numerical-integration error.  
We manufactured the surface current density, which yielded a source term that we could treat as a manufactured incident field in the EFIE.  Given the manufactured surface current, we were able to obtain an analytic expression for the magnetic current that did not require a source term in the slot equation.

We isolated and measured the solution-discretization error by integrating exactly over the domain. To integrate exactly, we manufactured the Green’s function in terms of even powers of the distance between the test and source points.  
On each surface, the interaction between the wire and the surface introduced a line discontinuity, which contaminated convergence studies.  We mitigated this problem by removing the discontinuity using other entries from the matrix that undergo code verification.  We additionally kept the discontinuity and varied the interaction between the discretization errors to demonstrate the implications.

To isolate the numerical-integration error, we removed the solution-discretization error by canceling the basis-function contribution.   We demonstrated the ability to detect a coding error on both sides of the equations. 

For both approaches, we performed \reviewerTwo{convergence} studies for a variety of cases for which we achieved the expected orders of accuracy.

%===============================================================================
\section*{Acknowledgments} %====================================================
%===============================================================================
\label{sec:acknowledgments}

This article has been authored by employees of National Technology \& Engineering Solutions of Sandia, LLC under Contract No.~DE-NA0003525 with the U.S.~Department of Energy (DOE). The employees own all right, title, and interest in and to the article and are solely responsible for its contents. The United States Government retains and the publisher, by accepting the article for publication, acknowledges that the United States Government retains a non-exclusive, paid-up, irrevocable, world-wide license to publish or reproduce the published form of this article or allow others to do so, for United States Government purposes. The DOE will provide public access to these results of federally sponsored research in accordance with the DOE Public Access Plan \url{https://www.energy.gov/downloads/doe-public-access-plan}.

\addcontentsline{toc}{section}{\refname}
\bibliographystyle{elsarticle-num}
\bibliography{../quadrature_manuscript/quadrature.bib}

\begin{thebibliography}{10}
\expandafter\ifx\csname url\endcsname\relax
  \def\url#1{\texttt{#1}}\fi
\expandafter\ifx\csname urlprefix\endcsname\relax\def\urlprefix{URL }\fi
\expandafter\ifx\csname href\endcsname\relax
  \def\href#1#2{#2} \def\path#1{#1}\fi

\bibitem{graglia_1993}
R.~D. Graglia, On the numerical integration of the linear shape functions times
  the {3-D} {Green's} function or its gradient on a plane triangle, {IEEE}
  Transactions on Antennas and Propagation 41~(10) (1993) 1448--1455.
\newblock \href {https://doi.org/10.1109/8.247786}
  {\path{doi:10.1109/8.247786}}.

\bibitem{wilton_1984}
D.~{Wilton}, S.~{Rao}, A.~{Glisson}, D.~{Schaubert}, O.~{Al-Bundak},
  C.~{Butler}, Potential integrals for uniform and linear source distributions
  on polygonal and polyhedral domains, {IEEE} Transactions on Antennas and
  Propagation 32~(3) (1984) 276--281.
\newblock \href {https://doi.org/10.1109/TAP.1984.1143304}
  {\path{doi:10.1109/TAP.1984.1143304}}.

\bibitem{rao_1982}
S.~{Rao}, D.~{Wilton}, A.~{Glisson}, Electromagnetic scattering by surfaces of
  arbitrary shape, {IEEE} Transactions on Antennas and Propagation 30~(3)
  (1982) 409--418.
\newblock \href {https://doi.org/10.1109/TAP.1982.1142818}
  {\path{doi:10.1109/TAP.1982.1142818}}.

\bibitem{khayat_2005}
M.~A. Khayat, D.~R. Wilton, Numerical evaluation of singular and near-singular
  potential integrals, {IEEE} Transactions on Antennas and Propagation 53~(10)
  (2005) 3180--3190.
\newblock \href {https://doi.org/10.1109/TAP.2005.856342}
  {\path{doi:10.1109/TAP.2005.856342}}.

\bibitem{fink_2008}
P.~W. Fink, D.~R. Wilton, M.~A. Khayat, Simple and efficient numerical
  evaluation of near-hypersingular integrals, {IEEE} Antennas and Wireless
  Propagation Letters 7 (2008) 469--472.
\newblock \href {https://doi.org/10.1109/LAWP.2008.2000788}
  {\path{doi:10.1109/LAWP.2008.2000788}}.

\bibitem{khayat_2008}
M.~A. Khayat, D.~R. Wilton, P.~W. Fink, An improved transformation and
  optimized sampling scheme for the numerical evaluation of singular and
  near-singular potentials, {IEEE} Antennas and Wireless Propagation Letters 7
  (2008) 377--380.
\newblock \href {https://doi.org/10.1109/LAWP.2008.928461}
  {\path{doi:10.1109/LAWP.2008.928461}}.

\bibitem{vipiana_2011}
F.~Vipiana, D.~R. Wilton, Optimized numerical evaluation of singular and
  near-singular potential integrals involving junction basis functions, {IEEE}
  Transactions on Antennas and Propagation 59~(1) (2011) 162--171.
\newblock \href {https://doi.org/10.1109/TAP.2010.2090464}
  {\path{doi:10.1109/TAP.2010.2090464}}.

\bibitem{vipiana_2012}
F.~Vipiana, D.~R. Wilton, Numerical evaluation via singularity cancellation
  schemes of near-singular integrals involving the gradient of {Helmholtz}-type
  potentials, {IEEE} Transactions on Antennas and Propagation 61~(3) (2013)
  1255--1265.
\newblock \href {https://doi.org/10.1109/TAP.2012.2227922}
  {\path{doi:10.1109/TAP.2012.2227922}}.

\bibitem{botha_2013}
M.~M. Botha, A family of augmented {D}uffy transformations for near-singularity
  cancellation quadrature, {IEEE} Transactions on Antennas and Propagation
  61~(6) (2013) 3123--3134.
\newblock \href {https://doi.org/10.1109/TAP.2013.2252137}
  {\path{doi:10.1109/TAP.2013.2252137}}.

\bibitem{rivero_2019}
J.~{Rivero}, F.~{Vipiana}, D.~R. {Wilton}, W.~A. {Johnson}, Hybrid integration
  scheme for the evaluation of strongly singular and near-singular integrals in
  surface integral equations, {IEEE} Transactions on Antennas and Propagation
  67~(10) (2019).
\newblock \href {https://doi.org/10.1109/TAP.2019.2920333}
  {\path{doi:10.1109/TAP.2019.2920333}}.

\bibitem{vipiana_2013}
F.~Vipiana, D.~R. Wilton, W.~A. Johnson, Advanced numerical schemes for the
  accurate evaluation of {4-D} reaction integrals in the method of moments,
  {IEEE} Transactions on Antennas and Propagation 61~(11) (2013) 5559--5566.
\newblock \href {https://doi.org/10.1109/TAP.2013.2277864}
  {\path{doi:10.1109/TAP.2013.2277864}}.

\bibitem{polimeridis_2013}
A.~G. {Polimeridis}, F.~{Vipiana}, J.~R. {Mosig}, D.~R. {Wilton}, {DIRECTFN}:
  Fully numerical algorithms for high precision computation of singular
  integrals in {Galerkin} {SIE} methods, {IEEE} Transactions on Antennas and
  Propagation 61~(6) (2013) 3112--3122.
\newblock \href {https://doi.org/10.1109/TAP.2013.2246854}
  {\path{doi:10.1109/TAP.2013.2246854}}.

\bibitem{wilton_2017}
D.~R. {Wilton}, F.~{Vipiana}, W.~A. {Johnson}, Evaluation of {4-D} reaction
  integrals in the method of moments: Coplanar element case, {IEEE}
  Transactions on Antennas and Propagation 65~(5) (2017) 2479--2493.
\newblock \href {https://doi.org/10.1109/TAP.2017.2677916}
  {\path{doi:10.1109/TAP.2017.2677916}}.

\bibitem{rivero_2019b}
J.~Rivero, F.~Vipiana, D.~R. Wilton, W.~A. Johnson, Evaluation of {4-D}
  reaction integrals via double application of the divergence theorem, {IEEE}
  Transactions on Antennas and Propagation 67~(2) (2019) 1131--1142.
\newblock \href {https://doi.org/10.1109/TAP.2018.2882589}
  {\path{doi:10.1109/TAP.2018.2882589}}.

\bibitem{freno_em}
B.~A. Freno, W.~A. Johnson, B.~F. Zinser, D.~F. Wilton, F.~Vipiana,
  S.~Campione, Characterization and integration of the singular test integrals
  in the method-of-moments implementation of the electric-field integral
  equation, Engineering Analysis with Boundary Elements 124 (2021) 185--193.
\newblock \href {https://doi.org/10.1016/j.enganabound.2020.12.015}
  {\path{doi:10.1016/j.enganabound.2020.12.015}}.

\bibitem{butler_1978}
C.~M. Butler, Y.~Rahmat-Samii, R.~Mittra, Electromagnetic penetration through
  apertures in conducting surfaces, {IEEE} Transactions on Electromagnetic
  Compatibility 20 (1978).
\newblock \href {https://doi.org/10.1109/TEMC.1978.303696}
  {\path{doi:10.1109/TEMC.1978.303696}}.

\bibitem{balanis_2012}
C.~A. Balanis, Advanced Engineering Electromagnetics, John Wiley \& Sons, Inc.,
  2012.

\bibitem{schelkunoff_1952}
S.~A. Schelkunoff, H.~T. Friis, Antennas: Theory and Practice, John Wiley \&
  Sons, Inc., 1952.

\bibitem{cerri_1992}
G.~Cerri, R.~D. Leo, V.~M. Primiani, Theoretical and experimental evaluation of
  the electromagnetic radiation from apertures in shielded enclosure, {IEEE}
  Transactions on Electromagnetic Compatibility 34 (1992).
\newblock \href {https://doi.org/10.1109/15.179275}
  {\path{doi:10.1109/15.179275}}.

\bibitem{robinson_1998}
M.~P. Robinson, T.~M. Benson, C.~Christopoulos, J.~F. Dawson, M.~Ganley,
  A.~Marvin, S.~Porter, D.~W. Thomas, Analytical formulation for the shielding
  effectiveness of enclosures with apertures, {IEEE} Transactions on
  Electromagnetic Compatibility 40 (1998).
\newblock \href {https://doi.org/10.1109/15.709422}
  {\path{doi:10.1109/15.709422}}.

\bibitem{araneo_2008}
R.~Araneo, G.~Lovat, An efficient {MoM} formulation for the evaluation of the
  shielding effectiveness of rectangular enclosures with thin and thick
  apertures, {IEEE} Transactions on Electromagnetic Compatibility 50 (2008).
\newblock \href {https://doi.org/10.1109/TEMC.2008.919031}
  {\path{doi:10.1109/TEMC.2008.919031}}.

\bibitem{hill_2009}
D.~A. Hill, Electromagnetic Fields in Cavities: Deterministic and Statistical
  Theories, Wiley--{IEEE} Press, 2009.
\newblock \href {https://doi.org/10.1002/9780470495056}
  {\path{doi:10.1002/9780470495056}}.

\bibitem{pozar_2011}
D.~M. Pozar, Microwave Engineering, John Wiley \& Sons, Inc., 2011.

\bibitem{campione_2020}
S.~Campione, L.~K. Warne, W.~L. Langston, R.~A. Pfeiffer, N.~Martin, J.~T.
  Williams, R.~K. Gutierrez, I.~C. Reines, J.~G. Huerta, V.~Q. Dang,
  Penetration through slots in cylindrical cavities operating at fundamental
  cavity modes, {IEEE} Transactions on Electromagnetic Compatibility 62 (2020).
\newblock \href {https://doi.org/10.1109/TEMC.2020.2977600}
  {\path{doi:10.1109/TEMC.2020.2977600}}.

\bibitem{illescas_2023}
M.~Illescas, Improved experimental validation of an electromagnetic subcell
  model for narrow slots with depth, {Master's} thesis, University of New
  Mexico (May 2023).

\bibitem{warne_1990}
L.~Warne, K.~Chen, Slot apertures having depth and losses described by local
  transmission line theory, {IEEE} Transactions on Electromagnetic
  Compatibility 32~(3) (1990) 185--196.
\newblock \href {https://doi.org/10.1109/15.57112}
  {\path{doi:10.1109/15.57112}}.

\bibitem{warne_1992}
L.~Warne, K.~Chen, A simple transmission line model for narrow slot apertures
  having depth and losses, {IEEE} Transactions on Electromagnetic Compatibility
  34~(3) (1992) 173--182.
\newblock \href {https://doi.org/10.1109/15.155827}
  {\path{doi:10.1109/15.155827}}.

\bibitem{warne_1995}
L.~K. Warne, Eddy current power dissipation at sharp corners: Rectangular
  conductor examples, Electromagnetics 15~(3) (1995) 273--290.
\newblock \href {https://doi.org/10.1080/02726349508908419}
  {\path{doi:10.1080/02726349508908419}}.

\bibitem{johnson_2002}
W.~A. Johnson, L.~K. Warne, R.~E. Jorgenson, J.~D. Kotulski, H.~G. Hudson,
  S.~L. Stronach, Incorporation of slot subcell models into {EIGER} for
  treatment of high {Q} cavity coupling problems, Sandia Report SAND2002-2681J,
  Sandia National Laboratories (Jul. 2002).

\bibitem{roache_1998}
P.~J. Roache, Verification and Validation in Computational Science and
  Engineering, Hermosa Publishers, 1998.

\bibitem{knupp_2022}
P.~Knupp, K.~Salari, Verification of Computer Codes in Computational Science
  and Engineering, Chapman \& Hall/CRC, 2002.
\newblock \href {https://doi.org/10.1201/9781420035421}
  {\path{doi:10.1201/9781420035421}}.

\bibitem{oberkampf_2010}
W.~L. Oberkampf, C.~J. Roy, Verification and Validation in Scientific
  Computing, Cambridge University Press, 2010.
\newblock \href {https://doi.org/10.1017/cbo9780511760396}
  {\path{doi:10.1017/cbo9780511760396}}.

\bibitem{roache_2001}
P.~J. Roache, Code verification by the method of manufactured solutions,
  Journal of Fluids Engineering 124~(1) (2001) 4--10.
\newblock \href {https://doi.org/10.1115/1.1436090}
  {\path{doi:10.1115/1.1436090}}.

\bibitem{nishikawa_2022}
H.~Nishikawa, Analytical formulas for verification of aerodynamic force and
  moment computations, Journal of Computational Physics 466 (2022).
\newblock \href {https://doi.org/10.1016/j.jcp.2022.111408}
  {\path{doi:10.1016/j.jcp.2022.111408}}.

\bibitem{roy_2004}
C.~J. Roy, C.~C. Nelson, T.~M. Smith, C.~C. Ober, Verification of
  {Euler/Navier}--{Stokes} codes using the method of manufactured solutions,
  International Journal for Numerical Methods in Fluids 44~(6) (2004) 599--620.
\newblock \href {https://doi.org/10.1002/fld.660} {\path{doi:10.1002/fld.660}}.

\bibitem{bond_2007}
R.~B. Bond, C.~C. Ober, P.~M. Knupp, S.~W. Bova, Manufactured solution for
  computational fluid dynamics boundary condition verification, {AIAA} Journal
  45~(9) (2007) 2224--2236.
\newblock \href {https://doi.org/10.2514/1.28099} {\path{doi:10.2514/1.28099}}.

\bibitem{veluri_2010}
S.~Veluri, C.~Roy, E.~Luke, Comprehensive code verification for an unstructured
  finite volume {CFD} code, in: 48th {AIAA} Aerospace Sciences Meeting
  including the New Horizons Forum and Aerospace Exposition, American Institute
  of Aeronautics and Astronautics, 2010.
\newblock \href {https://doi.org/10.2514/6.2010-127}
  {\path{doi:10.2514/6.2010-127}}.

\bibitem{oliver_2012}
T.~Oliver, K.~Estacio-Hiroms, N.~Malaya, G.~Carey, Manufactured solutions for
  the {Favre}-averaged {Navier--Stokes} equations with eddy-viscosity
  turbulence models, in: 50th {AIAA} Aerospace Sciences Meeting including the
  New Horizons Forum and Aerospace Exposition, American Institute of
  Aeronautics and Astronautics, 2012.
\newblock \href {https://doi.org/10.2514/6.2012-80}
  {\path{doi:10.2514/6.2012-80}}.

\bibitem{eca_2016}
L.~E\c{c}a, C.~M. Klaij, G.~Vaz, M.~Hoekstra, F.~Pereira, On code verification
  of {RANS} solvers, Journal of Computational Physics 310 (2016) 418--439.
\newblock \href {https://doi.org/10.1016/j.jcp.2016.01.002}
  {\path{doi:10.1016/j.jcp.2016.01.002}}.

\bibitem{hennink_2021}
A.~Hennink, M.~Tiberga, D.~Lathouwers, A pressure-based solver for {low-Mach}
  number flow using a discontinuous {Galerkin} method, Journal of Computational
  Physics 425 (2022).
\newblock \href {https://doi.org/10.1016/j.jcp.2020.109877}
  {\path{doi:10.1016/j.jcp.2020.109877}}.

\bibitem{freno_2021}
B.~A. Freno, B.~R. Carnes, V.~G. Weirs, Code-verification techniques for
  hypersonic reacting flows in thermochemical nonequilibrium, Journal of
  Computational Physics 425 (2021).
\newblock \href {https://doi.org/10.1016/j.jcp.2020.109752}
  {\path{doi:10.1016/j.jcp.2020.109752}}.

\bibitem{chamberland_2010}
{\'E}.~Chamberland, A.~Fortin, M.~Fortin, Comparison of the performance of some
  finite element discretizations for large deformation elasticity problems,
  Computers \& Structures 88~(11) (2010) 664 -- 673.
\newblock \href {https://doi.org/10.1016/j.compstruc.2010.02.007}
  {\path{doi:10.1016/j.compstruc.2010.02.007}}.

\bibitem{etienne_2012}
S.~{\'E}tienne, A.~Garon, D.~Pelletier, Some manufactured solutions for
  verification of fluid--structure interaction codes, Computers \& Structures
  106--107 (2012) 56--67.
\newblock \href {https://doi.org/10.1016/j.compstruc.2012.04.006}
  {\path{doi:10.1016/j.compstruc.2012.04.006}}.

\bibitem{bukac_2023}
M.~Buka{\v c}, G.~Fu, A.~Seboldt, C.~Trenchea, Time-adaptive partitioned method
  for fluid--structure interaction problems with thick structures, Journal of
  Computational Physics 473 (2023).
\newblock \href {https://doi.org/10.1016/j.jcp.2022.111708}
  {\path{doi:10.1016/j.jcp.2022.111708}}.

\bibitem{veeraragavan_2016}
A.~Veeraragavan, J.~Beri, R.~J. Gollan, Use of the method of manufactured
  solutions for the verification of conjugate heat transfer solvers, Journal of
  Computational Physics 307 (2016) 308--320.
\newblock \href {https://doi.org/10.1016/j.jcp.2015.12.004}
  {\path{doi:10.1016/j.jcp.2015.12.004}}.

\bibitem{brady_2012}
P.~T. Brady, M.~Herrmann, J.~M. Lopez, Code verification for finite volume
  multiphase scalar equations using the method of manufactured solutions,
  Journal of Computational Physics 231~(7) (2012) 2924--2944.
\newblock \href {https://doi.org/10.1016/j.jcp.2011.12.040}
  {\path{doi:10.1016/j.jcp.2011.12.040}}.

\bibitem{lovato_2021}
S.~Lovato, S.~L. Toxopeus, J.~W. Settels, G.~H. Keetels, G.~Vaz, Code
  verification of non-{Newtonian} fluid solvers for single- and two-phase
  laminar flows, Journal of Verification, Validation and Uncertainty
  Quantification 6~(2) (2021).
\newblock \href {https://doi.org/10.1115/1.4050131}
  {\path{doi:10.1115/1.4050131}}.

\bibitem{mcclarren_2008}
R.~G. McClarren, R.~B. Lowrie, Manufactured solutions for the $p_1$
  radiation-hydrodynamics equations, Journal of Quantitative Spectroscopy and
  Radiative Transfer 109~(15) (2008) 2590--2602.
\newblock \href {https://doi.org/10.1016/j.jqsrt.2008.06.003}
  {\path{doi:10.1016/j.jqsrt.2008.06.003}}.

\bibitem{riva_2017}
F.~Riva, C.~F. Beadle, P.~Ricci, A methodology for the rigorous verification of
  particle-in-cell simulations, Physics of Plasmas 24 (2017).
\newblock \href {https://doi.org/10.1063/1.4977917}
  {\path{doi:10.1063/1.4977917}}.

\bibitem{tranquilli_2022}
P.~Tranquilli, L.~Ricketson, L.~Chac\'{o}n, A deterministic verification
  strategy for electrostatic particle-in-cell algorithms in arbitrary spatial
  dimensions using the method of manufactured solutions, Journal of
  Computational Physics 448 (2022).
\newblock \href {https://doi.org/10.1016/j.jcp.2021.110751}
  {\path{doi:10.1016/j.jcp.2021.110751}}.

\bibitem{rueda_2023}
A.~M. Rueda-Ram{\'i}rez, F.~J. Hindenlang, J.~Chan, G.~J. Gassner,
  Entropy-stable {Gauss} collocation methods for ideal magneto-hydrodynamics,
  Journal of Computational Physics 475 (2023).
\newblock \href {https://doi.org/10.1016/j.jcp.2022.111851}
  {\path{doi:10.1016/j.jcp.2022.111851}}.

\bibitem{rudi_2024}
J.~Rudi, M.~Heldman, E.~M. Constantinescu, Q.~Tang, X.-Z. Tang, Scalable
  implicit solvers with dynamic mesh adaptation for a relativistic
  drift-kinetic {Fokker}--{Planck}--{Boltzmann} model, Journal of Computational
  Physics 507 (2024).
\newblock \href {https://doi.org/10.1016/j.jcp.2024.112954}
  {\path{doi:10.1016/j.jcp.2024.112954}}.

\bibitem{amormartin_2021}
A.~Amor-Martin, L.~E. Garcia-Castillo, J.-F. Lee, Study of accuracy of a
  non-conformal finite element domain decomposition method, Journal of
  Computational Physics 429 (2021).
\newblock \href {https://doi.org/10.1016/j.jcp.2020.109989}
  {\path{doi:10.1016/j.jcp.2020.109989}}.

\bibitem{amar_2008}
A.~J. Amar, B.~F. Blackwell, J.~R. Edwards, One-dimensional ablation using a
  full {Newton's} method and finite control volume procedure, Journal of
  Thermophysics and Heat Transfer 22~(1) (2008) 71--82.
\newblock \href {https://doi.org/10.2514/1.29610} {\path{doi:10.2514/1.29610}}.

\bibitem{amar_2009}
A.~J. Amar, B.~F. Blackwell, J.~R. Edwards, Development and verification of a
  one-dimensional ablation code including pyrolysis gas flow, Journal of
  Thermophysics and Heat Transfer 23~(1) (2009) 59--71.
\newblock \href {https://doi.org/10.2514/1.36882} {\path{doi:10.2514/1.36882}}.

\bibitem{amar_2011}
A.~Amar, N.~Calvert, B.~Kirk, Development and verification of the charring
  ablating thermal protection implicit system solver, in: 49th AIAA Aerospace
  Sciences Meeting including the New Horizons Forum and Aerospace Exposition,
  2011.
\newblock \href {https://doi.org/10.2514/6.2011-144}
  {\path{doi:10.2514/6.2011-144}}.

\bibitem{freno_ablation}
B.~A. Freno, B.~R. Carnes, N.~R. Matula, Nonintrusive manufactured solutions
  for ablation, Physics of Fluids 33~(1) (2021).
\newblock \href {https://doi.org/10.1063/5.0037245}
  {\path{doi:10.1063/5.0037245}}.

\bibitem{freno_ablation_2022}
B.~A. Freno, B.~R. Carnes, V.~E. Brunini, N.~R. Matula, Nonintrusive
  manufactured solutions for non-decomposing ablation in two dimensions,
  Journal of Computational Physics 463 (2022).
\newblock \href {https://doi.org/10.1016/j.jcp.2022.111237}
  {\path{doi:10.1016/j.jcp.2022.111237}}.

\bibitem{marchand_2013}
R.~G. Marchand, The method of manufactured solutions for the verification of
  computational electromagnetic codes, {PhD} dissertation, Stellenbosch
  University (Mar. 2013).

\bibitem{marchand_2014}
R.~G. Marchand, D.~B. Davidson, Verification of the method-of-moment codes
  using the method of manufactured solutions, {IEEE} Transactions on
  Electromagnetic Compatibility 56~(4) (2014) 835--843.
\newblock \href {https://doi.org/10.1109/TEMC.2014.2325826}
  {\path{doi:10.1109/TEMC.2014.2325826}}.

\bibitem{freno_em_mms_2020}
B.~A. Freno, N.~R. Matula, W.~A. Johnson, Manufactured solutions for the
  method-of-moments implementation of the electric-field integral equation,
  Journal of Computational Physics 443 (2021).
\newblock \href {https://doi.org/10.1016/j.jcp.2021.110538}
  {\path{doi:10.1016/j.jcp.2021.110538}}.

\bibitem{freno_em_mms_quad_2021}
B.~A. Freno, N.~R. Matula, J.~I. Owen, W.~A. Johnson, Code-verification
  techniques for the method-of-moments implementation of the electric-field
  integral equation, Journal of Computational Physics 451 (2022).
\newblock \href {https://doi.org/10.1016/j.jcp.2021.110891}
  {\path{doi:10.1016/j.jcp.2021.110891}}.

\bibitem{freno_mfie_2022}
B.~A. Freno, N.~R. Matula, Code-verification techniques for the
  method-of-moments implementation of the magnetic-field integral equation,
  Journal of Computational Physics 478 (2023).
\newblock \href {https://doi.org/10.1016/j.jcp.2023.111959}
  {\path{doi:10.1016/j.jcp.2023.111959}}.

\bibitem{freno_cfie_2023}
B.~A. Freno, N.~R. Matula, Code-verification techniques for the
  method-of-moments implementation of the combined-field integral equation,
  Journal of Computational Physics 488 (2023).
\newblock \href {https://doi.org/10.1016/j.jcp.2023.112231}
  {\path{doi:10.1016/j.jcp.2023.112231}}.

\bibitem{warnick_2008}
K.~F. Warnick, Numerical Analysis for Electromagnetic Integral Equations,
  Artech House, 2008.

\bibitem{dangelo_2012}
C.~D'Angelo, Finite element approximation of elliptic problems with {Dirac}
  measure terms in weighted spaces: Applications to one- and three-dimensional
  coupled problems, {SIAM} Journal on Numerical Analysis 50~(1) (2012).
\newblock \href {https://doi.org/10.1137/100813853}
  {\path{doi:10.1137/100813853}}.

\bibitem{li_2021}
H.~Li, X.~Wan, P.~Yin, L.~Zhao, Regularity and finite element approximation for
  two-dimensional elliptic equations with line {Dirac} sources, Journal of
  Computational and Applied Mathematics 393 (2021).
\newblock \href {https://doi.org/10.1016/j.cam.2021.113518}
  {\path{doi:10.1016/j.cam.2021.113518}}.

\bibitem{lyness_1975}
J.~N. Lyness, D.~Jespersen, Moderate degree symmetric quadrature rules for the
  triangle, {IMA} Journal of Applied Mathematics 15~(1) (1975) 19--32.
\newblock \href {https://doi.org/10.1093/imamat/15.1.19}
  {\path{doi:10.1093/imamat/15.1.19}}.

\bibitem{dunavant_1985}
D.~A. Dunavant, High degree efficient symmetrical {G}aussian quadrature rules
  for the triangle, International Journal for Numerical Methods in Engineering
  21~(6) (1985) 1129--1148.
\newblock \href {https://doi.org/10.1002/nme.1620210612}
  {\path{doi:10.1002/nme.1620210612}}.

\bibitem{kahaner_1989}
D.~Kahaner, C.~Moler, S.~Nash, Numerical Methods and Software, Prentice Hall,
  1989.

\end{thebibliography}
%\end{document}

%\clearpage
%\appendix
%\renewcommand{\thesection}{Appendix~\Alph{section}}
%\input{appendix.tex}
%\input{questions.tex}
%%
%\input{slot_equations.tex}

\end{document}